\documentclass[twocolumn,pra,aps,superscriptaddress]{revtex4-1}

\usepackage{mathptmx}
\usepackage{subfigure}
\usepackage{dcolumn}
\usepackage{amsmath,amssymb}
\usepackage{bm}
\usepackage{color}
\usepackage{latexsym}
\usepackage{epstopdf}
\usepackage{color}
\usepackage[english]{babel}
\usepackage{latexsym}
\usepackage{stmaryrd}

\usepackage{psfrag,graphicx} 
\usepackage{epsf} 
\usepackage{subfigure} 
\usepackage{amsmath} 
\usepackage{amssymb} 
\usepackage{amsfonts}
\usepackage{bm}
\usepackage{natbib}
\usepackage{epstopdf}
\usepackage{appendix}

\definecolor{mygrey}{gray}{0.}
\definecolor{myblue}{rgb}{0.2,0.2,0.8}
\definecolor{myzard}{cmyk}{0,0,0.05,0}
\definecolor{mywhite}{rgb}{1,1,1}
\definecolor{myred}{rgb}{1,0.,0.3}

\usepackage[colorlinks=true,citecolor=myblue,linkcolor=myred]{hyperref}
\def\be{\begin{equation}}
\def\ee{\end{equation}}
\def\ba{\begin{align}}
\def\enda{\end{align}}
\def\bi{\begin{itemize}}
\def\ei{\end{itemize}}

 \def\ee{\mathord{\rm e}}
 
 \def\ii{\mathord{\rm i}}

\def\half{\textstyle\frac{1}{2}}

 \def\ee{\mathord{\rm e}}
 
 \def\ii{\mathord{\rm i}}

\def\half{\textstyle\frac{1}{2}}

\renewcommand{\ii}{{\rm i}}
\renewcommand{\ee}{{\rm e}}

\def\beq{\begin{equation}}
\def\beq{\begin{equation}}
\def\eeq{\end{equation}}

 \newcommand{\ket}[1]{|#1\rangle}
 \newcommand{\bra}[1]{\langle #1|}

\begin{document}

\title[Short Title]{Quantum Magnetism of Spin-Ladder Compounds with Trapped-Ion Crystals}

\author{A.~Bermudez}
\author{J.~Almeida}
\affiliation{Institut f\"ur Theoretische Physik, Albert-Einstein Alle 11, Universit\"at Ulm, 89069 Ulm, Germany}

\author{K.~Ott}
\author{H.~Kaufmann}
\author{S.~Ulm}
\author{U.~Poschinger}
\author{F.~Schmidt-Kaler}
\affiliation{Institut f\"ur Physik, Staudingerweg 7, Johannes Gutenberg-Universit\"at Mainz, 55099 Mainz, Germany}

\author{A.~Retzker}
\affiliation{Institut f\"ur Theoretische Physik, Albert-Einstein Alle 11, Universit\"at Ulm, 89069 Ulm, Germany}
\author{M. B.~Plenio}
\affiliation{Institut f\"ur Theoretische Physik, Albert-Einstein Alle 11, Universit\"at Ulm, 89069 Ulm, Germany}


\begin{abstract}
The quest for experimental platforms that allow for the exploration, and even control, of the interplay of low dimensionality and frustration is a fundamental challenge in several fields of quantum many-body physics, such as quantum magnetism. Here, we propose the use of cold  crystals of trapped ions  to study a variety of  frustrated quantum spin ladders. By optimizing the trap geometry, we show how to tailor the low dimensionality of the  models by changing the number of legs of the ladders. Combined with a method for selectively hiding of ions  provided by  laser addressing, it becomes possible to synthesize stripes of both triangular and Kagome lattices. Besides,  the degree of frustration of the phonon-mediated spin interactions  can be controlled by shaping the trap frequencies. We  support our theoretical considerations by initial experiments with planar ion crystals, where a high and tunable anisotropy of the radial trap frequencies is demonstrated.   We take into account an extensive list of  possible error sources under typical experimental conditions, and describe explicit regimes that guarantee the validity of our scheme.

\end{abstract}

\maketitle

\begingroup
\hypersetup{linkcolor=black}
\tableofcontents
\endgroup

\section{Introduction}

The transition from single-particle to many-body quantum  systems yields one of the most interesting concepts of physics: {\it emergence}. As emphasized by P. W. Anderson~\cite{emergence}, the laws that describe the collective behavior of interacting many-body systems can be fundamentally different from those that govern each of the individual particles. This effect leads to some of the most exotic phenomena  of condensed-matter physics in the last decades~\cite{wen_book}. Unfortunately, the  high complexity of  many-body systems turns our endeavor to understand such emergent phenomena into a fundamental challenge for both experimental and theoretical physics. Since exact analytical solutions  seldom exist, even for oversimplified models,  one is urged to develop efficient numerical methods. An alternative to this approach are the so-called {\it quantum simulations}~\cite{qs}, which make use of a particular quantum system that can be experimentally controlled to a high extent, in order to unveil the properties of a complicated interacting many-body model.

 There are two different strategies for the quantum simulation of many-body systems~\cite{qs_review}, the so-called analog and digital quantum simulators (QSs). On the one hand, an ideal analog QS would be a dedicated experimental device where one can prepare the quantum state, engineer a Hamiltonian of interest, and measure its properties. On the other hand, a digital QS  aims at  reproducing the dynamics of any given Hamiltonian  by concatenating a set of available quantum gates. Independently of  their particular experimental implementation, these types of QSs  would be capable of exploring  models from very different areas of physics,  ranging from condensed-matter to high-energy physics. Let us remark that this enterprise benefits directly from the  development of architectures for quantum-information processing~\cite{qs_insight}. In this work, we focus on one of these architectures: laser-cooled  atomic ions confined in radio-frequency traps~\cite{ion_trap_reviews}. Here,   some remarkable quantum simulations have already been accomplished in experiments, either in the digital~\cite{digital_qs} or analog~\cite{ising_ions,frustration_monroe,ising_monroe,dirac_gerritsma,ising_bollinger} versions. In particular, we shall concentrate on  analog QSs, where a number of theoretical proposals already exist~\cite{ion_qsimulator_reviews}. Some of these proposals target the quantum simulation of relativistic effects~\cite{relativistic_effects},  quantum spin models~\cite{ising_porras,spin_models_ti,frustration_duan,frustration_ulm}, many-body boson systems~\cite{bosons_ti}, spin-boson models~\cite{spin_boson_ti}, or theories of quantum transport and friction~\cite{frenkel_kontorova}.

A direction of research that is being actively explored is the QS of {\it magnetism}~\cite{ising_porras}, which yields a playground for a variety of cooperative quantum many-body effects. Even though the phenomenon of magnetic ordering was already known to the ancient greeks, the understanding of its microscopic origin could only be achieved after the development of quantum mechanics.    A  particular topic of recurring interest is the fate of the magnetically-ordered phases  in the presence of quantum fluctuations~\cite{lechemitant_book},  which get more pronounced as the dimensionality of the system is reduced.  Such quantum fluctuations may destroy the long-range order, favoring paramagnetic phases  and triggering the so-called {\it quantum phase transitions}~\cite{sachdev_book}. Alternatively, they may stabilize more interesting phases, such as the long-sought two-dimensional {\it quantum spin liquids}, which have connections with the theory of high-temperature superconductivity~\cite{rvb_anderson}. The  advent of a QS for quantum magnetism would allow for an unprecedented experimental realization of these phases, addressing  puzzles that  remain unsolved due to their great  complexity.

 A representative model that exemplifies   the usefulness of analog QS  is the one-dimensional quantum Ising model~\cite{tfi}, which describes a chain of interacting spins   subjected to a transverse magnetic field. This model has traditionally been considered as a cornerstone in the theory of quantum phase transitions~\cite{sachdev_book}. However,  the vast majority of low-dimensional materials realize instances of the so-called Heisenberg model (see e.g.~\cite{heisenberg_materials}). In fact, the demanding  requirements to explore the Ising magnetism in real materials (e.g. precise one-dimensionality, strong Ising anisotropy, and weak exchange couplings  matching the available magnetic fields) have postponed its observation until the  recent experiments with  CoNb$_2$O$_6$~\cite{tfi_solid,note,cobalt_balents,cobalt_moore}. We stress that prior to this experiment,  trapped-ion QSs  had already targeted the  onset of  Ising criticality~\cite{ising_ions,ising_monroe}. More recently, Ising interactions in a large collection of ions have been observed  in Penning traps~\cite{ising_bollinger}, which in combination with the recent results for neutral atoms~\cite{tfi_atoms},  yield the unique possibility of tailoring the microscopic properties of the quantum Ising magnet (e.g. couplings, dimensionality, geometry).

In this manuscript, we investigate  the capabilities of  trapped ions as QSs of frustrated quantum Ising models (FQIMs). Here,  frustration arises  from  the impossibility of minimizing simultaneously a set of competing commuting interactions (i.e. classical frustration)~\cite{ising_af_triangular}, whereas the quantum fluctuations are introduced by a transverse magnetic field. We remark that, with the exceptions of three-dimensional spin ice~\cite{spin_liquid}  and some disordered spin-glass compounds~\cite{itf_glass},  most of the  frustrated materials correspond once more to Heisenberg magnets~\cite{frustration_materials}. This would turn the extensive research on low-dimensional  frustrated Ising models~\cite{frustration_book,qim_frustration} into a pure theoretical enterprise. Fortunately, the   seminal experiment~\cite{frustration_monroe} has proved the contrary by demonstrating that the physics of small  frustrated networks~\cite{frustration_duan} can be explored in trapped-ion laboratories.  In a recent work~\cite{frustration_ulm}, we have studied a different approach to control the degree of frustration in ion crystals, which works independently of the number of ions and is thus amenable of being scaled to large systems. In this manuscript, we elaborate on that proposal to develop a versatile quantum simulator  for a  variety of spin ladder geometries.

A {\it quantum spin ladder} is an array of coupled  quantum spin chains. Each of these spin chains is usually known as a leg of the ladder, whereas the couplings between them define the  ladder rungs.  Let us note that the study of antiferromagnetic Heisenberg ladders has a long tradition inspired by the experiments with  insulating cuprate compounds~\cite{heisenberg_spin_ladders}, which become high-temperature superconductors after doping~\cite{superconductors}. 
These strongly-correlated systems  lead to  fascinating and thoroughly-studied phenomena~\cite{ladder_sierra}. In contrast, the {\it quantum Ising  ladders} have remained largely unexplored, probably  due to the absence of materials that realize them. Our work discusses a  proposal to fill in such gap, which is based on techniques of current trapped-ion experiments. 

  This article is organized as follows. In Sec.~\ref{section_model}, we describe the properties of the ladder compounds formed by a collection of ions confined in a radio-frequency trap. We discuss the experimental conditions leading to a particular vibrational band structure,  which shall be exploited to explore the physics of magnetic frustration. In section~\ref{section_zz}, we support the above conclusions  by the numerical simulation of a particular case: the  trapped-ion zigzag ladder. A  discussion of the experimental feasibility of our scheme with state-of-the-art setups is also presented in Sec.~\ref{experimental}. Here, we also present initial experiments regarding the appropriate trap design for the QS, and discuss possible sources of error.  In Sec.~\ref{section_scope}, we list the many-body phenomena that can be addressed with the proposed QS, ranging from ordered phases of quantum dimer models, to disordered quantum spin liquids. In Sec.~\ref{section_j1_j2}, we address in detail the  phase diagram of the dipolar $J_1$-$J_2$ quantum Ising model. Finally, we present our conclusions  in Sec.~\ref{section_conclusions}. In the Appendixes~\ref{appendix_dipole},~\ref{appendix_micromotion} and~\ref{heisenberg}, we discuss  technical details regarding both the spin-dependent dipole forces and spontaneous decay, the  micromotion, and the thermal secular motion. We describe conditions under which  they do not affect  the proposed QS.

\section{Trapped-Ion Quantum Ising Ladders} 
\label{section_model}

\subsection{ Geometry of the trapped-ion ladders}

 An ensemble of $N$ atomic ions of mass $m$, and charge $e$, can be trapped in a microscopic region of space by means of radio-frequency fields~\cite{ion_trap_reviews}. This system is described by 
\begin{equation}
\label{coulomb_ham}
H\hspace{-0.05cm}=\hspace{-0.05cm}\sum_{i=1}^{N}\sum_{\alpha=x,y,z}\!\!\left(\frac{1}{2m}{ p}_{i\alpha}^2+\frac{1}{2}m\omega_{\alpha}^2r_{i\alpha}^2\!\right)\!+\frac{e^2}{2}\sum_i\sum_{j\neq i}\frac{1}{|{\bf r}_i-{\bf r}_j|},
\end{equation}
where  $\{\omega_{\alpha}\}_{\alpha=x,y,z}$ represent the effective trapping frequencies (see also Sec.~\ref{experimental} and Appendix~\ref{appendix_micromotion}),  and $\{{\bf r}_i,{\bf p}_i\}_{i=1}^N$ are the positions and momenta  of the ions. Note that we use gaussian units and  $\hbar=1$ throughout this text. 

With the advent of laser cooling, it has become  possible to reduce the temperature of the ions  to such an extent  that they crystallize. This gives rise to one of the forms of condensed matter with the lowest attained density, ranging from clusters composed by two ions~\cite{coulomb_crystal_exps}, to crystals of  several thousands~\cite{other_ion_crystals}.  Interestingly, these crystals can be assembled sequentially by increasing the number of trapped ions one by one, so that the aforementioned transition to the many-body regime~\cite{emergence} can be studied in the laboratory. Besides, the geometry of the crystals can be controlled experimentally by tuning the anisotropy of the trapping frequencies~\cite{structural_theory}, which opens a vast amount of possibilities for the QS of magnetism.

\begin{figure}
\centering
\includegraphics[width=0.97\columnwidth]{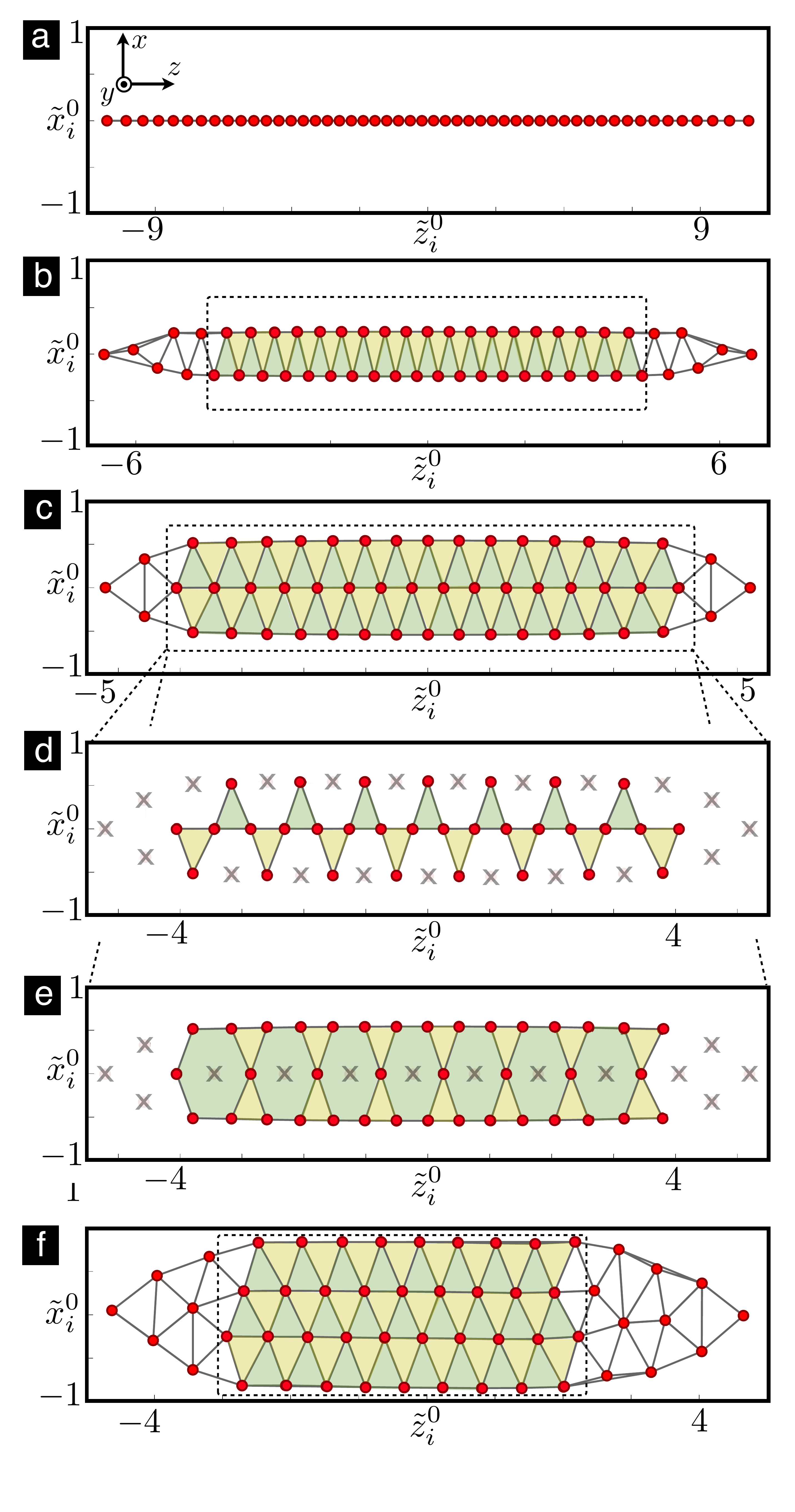}
\caption{ {\bf Trapped-ion ladders:} Self-assembled geometries for $N$ trapped ions as a function of the anisotropic trapping frequencies $\omega_y\gg\omega_x,\omega_z$: {\bf (a)} Linear chain for $N=50$ and $\kappa_x=4\cdot 10^{-4}$, {\bf (b)} Two-leg zigzag ladder for $N=51$ and $\kappa_x=1\cdot 10^{-2}$.  {\bf (c)} Three-leg  ladder of bond-sharing triangles for $N=52$ and $\kappa_x=4\cdot 10^{-2}$,  {\bf (d)} Two-leg  ladder of corner-sharing triangles for $N=52$.  From this geometry, it is possible to obtain the two following ladders by selectively hiding some of the ions (grey crosses) following the methods outlined in Sec.~\ref{experimental}. {\bf (e)} Three-leg ladder corresponding to a Kagome stripe of corner-sharing triangles and hexagons for  $N=52$,  {\bf (f)} Four-leg  ladder of corner-sharing triangles for $N=53$ and $\kappa_x=8\cdot 10^{-2}$.}
\label{ladder_geometries}
\end{figure}

The ion equilibrium positions $\{{\bf r}_i^0\}_{i=1}^{N}$ are determined  by the balance of the trapping forces and the Coulomb repulsion. Formally, they are given by  $\boldsymbol{\nabla}_{{\bf r}_{i}}V={\bf 0}$, where
$V$
 contains the trapping and Coulomb potentials. By introducing the unit of length  $l_z=(e^2/m\omega_z^2)^{1/3}$, the dimensionless equilibrium positions $\tilde{{\bf r}}_i^0={\bf r}_i^0/l_z$ follow from the solution of the system 
 \begin{equation}
 \label{system}
 {\tilde{r}}_{i\alpha}^0-\kappa_{\alpha}\sum_{j\neq i}\frac{ {\tilde{r}}^0_{i\alpha}- {\tilde{r}}^0_{j\alpha}}{| {\tilde{\bf r}}^0_i- {\tilde{\bf r}}^0_j|^3}=0,\hspace{2ex} i=1\dots N,\hspace{0.5ex}\alpha=x,y,z,
 \end{equation}
 where we have introduced the anisotropy parameters $\kappa_{\alpha}=(\omega_{z}/\omega_{\alpha})^2$. As shown in the experiments~\cite{exps_zig_zag}, by increasing  the parameters $\kappa_x=\kappa_y$ towards a maximum value $\kappa_{\rm max}=1$, the geometry of the ion crystal undergoes a series of phase transitions starting from a linear chain, via a two-dimensional zigzag ladder,  to a three-dimensional helix. We remark that these structural phase transitions have been the subject of considerable interest on their own~\cite{zig_zag_theory,zigzag_inhomogeneous}. 
 
 We have recently investigated the possibility of setting a large anisotropy between the radial trap frequencies $1\geq\kappa_x\gg\kappa_y$, such that the ion  crystals get pinned to the $xz$-plane~\cite{frustration_ulm}. This requires a modification of the symmetric electrode configuration of the usual linear traps  (see the details in Sec.~\ref{experimental}). Besides, the  parameter $\kappa_x$ can be increased by  a bias voltage,  yielding a variety of geometries. For instance, above a certain value $\kappa_{\rm c,2}$, the linear chain  transforms onto a $2$-leg ladder  (see Figs.~\ref{ladder_geometries}{\bf (a)-(b)}). By solving numerically the system of equations~\eqref{system}, we observe that the 2-leg zigzag  ladder  first evolves into a 3-leg ladder as $\kappa_x$ is increased, then yields a 4-leg ladder [see Figs.~\ref{ladder_geometries}{\bf (c)} and~\ref{ladder_geometries}{\bf (f)}], and so on. Hence, there should be a sequence of structural transitions at the critical values $\kappa_{\rm c,2}<\kappa_{\rm c,3}<\kappa_{\rm c,4}<\dots<\kappa_{{\rm c},n_{\rm l}}$, where $n_{\rm l}$ determines the number of legs of the  trapped-ion ladders. The rungs of these ladders correspond to the diagonal links, yielding a collection of bond-sharing triangular plaquettes. In Sec.~\ref{experimental}, we describe a method to build ladders from  corner-sharing triangles (see Figs.~\ref{ladder_geometries}{\bf (d)-(e)}), which widens the applicability of our QS. 
 
 Eventually, when $\kappa_x\to1$, the crystal must correspond to a  two-dimensional bond-sharing triangular lattice, or corner-sharing Kagome lattice, with ellipsoidal boundaries. Accordingly, not only can we control the number of legs of the trapped-ion ladder, but also explore the crossover from quasi-one-dimensional physics to the two-dimensional realm. With respect to the FQIMs, this two-dimensional limit is very appealing due to its connection to quantum spin liquids~\cite{rvb_anderson}, and quantum dimer models~\cite{qim_frustration}. However, as happens for Heisenberg magnets~\cite{heisenberg_spin_ladders}, the ladder compounds may already contain fascinating phenomenology. Let us remark that the precise knowledge of the crystal plane is a fundamental advantage for the quantum simulation of FQIM~\cite{frustration_ulm}.  Prior to the discussion of the effective FQIM, we will describe the collective vibrational modes in these ladder compounds, since they shall act as mediators of the magnetic Ising-type interaction.

At this point, it is worth commenting on the possibility of engineering the lattice positions of the ion crystal by means of  micro-fabricated electrode arrangements, (e.g. surface traps as proposed in~\cite{surface_traps}). If the technical problems related to the  anomalous heating observed close to the electrodes are overcome, these new generation of traps could be combined with the proposed QS to study  frustrated spin models in arbitrary lattices without the complication of micromotion. However, to keep within reach of the current technology, we focus on the more standard Paul traps where experiments on quantum magnetism have already been performed~\cite{ising_ions,frustration_monroe,ising_monroe}.

\subsection{ Collective vibrational modes of the ion ladder}

 Due to the Coulomb interaction, the vibrations of the ions around the equilibrium positions, ${\bf r}_i={\bf r}_i^{0}+ \Delta{\bf r}_i$, become coupled. By expanding the Hamiltonian~\eqref{coulomb_ham} to second order in the displacements $  \Delta{\bf r}_i$, one obtains a model of coupled harmonic oscillators corresponding to the  harmonic approximation
\begin{equation}
\label{vib}
H=\!\sum_{i,\alpha}\!\left(\frac{1}{2m}{ p}_{i\alpha}^2+\frac{1}{2}m{\omega}_{\alpha}^2\Delta r_{i\alpha}^2\!\right)\!+\!\frac{1}{2}\sum_{i, j,\alpha,\beta}\mathcal{V}_{ij}^{\alpha\beta}\Delta r_{i\alpha}\Delta r_{j\beta}.
\end{equation}
In this expression, the coupling matrix between the harmonic oscillators can be expressed in terms of the mutual ion distance ${\bf r}^0_{ij}={\bf r}^0_i-{\bf r}^0_j$, and the  quadrupole moment for each pair of ions $Q^{\alpha\beta}_{ij}=-e[3({\bf r}^0_{ij})_{\alpha}({\bf r}^0_{ij})_{\beta}-\delta_{\alpha\beta}({\bf r}^0_{ij})^2]$ as follows
\begin{equation}
\label{vs}
\mathcal{V}_{ij}^{\alpha\beta}=\frac{ e Q^{\alpha\beta}_{ij}}{|{\bf r}^0_{ij}|^5}-\left(\sum_{l\neq i}\frac{ e Q^{\alpha\beta}_{il}}{|{\bf r}^0_{il}|^5}\right)\delta_{ij}.
\end{equation}
For the particular ladder geometries in Fig.~\ref{ladder_geometries}, we identify two types of vibrational excitations,  the so-called {\it transverse modes} that correspond to the vibrations of the ions perpendicular to the ladder (i.e. $y$-axis), and the {\it planar} modes that account for the coupled vibrations within the ladder  (i.e. $x,z$ axes). The transverse modes are decoupled from the planar vibrations, and  are described by a set of coupled oscillators
\begin{equation}
\label{transverse}
H_{\bot}\!=\!\sum_{i}\left(\frac{1}{2m}{ p}_{iy}^2+\frac{1}{2}m\tilde{{\omega}}_{iy}^2\Delta r_{iy}^2\right)+ \frac{e^2}{2}\sum_i\sum_{j\neq i}\frac{1}{|{\bf r}_{ij}^0|^3}\Delta r_{iy}\Delta r_{jy},
\end{equation}
where $\omega_y^2\to\tilde{{\omega}}_{iy}^2={\omega}_{y}^2(1-\kappa_y\sum_{l\neq i}|{\bf \tilde{r}}_l^0-{\bf \tilde{r}}_i^0|^{-3})$. This term amounts to the Einstein model of individual lattice vibrations,    where the renormalization of the  frequencies  is caused by the mean-field-type  interaction of one ion with the rest of the ion ensemble. In our case, we must also consider the coupling between distant oscillators, whose magnitude relative to the trapping frequencies scales as $(e^2/l_z^3)/(m\omega_y^2)=\kappa_y\ll1$. Hence, the transverse vibrations  are described by  a set of harmonic oscillators with  weak couplings that decay with a dipolar law.

The planar vibrations are more complex since the motion along both axes, $x$ and $z$, becomes coupled through the Coulomb interaction. In this case, 
the Hamiltonian is
\begin{equation}
\label{in_plane}
\begin{split}
H_{\parallel}\!=\!&\sum_{i}\left(\frac{1}{2m}{ p}_{ix}^2+\frac{1}{2}m\tilde{{\omega}}_{ix}^2\Delta r_{ix}^2\right)\!+ \!\frac{1}{2}\sum_i\sum_{j\neq i}\frac{eQ^{xx}_{ij}}{|{\bf r}_{ij}^0|^5}\Delta r_{ix}\Delta r_{jx}\\
+&\sum_{i}\left(\frac{1}{2m}{ p}_{iz}^2+\frac{1}{2}m\tilde{{\omega}}_{iz}^2\Delta r_{iz}^2\right)\!+\! \frac{1}{2}\sum_i\sum_{j\neq i}\frac{eQ^{zz}_{ij}}{|{\bf r}_{ij}^0|^5}\Delta r_{iz}\Delta r_{jz}\\
+& \sum_i\sum_{j\neq i}\frac{eQ^{xz}_{ij}}{|{\bf r}_{ij}^0|^5}\Delta r_{ix}\Delta r_{jz}+\sum_i\sum_{j\neq i}\frac{eQ^{zx}_{ij}}{|{\bf r}_{ij}^0|^5}\Delta r_{iz}\Delta r_{jx},
\end{split}
\end{equation}
where   $\tilde{{\omega}}_{ix}^2={\omega}_{x}^2(1+\kappa_x\sum_{l\neq i}\tilde{Q}^{xx}_{il}|{\bf \tilde{r}}_i^0-{\bf \tilde{r}}_l^0|^{-5})$ (equivalently for $z$), and $\tilde{Q}^{xx}_{il}$ is expressed in terms of the dimensionless equilibrium positions. Since we have assumed that $\kappa_x\gg\kappa_y$, the planar vibrations are coupled more strongly than the transverse ones. 
 In the last term of this Eq.~\eqref{in_plane}, we observe how the non-diagonal terms of the quadrupole, $Q^{xz}_{ij}\neq 0$, are responsible for the coupling of the motion along the $x$ and $z$ axes of the ladder. 

So far, we have reduced the original Hamiltonian~\eqref{coulomb_ham} to a pair of quadratic boson models~\eqref{transverse}-\eqref{in_plane}. These can  be exactly solved by introducing the canonical transformations 
\begin{equation}
\label{phonons}
\begin{split}
\Delta r_{iy}=&\sum_{n=1}^N\frac{1}{\sqrt{2m\Omega^{\bot}_n}}\mathcal{M}^{\bot}_{i,n}(a_n^{\phantom \dagger}+a_n^{\dagger}),\\
\Delta r_{ix}=&\sum_{n=1}^{2N}\frac{1}{\sqrt{2m\Omega^{\shortparallel}_n}}\mathcal{M}^{\shortparallel}_{i,n}(b_n^{\phantom \dagger}+b_n^{\dagger}),\\
\Delta r_{iz}=&\sum_{n=1}^{2N}\frac{1}{\sqrt{2m\Omega^{\shortparallel}_n}}\mathcal{M}^{\shortparallel}_{N+i,n}(b_n^{\phantom \dagger}+b_n^{\dagger}),\\
\end{split}
\end{equation}
where $a_n^{\dagger},a_n^{\phantom \dagger}$ stand for the $N$ creation-annihilation operators for the quantized excitations (i.e. phonons) of the transverse modes, and $b_n^{\dagger},b_n^{\phantom \dagger}$ are the the corresponding $2N$ operators for the planar modes. 
In these expressions, $\mathcal{M}_{i,n}^{\bot}$ ($\mathcal{M}_{i,n}^{\shortparallel}$) determines the amplitude of the transverse (planar) oscillations of the ion at site $i$ due to the collective vibrational mode labeled by $n$, such that $\Omega^{\bot}_n(\Omega^{\shortparallel}_n)$ are the corresponding normal-mode frequencies. Formally~\cite{feynman_lectures}, $\mathcal{M}_{i,n}^{\bot}$ ($\mathcal{M}_{i,n}^{\shortparallel}$) are given by the orthogonal matrices that diagonalize the second order Coulomb couplings in Eqs.~\eqref{transverse}-\eqref{in_plane}, or equivalently
\begin{equation}
\label{diag}
\begin{split}
&\sum_{i,j=1}^N\mathcal{M}_{in}^{\bot}\left(\mathbb{I}+\kappa_y\mathcal{\tilde{V}}^{yy}\right)_{ij}\mathcal{M}_{jm}^{\bot}=\left(\frac{\Omega^{\bot}_n}{\omega_y}\right)^2\delta_{nm},\\
&\sum_{i,j=1}^{2N}\mathcal{M}_{in}^{\shortparallel}\left(\begin{array}{cc}\mathbb{I}+\kappa_x\mathcal{\tilde{V}}^{xx} & \kappa_x\mathcal{\tilde{V}}^{xz}\\ \kappa_x\mathcal{\tilde{V}}^{zx} & \kappa_x\mathbb{I}+\kappa_x\mathcal{\tilde{V}}^{zz}\end{array}\right)_{ij} \mathcal{M}_{jm}^{{\shortparallel}}=\left(\frac{\Omega^{\shortparallel}_n}{\omega_x}\right)^2\delta_{nm},
\end{split}
\end{equation}
where $\mathcal{\tilde{V}}_{ij}^{\alpha\beta}=\mathcal{{V}}_{ij}^{\alpha\beta}/(e^2/l_z^3)$ are dimensionless couplings. These equations~\eqref{diag} must be solved numerically with the previous knowledge of the equilibrium positions~\eqref{system}, and yield 
the following quadratic phonon Hamiltonian 
\begin{equation}
\label{phonon_ham}
H_{\rm p}=\sum_{n=1}^{N}\Omega_n^{\bot}\left(a_n^{\dagger}a_n^{\phantom \dagger}+\frac{1}{2}\right)+\sum_{n=1}^{2N}\Omega_{n}^{\shortparallel}\left(b_{n}^{\dagger}b_n^{\phantom \dagger}+\frac{1}{2}\right),
\end{equation}
where the index $n$ labels the normal modes with an increasing vibrational frequency $\Omega_{n+1}^{\bot}>\Omega_{n}^{\bot},\Omega_{n+1}^{\shortparallel}>\Omega_{n}^{\shortparallel}$. 

\begin{figure}

\centering
\includegraphics[width=1\columnwidth]{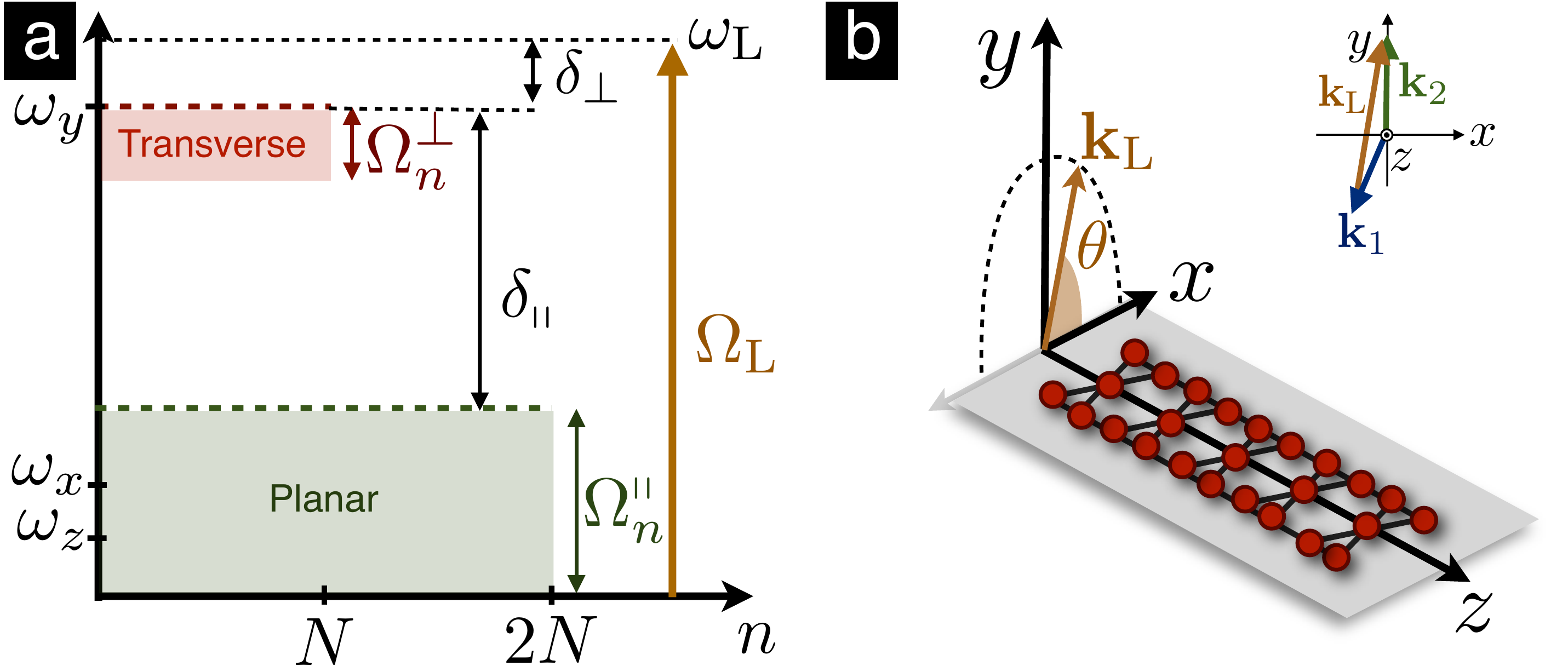}
\caption{ {\bf Vibrational modes of the ladder compounds:} {\bf (a)} Schematic representation of the phonon branches, whereby the transverse vibrational frequencies $\Omega_n^{\bot}$ span around the trap frequency $\omega_y$, and are situated far away from the planar phonon branch $\Omega_n^{\shortparallel}$, which spans around $\omega_x,\omega_z$. {\bf (b)} Laser beam arrangement (inset) for a spin-dependent dipole force with an effective wavevector ${\bf k}_{\rm L}$ in the $xy$-plane slightly tilted from the $y$-axis. The effective frequency $\omega_{\rm L}$ is tuned above the resonance of the transverse modes (see {\bf (a)}), such that the detunings fulfill $\delta_{\shortparallel}\gg\delta_{\bot}$.  }
\label{ladder_scheme_phonons}
\end{figure}

We now  discuss a  qualitative picture for the phonon branches valid for all the different ladders. As argued above, the transverse phonon modes are weakly coupled due to the small parameter $\kappa_y\ll1$. By inspecting Eq.~\eqref{diag}, one realizes that this condition will lead to a transverse phonon branch  $\Omega^{\bot}_n\in[\Omega_1^{\bot},\Omega_N^{\bot}]$ with a small width around the trap frequency $\omega_y$ (see Fig.~\ref{ladder_scheme_phonons}{\bf (a)}). This property is not fulfilled by the planar modes $\Omega^{\shortparallel}_n\in[\Omega_1^{\shortparallel},\Omega_{2N}^{\shortparallel}]$, which are more strongly coupled $\kappa_x\gg\kappa_y$, and will thus present a wider branch around $\omega_x,\omega_z$.  However, due to the frequency anisotropy $\omega_y\gg\omega_x,\omega_z$, the planar phonon branch  will always be situated far away from the transverse-mode frequencies, even if its width is considerably larger.  This property will turn out to be essential for the QS of frustrated  spin models~\cite{frustration_ulm}, whereby the role of the spin is played by two electronic levels of the ion, and the interactions are mediated by the collective vibrational excitations. 

The main idea is that such a clustered phonon spectrum will allow us to  use a pair of laser beams that only couple the spins  to the transverse phonon modes, even when the radiation resulting from the interference also propagates along the plane of the ladder (Fig.~\ref{ladder_scheme_phonons}{\bf (b)}).  The interest of this idea is two-fold. On the one hand, the transverse phonon modes are the ideal mediators of the spin-spin interactions due to their higher insensitivity to ion-heating and their lower contributions to the thermal noise in the QS, as compared to the planar modes.  On the other hand, by using a laser configuration with a component along the rungs of the ladder, we can exploit the ratio of its effective wavelength with the ion mutual distances in order to tailor the sign (ferromagnetic/antiferromagnetic) and the magnitude  of the spin-spin interactions anisotropically (i.e. depending on the direction joining the pair of  ions). As shown below, this opens the possibility of realizing a versatile QS of  frustrated quantum spin ladders, which  is amenable of being scaled to larger ion numbers and a variety of geometries.

\subsection{ State-dependent dipole forces}

The use of the collective vibrational modes as a common
data bus to perform  two-qubit gates can also be understood in terms of  phonon-mediated spin-spin interactions~\cite{ions_ising_interaction}, which opens the route towards the QS of quantum magnetism~\cite{ising_porras,spin_models_ti}. We now introduce the key ingredient for such  QSs, namely, a spin-phonon coupling that originates from a laser-induced state-dependent dipole force~\cite{sd_force,sd_force_review}.

 So far, our discussion  applies to all  ion species and radio-frequency traps. However,   in order to exploit the phonons as mediators of a  spin-spin interaction that is sufficiently strong, the trap frequencies for each particular  ion species must be tuned such that $l_z=(e^2/m\omega_z^2)^{1/3}\approx1$-10 $\mu$m, and $\omega_y\gg\omega_x\geq\omega_z$. This typically constraints  the order of magnitude of the trap frequencies to the range  $\omega_{z}/2\pi,\omega_x/2\pi\approx$ 0.1-1 MHz, and $\omega_y/2\pi\approx$ 1-10 MHz. Moreover, since we aim  at a flexible control of the anisotropy of the spin couplings, we shall focus on the ion species that allow for a two-photon lambda scheme to implement the  spin-phonon coupling. Hence, our discussion will be specific to singly ionized alkaline-earth ions, either with a hyperfine structure  $^{9}{\rm Be}^+$, $^{25}{\rm Mg}^+$, and $^{43}{\rm Ca}^+$, or with a pair of Zeeman-split levels such as $^{40}{\rm Ca}^+$ and $^{24}{\rm Mg}^+$~\cite{note_optical}.

\begin{figure}

\centering
\includegraphics[width=1\columnwidth]{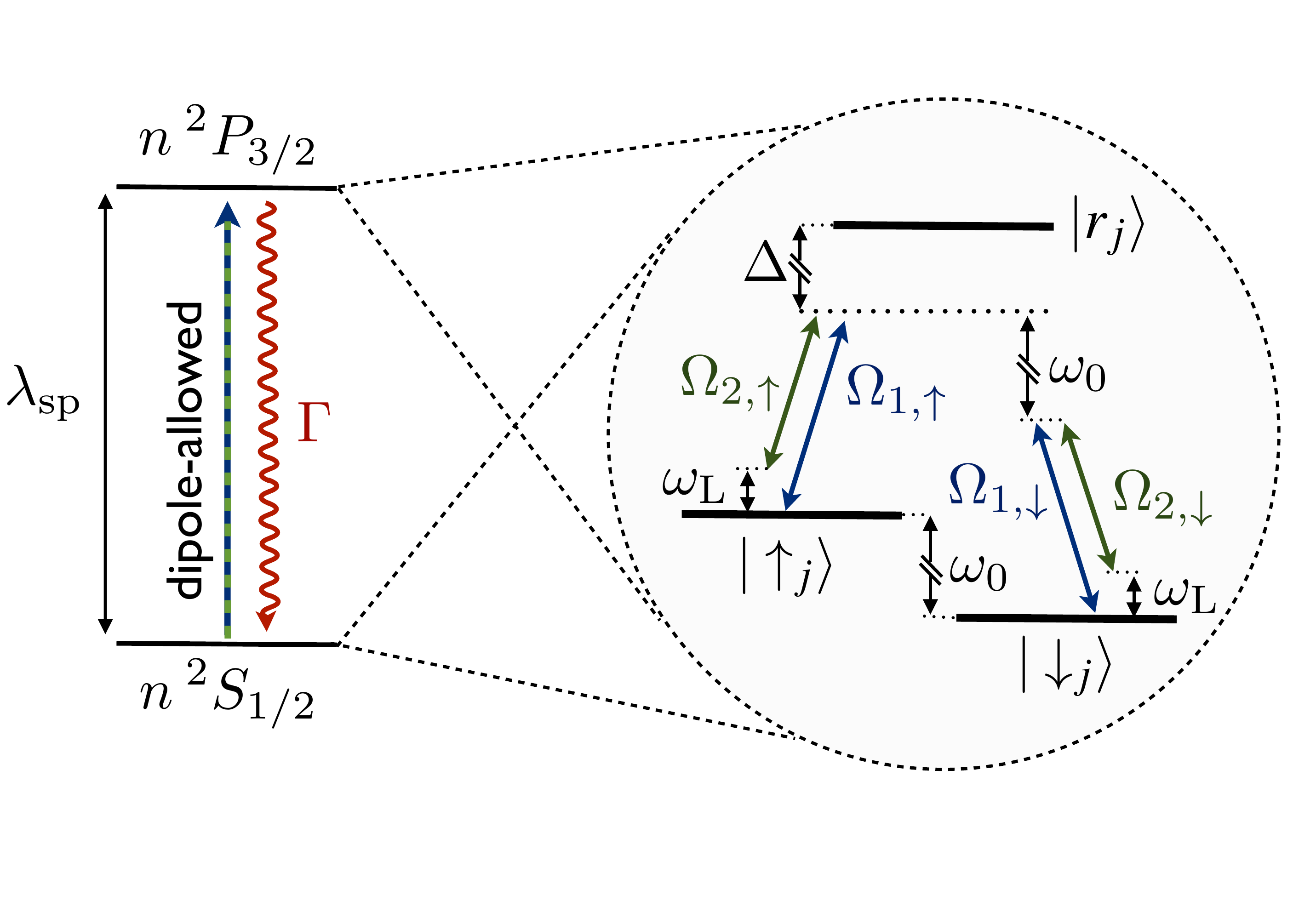}
\caption{ {\bf Lambda scheme for the dipole force:} Two electronic states $\{\ket{\uparrow_j},\ket{\downarrow_j}\}$ with an energy difference $\omega_0$ are selected from the ground-state manifold $n^{2}S_{1/2}$ to form an effective spin-1/2.  These states can be manipulated by a pair of laser beams with Rabi frequencies $\Omega_{1,\sigma},\Omega_{2,\sigma}$ that induce a transition to an excited state $\ket{r_j}$ in the $n^2P_{3/2}$ manifold from the spin state $s=\uparrow,\downarrow$ in  $n^{2}S_{1/2}$. Note that the transition wavelength $\lambda_{\rm sp}$ lies in the optical regime. When the beatnote $\omega_{\rm L}=\omega_1-\omega_2$ is tuned close to the trap frequency $\omega_y$, such that the detuning $\Delta$  is much larger than the spontaneous decay rate $\Gamma$ of the transition and the Rabi frequencies, one obtains the desired state-dependent dipole force. }
\label{level_scheme}
\end{figure}

In Fig.~\ref{level_scheme}, we represent schematically the atomic energy levels of such ions, which have a single valence electron in the orbital $n\hspace{0.5ex}^2S_{1/2}$ that can be optically excited to  $n\hspace{0.5ex}^2P_{1/2},n\hspace{0.5ex}^2P_{3/2}$ via a dipole-allowed transition. Here, we use the standard notation $n\hspace{0.5ex}^{2S+1}L_J$, with $n$ as the principal quantum number, and $S, L ,J$  as the  spin, orbital, and total electronic  angular momentum. Depending on the nuclear spin $I$ of the particular ion, the ground-state manifold will be split into a pair of Zeeman states ($I=0$), or a set of hyperfine levels ($I\neq 0$). We select two of such states $\ket{{\uparrow_i}},\ket{{\downarrow_i}}$, which are separated by an energy gap $\omega_0$, to form the effective spins of our QS (see the inset of Fig.~\ref{level_scheme}). For hyperfine spins, the energy splitting is $\omega_0/2\pi\approx$ 1-10 GHz, whereas for Zeeman spins its order of magnitude depends on the external magnetic field $\omega_0/B_0\approx 2\pi\times$10 GHz/T. 

Accordingly, the resonance frequencies usually lie in the radio-frequency/microwave regime. However, it is not customary to use  radio-frequency/microwave radiation to couple the spins to the collective vibrational modes, since its wavelength is much larger than the typical ion oscillations (but see the recent experimental progress~\cite{microwaves}). A possible alternative is to use a pair of laser beams with optical frequencies $\omega_1,$ $\omega_2$, a "moving standing wave", to induce a two-photon  stimulated Raman transition  through an excited state $\ket{r_i}$ (see Fig.~\ref{level_scheme}). When the laser beams are far off-resonant with respect to this dipole-allowed transition, such that the detuning  $\Delta$ is much larger than the decay rate of the excited state $\Gamma$ and the laser Rabi frequencies $\Omega_{1,s},\Omega_{2,s}$, where $s=\uparrow,\downarrow$, it is possible to eliminate the excited state from the dynamics and obtain a Hamiltonian that only involves the spins and the phonons (see Appendix~\ref{appendix_dipole} for the effective master equation). In particular, when the laser beatnote  is $\omega_{\rm L}=\omega_1-\omega_2\approx\omega_y\ll \omega_0\ll\Delta$ (Fig.~\ref{ladder_scheme_phonons}{\bf (a)}), one obtains the following ion-laser Hamiltonian
\begin{equation}
\label{dipole_force}
H_{\rm d}=\frac{\Omega_{\rm L}}{2}\sum_i\sigma_i^z\ee^{\ii{\bf k}_{\rm L}\cdot {\bf r}^0_i}\ee^{\ii({\bf k}_{\rm L}\cdot \Delta{\bf r}_i-\omega_{\rm L} t)}+\text{H.c.},
\end{equation}
which can be interpreted as a time-dependent differential ac-Stark shift due to the interference of the two laser beams. Here, we have introduced  $\sigma_i^z=\ket{{\uparrow_i}}\bra{{\uparrow_i}}-\ket{{\downarrow_i}}\bra{{\downarrow_i}}$, and the two-photon differential Rabi frequency $\Omega_{\rm L}=(\Omega_{\rm 1,\downarrow}^{\phantom *}\Omega_{\rm 2,\downarrow}^{*}-\Omega_{\rm 1,\uparrow}^{\phantom *}\Omega_{\rm 2,\uparrow}^{*})/2\Delta$.  The corrections due to the spontaneous decay from the excited level  contribute to the Rabi frequency with $\Omega_{\rm L}\to\Omega_{\rm L}(1+(\Gamma/\Delta)^2)$, and can be thus neglected if $\Gamma\ll\Delta$.

The idea is that by controlling experimentally the polarizations, intensities and detunings  of the laser beams, one can find regimes where the effective Rabi frequency $\Omega_{\rm L}$ does not vanish and leads to a spin-dependent dipole force. Let us note that the adiabatic elimination also leads to standard ac-Stark shifts and to a running wave that only couples to the vibrational excitations.  These terms must be  compensated by carefully selecting the laser-beam parameters. Let us finally emphasize that the effective Raman wavevector  ${\bf k}_{\rm L}={\bf k}_1-{\bf k}_2=k_{\rm L}\cos\theta{\bf e}_x+k_{\rm L}\sin\theta{\bf e}_y$  has a component along the ladder plane (see Fig.~\ref{ladder_scheme_phonons}{\bf (b)}), and in principle couples to both the planar and transverse phonons. The reason for the announced decoupling from the planar vibrational modes is that the beatnote of the laser beams, which is near the resonance of the transverse phonons, will lie far off-resonance with respect to the planar vibrational modes (Fig.~\ref{ladder_scheme_phonons}{\bf (a)}). This qualitative argument will be quantified below, and supported  numerically in Sec~\ref{section_zz}.

After introducing the phonon operators in Eq.~\eqref{phonons}, we perform a Taylor expansion of Eq.~\eqref{dipole_force} for the small transverse and planar Lamb-Dicke parameters 
\begin{equation}
\label{ld}
\eta_{n\bot}=\frac{k_{{\rm L}}\sin\theta}{\sqrt{2m\Omega_{n}^{\bot}}}\ll1,\hspace{2ex}\eta_{n\shortparallel}=\frac{k_{{\rm L}}\cos\theta}{\sqrt{2m\Omega_{n}^{\shortparallel}}}\ll1.
\end{equation}
 By setting $\omega_{\rm L}\gtrsim \omega_y$, such that the bare detuning fulfills $|\delta_y|=|\omega_y-\omega_{\rm L}|\ll |\delta_x|=|\omega_x-\omega_{L}|,\omega_y$, one can neglect all non-resonant terms apart from a state-dependent dipole force that couples the spins to the transverse phonons. In the interaction picture with respect to the phonon Hamiltonian~\eqref{phonon_ham}, we get
 \begin{equation}
 \label{transverse_pushing}
 H_{\rm d}=\frac{\Omega_{\rm L}}{2}\sum_{i,n}\ii\ee^{\ii{\bf k}_{\rm L}\cdot {\bf r}_i^0}\eta_{n\bot}\mathcal{M}_{in}^{\bot}\sigma_i^za_{n}^{\dagger}\ee^{\ii\delta_{n\bot}t}+\text{H.c.},
 \end{equation}
 where $\delta_{n\bot}=\Omega_{n}^{\bot}-\omega_{\rm L}$. We note that,  in order to neglect all the remaining terms of the Taylor expansion, a rotating wave approximation (RWA) must be performed, provided that
 \begin{equation}
 \label{conditions}
 \frac{\Omega_{\rm L}}{\omega_{\rm L}}\ll 1,\hspace{2ex}\frac{\eta_{n\shortparallel}\Omega_{\rm L}}{|\Omega_{n}^{\shortparallel}-\omega_{\rm L}|}\ll \frac{\eta_{n\bot}\Omega_{\rm L}}{|\Omega_{n}^{\bot}-\omega_{\rm L}|}.
 \end{equation}
 The first condition is required to neglect the off-resonant contributions to the ac-Stark shift of the energy levels. Additionally, by virtue of Eq.~\eqref{ld}, this condition ensures that the spin-phonon couplings involving higher-order powers of the transverse phonon operators can also be neglected.   For the regime considered in this work, namely $\omega_{\rm L}\approx\omega_y\gg\omega_x\geq\omega_z$, it will suffice to consider $\Omega_{\rm L}/2\pi\leq$ 0.1-1 MHz to accomplish this constraint. The second condition in~\eqref{conditions} is necessary to avoid that the dipole force also couples the spins to the planar phonons. Hence, it characterizes the parameter regime where our previous qualitative discussion about the decoupling of the planar vibrational modes holds. The fulfillment of this condition relies on the large gap between the frequencies of the transverse and planar normal modes. Besides, by setting  the  laser-beam arrangement such that  $\theta\approx \pi/2$, one obtains $\eta_{n\shortparallel}\ll1$,  which warrants the fulfillment of the last condition. In Sec.~\ref{section_zz}, we will confirm the validity of these constraints numerically for the particular case of a zigzag ion ladder. 
 
Depending on the spin state, the dipole force in Eq.~\eqref{transverse_pushing} pushes the ions in opposite directions transversally to the ladder plane.
 Besides, the phase of this pushing force  depends on the ratio of the ion equilibrium positions and the effective wavelength of the interfering laser beams. As shown below, this is precisely the parameter that shall allow us to control the anisotropy of the effective spin-spin couplings.
 
 \subsection{Spin models with  tunable anisotropy and frustration}
 
  There are numerous situations in nature where  interactions are mediated by  the exchange of particles. Of particular relevance to the field of magnetism is the so-called  Ruderman-Kittel-Kasuya-Yosida (RRKY) mechanism~\cite{rkky}, whereby  a Heisenberg coupling between distant nuclear spins is mediated by electrons from the conduction band of a metal. The sign of the Heisenberg couplings alternates between ferromagnetic/antiferromagnetic as a function of the ratio between the Fermi wavelength and the mutual  spin distance. We have recently shown~\cite{frustration_ulm} that a similar phenomenon occurs for trapped-ion ladders subjected to the dipole force in Eq.~\eqref{transverse_pushing}. Instead of a Heisenberg-type interaction between the nuclear spins, one obtains a periodically-modulated Ising-type coupling between the spins formed by two electronic states of the ion. Interestingly enough, the modulation can be experimentally tailored by controlling the direction of propagation of the interfering laser
 beams providing the dipole force.
 
 The  Ising interaction between the effective spins of two distant ions can be understood as a consequence of the virtual phonon exchange between these ions. The dipole force~\eqref{transverse_pushing} pushes the ions transversally, exciting thus the transverse vibrational modes. These phonon excitations, being collective, can be reabsorbed elsewhere in the ion crystal, providing a mechanism to couple the distant spins. The exact expression  can be obtained by  a Lang-Firsov-type transformation~\cite{comment, polaron_book, kanamori,elliot,spin_displacement_wunderlich,ising_porras} that decouples the spins from the phonons
\begin{equation}
\label{lang_firsov}
U_{S}=\ee^{S},\hspace{2ex}S=\frac{\Omega_{\rm L}}{2}\sum_{i,n}\ii\ee^{\ii{\bf k}_{\rm L}\cdot{\bf r}_i^0}\frac{\eta_{n\bot}}{\delta_n^{\bot}}\mathcal{M}_{in}^{\bot}\sigma_i^za_{n}^{\dagger}-\text{H.c.}
\end{equation}
 This canonical transformation leads to an effective Hamiltonian $U^{\dagger}_S(\tilde{H}_{\rm p}+\tilde{H}_{\rm d})U_S\approx \tilde{H}_{\rm eff}+\tilde{H}_{\rm p}$, where we  consider a picture such that the vibrational Hamiltonian $H_{\rm p}$~\eqref{phonon_ham}  absorbs the time-dependence of the dipole force $H_{\rm d}$~\eqref{transverse_pushing}, namely $ \ket{\tilde{\psi}(t)}=U(t)\ket{\psi (t)}$, where $U(t)={\rm exp}({\ii t\sum_n\omega_{\rm L}a_n^{\dagger}a_n^{\phantom \dagger}})$. This leads to the following representation of the total Hamiltonian
 \begin{equation}
 \label{time_indep}
 \tilde{H}_{\rm p}+\tilde{H}_{\rm d}=\sum_n\delta_n^{\bot}a_n^{\dagger}a_n^{\phantom \dagger}+\frac{\Omega_{\rm L}}{2}\sum_{i,n}(\ii\ee^{\ii{\bf k}_{\rm L}\cdot {\bf r}_i^0}\eta_{n\bot}\mathcal{M}_{in}^{\bot}\sigma_i^za_{n}^{\dagger}+\text{H.c.}).
 \end{equation}
After applying the above Lang-Firsov-type transformation, and moving back to the original Schr\"odinger picture, we obtain   the following effective Ising model 
\begin{equation}
\label{ising}
{H}_{\rm eff}=\sum_i\sum_{j\neq i}J_{ij}^{\rm eff}\sigma_i^z\sigma_j^z.\\
\end{equation} 
 The effective spin couplings have the following expression
\begin{equation}
\label{couplings}
J_{ij}^{\rm eff}=-\sum_{n}\frac{\Omega_{\rm L}^2k_{\rm L}^2\sin^2\theta}{8m\Omega_n^{\bot}\delta_{n\bot}}\mathcal{M}_{in}^{\bot}\mathcal{M}_{jn}^{\bot}\cos({\bf k}_{\rm L}\cdot{\bf r}_{ij}^0)
\end{equation}
As announced previously, the phase dependence of the dipole force~\eqref{transverse_pushing} on the ratio of the ion equilibrium positions and the effective wavelength of the light has been translated in the particular  periodic modulation  of the interaction strengths $J_{ij}^{\rm eff}\propto\cos({\bf k}_{\rm L}\cdot {\bf r}_{ij}^0)$. This sign alternation  is similar to that found in RKKY metals, such that the transverse phonons play the role of the conduction electrons, and the  wavelength of the interfering laser beams acts as the Fermi wavelength. In the regime of interest for the ladders,  $\kappa_y\ll 1$, the couplings 
\begin{equation}
\label{ising_strengths}
J_{ij}^{\rm eff}=\frac{J_{\rm eff}\cos\phi_{ij}}{|{\bf \tilde{r}}_{i}^0-{\bf \tilde{r}}_{j}^0|^3},
\end{equation}
display a dipolar decay law, where we have introduced 
\begin{equation}
\label{eq19}
J_{\rm eff}=\frac{\Omega_{\rm L}^2\eta_y^2}{8\delta_y^2}\kappa_y\omega_y,\hspace{2ex}\phi_{ij}= 2\pi\frac{l_z(\tilde{x}_i^0-\tilde{x}_j^0)\cos\theta
}{\lambda_{\rm L}},\end{equation}
such that $\eta_y=k_{\rm L}\sin\theta/\sqrt{2m\omega_y}$ is the bare Lamb-Dicke parameter. From these expressions, it becomes apparent that the interactions between spins belonging to the same leg of the ion ladder ($\phi_{ij}=0$) correspond to  antiferromagnetic $J^{\rm eff}_{ij}>0$ Ising couplings. Conversely, the interactions between the spins from different legs of the ladder ($\phi_{ij}\neq0$) can be ferromagnetic $J^{\rm eff}_{ij}<0$ or antiferromagnetic $J^{\rm eff}_{ij}>0$ depending on the laser parameters.  We can thus tune the sign and magnitude of the spin-spin couplings anisotropically. Note that, even if the typical optical wavelengths are much smaller than the mutual ion distances $\lambda_{\rm L}\ll l_z\approx 1$-10 $\mu$m,  the angle $\theta$  can be tuned around $\theta\approx\frac{\pi}{2}$ so that $\phi_{ij}$ attains any desired value $\phi_{ij}\in[0,2\pi]$. Alternatively, it is also possible to maintain the laser-beam arrangement fixed, and modify the anisotropy ratio $\kappa_x=(\omega_z/\omega_x)^2$ in order to control the inter-ion distances $\tilde{x}_i^0-\tilde{x}_j^0$, attaining thus the desired value $\phi_{ij}\in[0,2\pi]$. Either of these two methods will turn out to be  essential  to explore the full phase diagram of the frustrated quantum spin ladders.

This  anisotropy of the model leads to frustration when 
\begin{equation}
\label{frustration_criteria}
\mathcal{F}_P=\text{sign}\left\{\prod_{(i,j)\in{ P}}-J^{\text{eff}}_{ij}\right\}=-1,
\end{equation}
where $P$ stands for the elementary plaquette of the lattice, which is a triangle in our case $\mathcal{F}_{\triangle}=-1$ (see  Figs.~\ref{ladder_geometries}{\bf (b)},{\bf (c)} and {\bf (e)}). Therefore, the Ising model is frustrated if there  is an  {\it odd} number of antiferromagnetic couplings per unit cell. We note that this standard criterion of geometric frustration~\cite{villain_frustration} can be extended to situations where quantum fluctuations and frustration have the same source~\cite{frustration_criteria}. In our case, however, the frustration is purely classical, interpolating between the antiferromagnetic frustration, and the frustration due to competing ferromagnetic and antiferromagnetic interactions.  In order to introduce  quantum fluctuations,  a microwave directly coupled to the atomic transition yields
 \begin{equation}
\label{quantum_ising}
{H}_{\rm eff}=\sum_i\sum_{j\neq i}J_{ij}^{\rm eff}\sigma_i^z\sigma_j^z-h\sum_i\sigma_i^x,
\end{equation} 
where we have introduced $\sigma_i^x=\ket{{\uparrow_i}}\bra{{\downarrow_i}}+\ket{{\downarrow_i}}\bra{{\uparrow_i}}$, and $h$ plays the role of an effective transverse field of strength $h$  due to the microwave, which is responsible for the quantum fluctuations. 

\begin{figure}

\centering
\includegraphics[width=1\columnwidth]{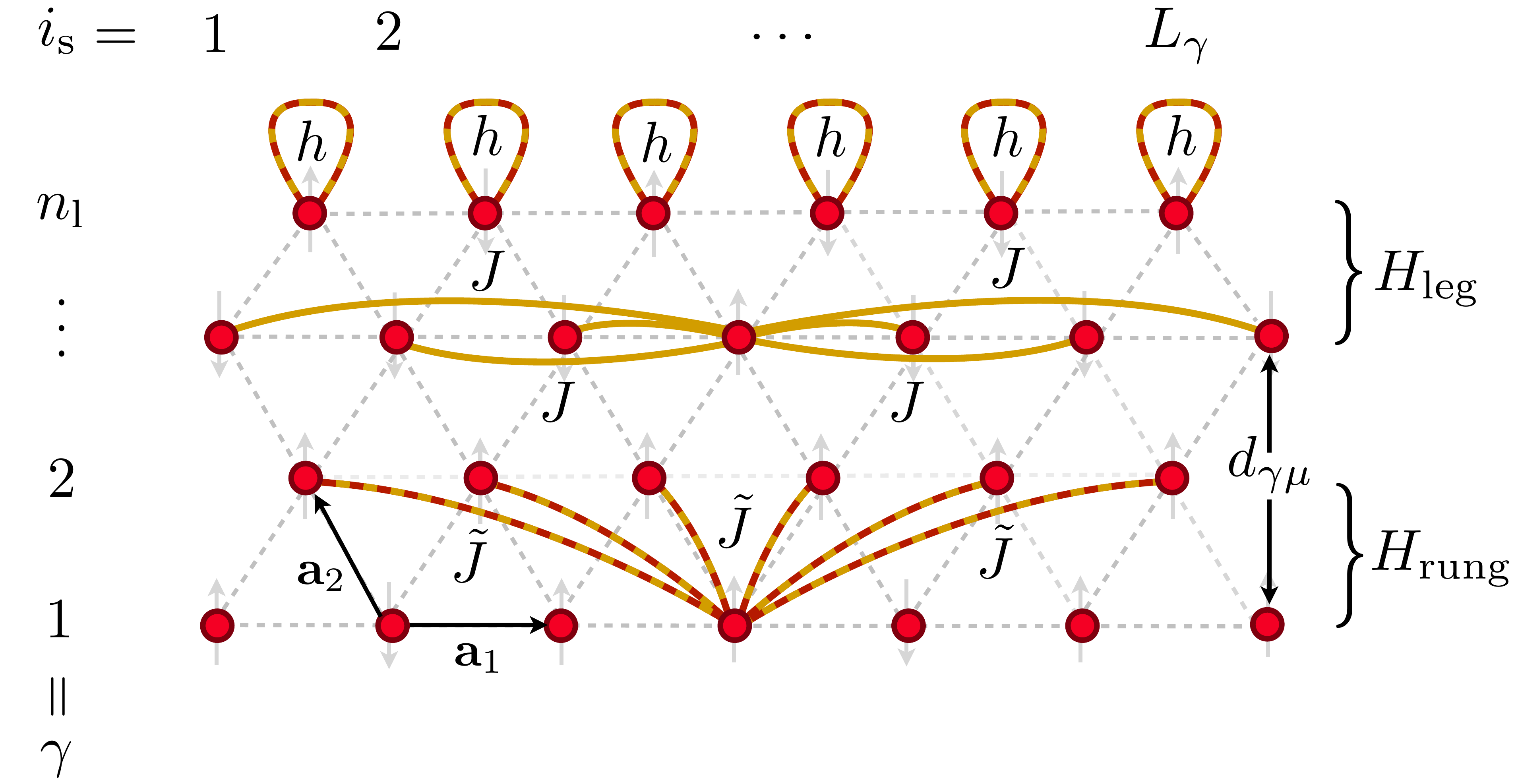}
\caption{ {\bf Anisotropic spin-spin interactions:} The legs of the trapped-ion ladder are labeled by the index $\gamma=1,\dots,n_{\rm l}$, whereas the spins in each leg correspond to $i_{\rm s}=1,\dots,L_{\gamma}$. The effective spin ladder hamiltonian is composed of two terms: $H_{\rm leg}$ contains single-spin terms corresponding to the transverse field $h\lessgtr 0$ (yellow-red laces) and the antiferromagnetic dipolar couplings $J>0$ (yellow bonds). $H_{\rm rung}$ contains the couplings between spins of different rungs, such that $\tilde{J}\lessgtr$ (yellow-red bonds) depends on the mutual rung distance $d_{\gamma\mu}$. We also show the unit vectors ${\bf a}_1,{\bf a}_2$ of the triangular lattice.  }
\label{qim_ladder}
\end{figure}

Let us close this section by rewriting the effective anisotropic quantum Ising model~\eqref{quantum_ising} in a notation that is more appropriate for quantum spin ladders (see Fig.~\ref{qim_ladder}). We substitute the label of the spins $i=1,\cdots, N$ for two new indices that account for the number of legs $\gamma=1,\cdots, n_{\rm l}$, and the number of spins in each of these legs  $i_{\rm s}=1,\cdots, L_{\gamma}$, such that $\sum_{\gamma}L_{\gamma}=N$. Then, the Hamiltonian can be rewritten as $H_{\rm eff}=H_{\rm leg}+H_{\rm rung}$, where the leg and rung Hamiltonians are 
\begin{equation}
\label{qi_ladder}
\begin{split}
H_{\rm leg}&=\sum_{\gamma}\sum_{i_{\rm s}\neq j_{\rm s}}J^{\gamma}_{i_{\rm s},j_{\rm s}}\sigma_{i_{\rm s}}^z(\gamma)\sigma_{j_{\rm s}}^z(\gamma)-h\sum_{\gamma}\sum_{i_{\rm s}}\sigma_{i_{\rm s}}^x(\gamma),\\
H_{\rm rung}&=\sum_{\gamma\neq\mu}\sum_{i_{\rm s}\neq j_{\rm s}}\tilde{J}_{i_{\rm s},j_{\rm s}}^{\gamma,\mu}\sigma_{i_{\rm s}}^z(\gamma)\sigma_{j_{\rm s}}^z(\mu).\\
\end{split}
\end{equation}
Here, we have introduced the intra- and inter-leg Ising coupling strengths, which are respectively
\begin{equation}
\begin{split}
J_{i_{\rm s},j_{\rm s}}^{\gamma}&= \frac{J_{\rm eff}}{|{\bf \tilde{r}}_{i_{\rm s}}^0(\gamma)-{\bf \tilde{r}}^0_{j_{\rm s}}(\gamma)|^3},\\
\tilde{J}_{i_{\rm s},j_{\rm s}}^{\gamma,\mu}&= \frac{\tilde{J}_{\rm eff}}{|{\bf \tilde{r}}_{i_{\rm s}}^0(\gamma)-{\bf \tilde{r}}^0_{j_{\rm s}}(\mu)|^3},
\end{split}
\end{equation} 
where  $\tilde{J}_{\rm eff}=J_{\rm eff}\cos\tilde{\phi}_{\gamma\mu}$, such that $\tilde{\phi}_{\gamma\mu}=k_{\rm L}\cos\theta d_{\gamma\mu}$, and $d_{\gamma\mu}$ is the inter-leg distance (see Fig.~\ref{qim_ladder}). These are the central equations of this manuscript, describing a general $n_{\rm l}$-leg quantum Ising ladder, whereby the each of the legs corresponds to an antiferromagnetic quantum Ising chain with long-range dipolar interactions. The one-dimensional chains are coupled by means of dipolar Ising pairwise interactions with a strength and sign that can be experimentally controlled. In particular, we shall be interested in a ferromagnetic coupling that competes with the intra-leg antiferromagnetic interactions, leading thus to the phenomenon of magnetic frustration.

At this point, it is worth commenting on the interesting recent proposal~\cite{arbitrary_lattice} for the simulation of  any network of $N$ interacting spins by using a linear crystal of ions. By exploiting $N$ different  laser beams  individually addressed to each ion, such that the detunings and  Rabi frequencies are fine tuned, it is possible to synthesize the connectivity of any desired network. Our approach is different since it exploits  the specific  geometry of self-organized planar ion crystals, and can be scaled  to large ion ensembles straightforwardly. Additionally, it benefits from the simplicity of using  a single laser-induced dipole force that lies  far off-resonance from the whole vibrational branch. This contrasts the proposal in~\cite{arbitrary_lattice}, where the forces lie within the vibrational branch, and resonance effects must be carefully avoided for  larger ion chains where the phonon branches become denser.  We note that the scheme in~\cite{arbitrary_lattice} has a higher flexibility in the simulated lattices, although the geometry of the ladders in our approach can be partially modified following the prescriptions of Sec.~\ref{experimental}.

In the following section, we support the validity of this analytical treatment with a numerical study of the zigzag  ladder.

\section{A Detailed Case: The Zigzag Ladder}
\label{section_zz}

\subsection{ Numerical support for the anisotropic Ising model}

 We consider the simplest scenario where the anisotropy of the  Ising model  can be tested, namely,  a three-ion chain in a zigzag configuration. We  consider the following guiding numbers for   the trap frequencies $\omega_y/\omega_z=20,\omega_x/\omega_z=1.43$, and  $\omega_z/2\pi=1$ MHz, although we emphasize that the scheme will equally work for different values as far as the above constraints are met. These values lead to the equilibrium positions  
\begin{equation}
\tilde{{\bf r}}_i^0\in\{(-0.22,0,-0.92),(0.44,0,0),(-0.22,0,0.92)\}.
\end{equation}
We  focus on the crucial assumption that allows us to derive the effective frustrated Hamiltonian~\eqref{quantum_ising}, namely the possibility to neglect the pushing force on the planar modes. To address the validity of this approximation, we start from the vibrational Hamiltonian in Eq.~\eqref{vib}, and  introduce the creation-annihilation operators for the local ion vibrations 
\begin{equation}
\Delta r_{i\alpha}=\frac{1}{\sqrt{2m\omega_{\alpha}}}\bigg(a_{i\alpha}^{\phantom \dagger}+a_{i\alpha}^{\dagger}\bigg),\hspace{1ex} p_{i\alpha}=\ii\sqrt{\frac{m\omega_{\alpha}}{2}}\bigg(a_{i\alpha}^{\dagger}-a_{i\alpha}^{\phantom \dagger}\bigg).
\end{equation}
Even if  the state-dependent dipole coupling in Eq.~\eqref{dipole_force} only acts along the $xy$ plane, the motion along the $z$-axis gets coupled through the Coulomb interaction~\eqref{in_plane}. Therefore, we must treat the complete vibrational Hamiltonian 
\begin{equation}
\label{hopping_nrw}
\begin{split}
H_1=&\sum_{i}\sum_{\alpha=x,y,z}\omega_{\alpha}\big(a_{i\alpha}^{\dagger}a_{i\alpha}^{\phantom \dagger}+\half\big)+\\
+&\frac{\omega_z}{4}\sum_{\alpha,\beta}\sum_{i,j}(\kappa_{\alpha}\kappa_{\beta})^{1/4}\mathcal{\tilde{V}}_{ij}^{\alpha\beta}(a_{i\alpha}^{\phantom \dagger}+a_{i\alpha}^{\dagger})(a_{j\beta}^{\phantom \dagger}+a_{j\beta}^{ \dagger}).
\end{split}
\end{equation}
The state-dependent dipole force can also be written in this local basis, yielding the following Hamiltonian
\begin{equation}
\label{dipole_force_local}
H_{2}=\ii\frac{\Omega_{\rm L}}{2}\sum_i\sum_{\alpha=x,y}\ee^{\ii{\bf k}_{\rm L}\cdot {\bf r}^0_i}\eta_{\alpha}\sigma_i^za^{\dagger}_{i\alpha}\ee^{-\ii\omega_{\rm L} t}+\text{H.c.},
\end{equation}
where $\eta_{\alpha}=k_{{\rm L\alpha}}/\sqrt{2m\omega_{\alpha}}$ are the bare Lamb-Dicke factors.

In order to integrate numerically the Schr\"odinger equation for the timescales of interest $t_{\rm f}\approx 1/|J_{\rm eff}|$, which are  three orders of magnitude larger than the timescale set by the dynamics of the dipole force $ 1/\omega_{\rm L}$, it would be desirable to work in a picture that absorbs the fast time-dependence of~\eqref{dipole_force_local}.  This is  possible if we neglect the counter-rotating terms of Eq.~\eqref{hopping_nrw} that correspond to phonon non-conserving processes, which is justified by a rotating-wave approximation when
\begin{equation}
\frac{\omega_z}{4}(\kappa_{\alpha}\kappa_{\beta})^{1/4}\mathcal{\tilde{V}}_{ij}^{\alpha\beta}\ll (\omega_{\alpha}+\omega_{\beta}).
\end{equation}
Then, it is possible to move to a picture where the creation-annihilation operators rotate with the laser frequency, $ \ket{\tilde{\psi}(t)}={\rm exp}({\ii t\sum_{\alpha i}\omega_{\rm L}a_{i\alpha}^{\dagger}a_{i\alpha}^{\phantom \dagger}})\ket{\psi (t)}$, and the Hamiltonian that will be numerically explored $\tilde{H}=\tilde{H}_{1}+\tilde{H}_2$ becomes time-independent, namely
 \begin{equation}
\label{hopping}
\begin{split}
\tilde{H}_1&=\sum_{i,\alpha}\delta_{\alpha}a_{i\alpha}^{\dagger}a_{i\alpha}^{\phantom \dagger}+\frac{\omega_z}{2}\sum_{\alpha,\beta}\sum_{i,j}(\kappa_{\alpha}\kappa_{\beta})^{1/4}\mathcal{\tilde{V}}_{ij}^{\alpha\beta}a_{i\alpha}^{\dagger}a_{j\beta}^{\phantom \dagger},\\
\tilde{H}_{2}&=\ii\frac{\Omega_{\rm L}}{2}\sum_i\sum_{\alpha=x,y}\ee^{\ii{\bf k}_{\rm L}\cdot {\bf r}^0_i}\eta_{\alpha}\sigma_i^za^{\dagger}_{i\alpha}+\text{H.c.}.
\end{split}
\end{equation}
Note that this unitary transformation does not affect the spin dynamics, and is thus well-suited to study the validity of our previous derivation of the effective Ising model. 

In order to account for two sources of noise that are usually the experimental limiting factors for spin-oriented QSs, we  include a fluctuation of the atomic resonance frequency
\begin{equation}
\label{noise_h}
\tilde{H}_3=\sum_i\half\Delta\epsilon(t)\sigma_i^z,
\end{equation}
where $\Delta\epsilon(t)$ is a stochastic process. This term may correspond to the Zeeman shift of  non-shielded fluctuating magnetic fields, or to a non-compensated ac-Stark shift caused by  fluctuating laser intensities. Its dynamics can be modeled as a stationary, Markovian, and Gaussian process~\cite{ou_process}, as follows
\begin{equation}
\label{noise}
\Delta\epsilon(t+\delta t)=\Delta\epsilon(t)\ee^{-\frac{\delta t}{\tau}}+\big[\textstyle{\frac{c\tau}{2}}(1-\ee^{-\frac{2\delta t}{\tau}})\big]^{\frac{1}{2}}n_{\rm g},
\end{equation}
where $n_{\rm g}$ is a unit Gaussian random variable, and $c,\tau$ characterize the diffusion constant and the correlation time of the noise.  For short correlation times $\tau\ll t$, one obtains an exponential damping of the coherences  with a typical time $T_2=2/c\tau^2$. 
We set $\tau=0.1 T_2,$ and  $T_2\approx 10$ ms, which is a reasonable estimate for the observations in experiments. Note that this dephasing timescale still allows for the observation of the faster coherent spin dynamics   $1/J^{\rm eff}_{ij}\approx 1$ ms $\ll T_2$.

 We  study numerically the time evolution under the  total Hamiltonian $\tilde{H}=\sum_m\tilde{H}_m$ in Eqs.~\eqref{hopping}-\eqref{noise_h}, and compare it to the effective description $H_{\rm eff}$ for a vanishing transverse-field in Eq.~\eqref{quantum_ising}.  We note that the following numerical simulations focus on a  ground-state-cooled crystal, where the nine vibrational modes  have one  excitation at most. To consider the dephasing noise, we integrate over $N=10^3$ different histories of the  fluctuating frequency~\eqref{noise}, and perform the statistical average of the dynamics. The effects of finite temperatures, together with other error sources, are addressed in Sec.~\ref{experimental}. To observe a neat hallmark of the anisotropy due to the  modulation of the interaction strengths  $J^{\rm eff}_{ij}\propto\cos({\bf k}_{\rm L}\cdot{\bf r}_{ij}^0)$, we study the dynamics of the spin state $\ket{\psi_{\rm s}}=\ket{+}_1\otimes\ket{-}_2\otimes\ket{-}_3$, where $\ket{\pm}=(\ket{\uparrow}\pm\ket{\downarrow})/\sqrt{2}$, for two  sets of parameters. 

\begin{figure}

\centering
\includegraphics[width=0.8\columnwidth]{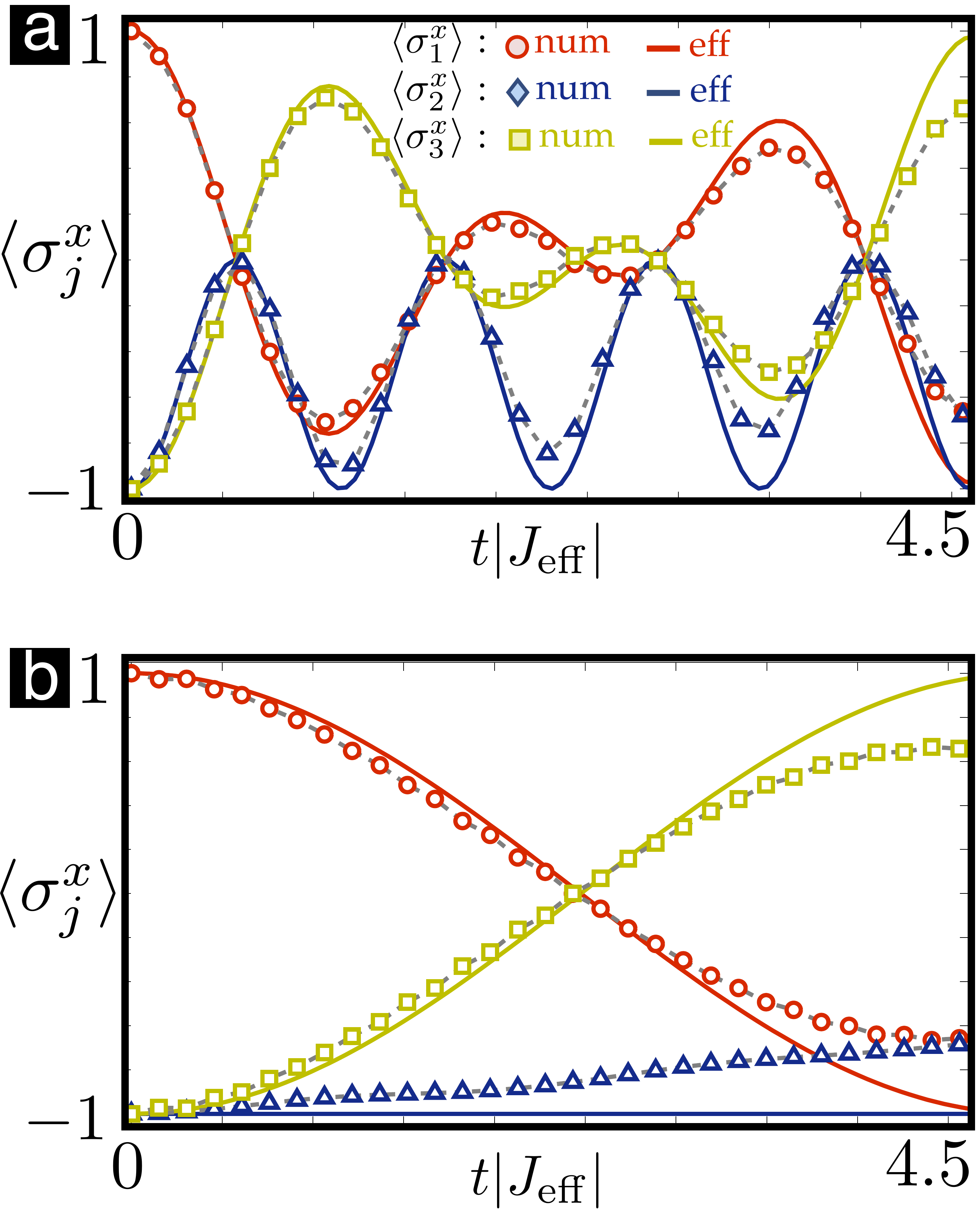}
\caption{ {\bf Spin dynamics in the triangular plaquette:}  {\bf (a)} Dynamics of an initial spin excitation $\ket{+}_1$ in the allowed-hopping regime. The dynamics of $\langle \sigma_j^x(t)\rangle$ obtained from the effective description $H_{\rm eff}$ ($\langle \sigma_1^x(t)\rangle$ red line, $\langle \sigma_2^x(t)\rangle$ blue line, $\langle \sigma_3^x(t)\rangle$ yellow line) show a remarkable agreement with the numerical simulation of the complete Hamiltonian $\tilde{H}$  ($\langle \sigma_1^x(t)\rangle$ red circles, $\langle \sigma_2^x(t)\rangle$ blue triangles, $\langle \sigma_3^x(t)\rangle$ yellow squares), and both describe the transport of the spin excitation around the plaquette. {\bf (b)} In the inhibited hopping regime, one also observes a complete agreement between both descriptions,  which describe how the spin excitation cannot occupy site $2$ as a consequence of the vanishing Ising interaction.   }
\label{spin_dynamics}
\end{figure}

{\it (i) Allowed spin hopping:} We set the following parameters $\eta_y=0.1=10\eta_x$,  $\omega_{\rm L}=1.1\omega_y$, and $\Omega_{\rm L}=0.15|\delta_y|/\eta_y$, and direct the  laser beams so that $\theta=\pi/2$ (i.e. ${\bf e}_x\cdot{\bf k}_{\rm L}=0$). In this case, the interfering radiation does not propagate along the triangle plane, and there is no modulation of the sign of the couplings. We observe that the initial spin excitation $\ket{+}$ located at site 1, can tunnel to the two remaining sites of the triangular plaquette as a consequence of the Ising-like coupling. In Fig.~\ref{spin_dynamics}{\bf (a)}, we compare the numerical results with the effective description. From this figure, we can conclude that the effective Ising Hamiltonian yields an accurate description of the spin dynamics in a timescale $ t_{\rm f}\approx1/J_{\rm eff}$- $2/J_{\rm eff}\approx1$-$2$ ms$<T_2\approx 10$ ms. For longer timescales, the dephasing leads to a larger deviation from the effective description. Note that this result  supports the validity of the isotropic Ising interaction, but we still have to address whether our scheme to control interaction anisotropy is also accurate. 

{\it (ii) Inhibited spin hopping:} We use the same parameters as before, but now consider that the interfering laser radiation also propagates along the plane defined by the triangle. In particular, we set the angle and the effective wavelength as follows $\lambda_{\rm L}/\cos\theta=4l_z(\tilde{x}_1^0-\tilde{x}_2^0)=4l_z(\tilde{x}_3^0-\tilde{x}_2^0)
$, which leads to the factors $\phi_{12}=\phi_{32}=\pi/2$, such that the effective Ising couplings between sites $1$-$2$ and $2$-$3$ completely vanish $J_{12}^{\rm eff}=J_{23}^{\rm eff}=0$. Therefore, we reach a highly anisotropic situation where  the spin excitation can only hop between sites $1\leftrightarrow 3$. This  is confirmed by the numerical results in Fig.~\ref{spin_dynamics}{\bf (b)}, where a  good  agreement with the effective description is displayed once more in the timescale $ t_{\rm f}\approx1$-$2$ ms.

 Let us finally stress that an experiment with three ions in this triangule would be a neat  proof-of-principle to show that anisotropic Ising models can be realized following our scheme. We also stress that this scheme is amenable of being scaled to larger systems, since the periodically modulated couplings are controlled globally by the ratio of the laser wavelength to the inter-leg equilibrium positions, and thus does not depend critically on the size of the Coulomb crystal.

\subsection{ Frustration by competing interactions}

\begin{figure}
\centering
\includegraphics[width=1\columnwidth]{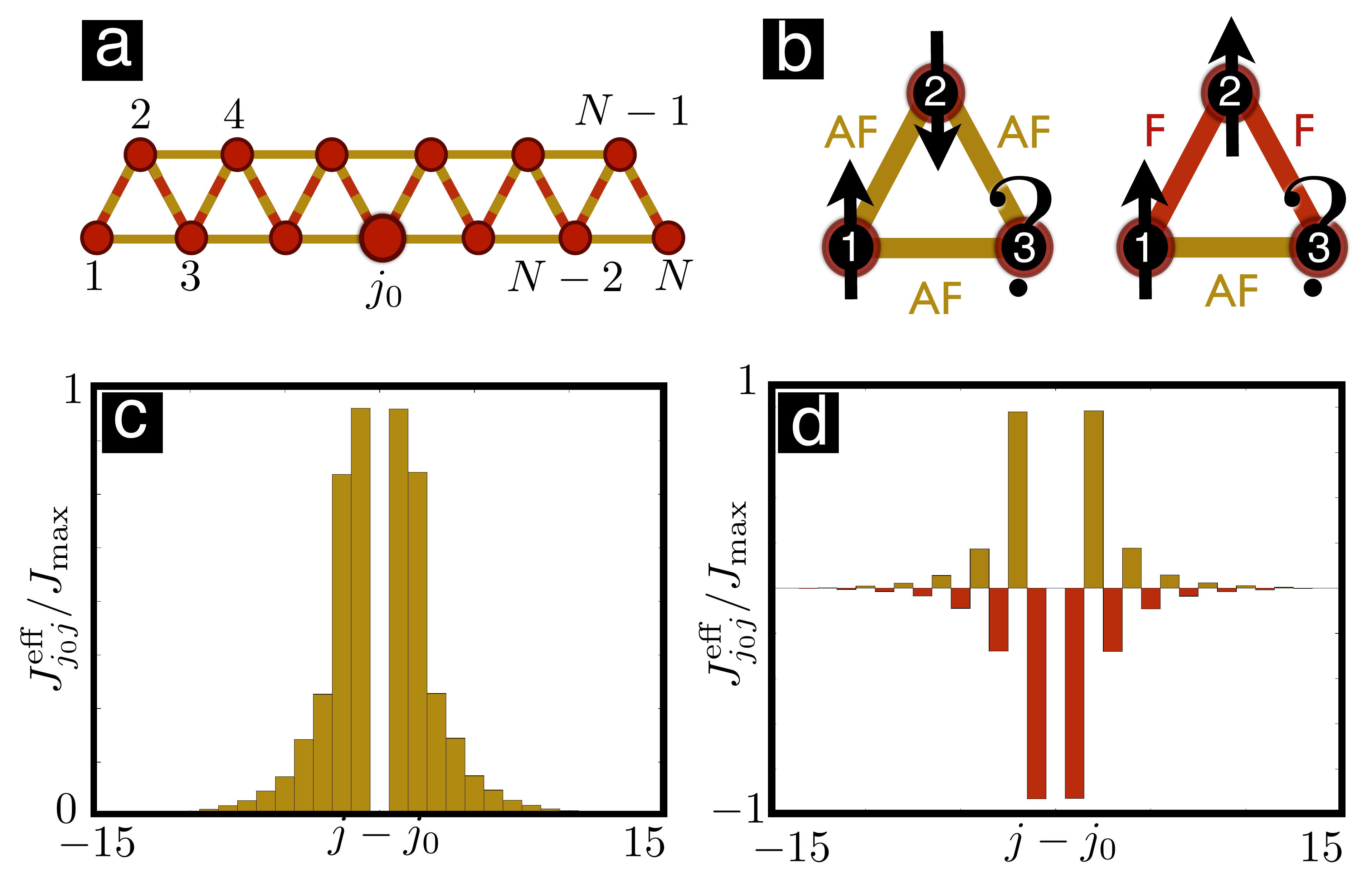}
\caption{ {\bf Spin frustration in the zig-zag ladder:}  {\bf (a)} Scheme of the  Ising interactions between neighboring ions in the zig-zag ladder, whereby the red links stand for a ferromagnetic coupling, whereas the yellow-red links can be  antiferromagnetic/ferromagnetic interactions. {\bf (b)} Frustration in a triangular plaquette  where one of the bonds cannot be satisfied by any spin configuration (left: purely antiferromagnet, right: competing interactions). {\bf (c)} Effective interaction strength between  the central ion $j_0=N/2$ and its neighbors, as obtained from the numerical solution of Eq.~\eqref{couplings} for $N=30$ ions in a  trap with  $\omega_y\gg \omega_x=6.1\omega_z$. For $\phi_{j_0,j_0+1}=0$, one  observes the dipolar decay of the frustrated antiferromagnetic interactions. {\bf (d)} For $\phi_{j_0,j_0+1}=\pi/2$, the sign alternation leads to competing interactions.   }
\label{spin_couplings}
\end{figure}

 Once the validity of the effective  Ising model~\eqref{quantum_ising} has been numerically supported, we can exploit the  anisotropic interactions~\eqref{couplings} to interpolate between a frustrated  Ising ladder due to antiferromagnetic couplings, or due to the competition of ferromagnetic and antiferromagnetic interactions. In Fig.~\ref{spin_couplings}{\bf (a)},  we present a scheme of the spin frustration. The intra-leg spin couplings in the ladder (yellow) correspond to antiferromagnetic interactions which, according to Eq.~\eqref{couplings}, cannot be modulated. On the other hand, the inter-leg couplings along the diagonal rungs of the ladder  may correspond to antiferromagnetic (yellow) or ferromagnetic (red) interactions depending on the value of $\phi_{ij}= k_{\rm L}\cos\theta (x_i^0-x_j^0)$. Both situations lead to  frustration (see Fig.~\ref{spin_couplings}{\bf (b)}), since only two of the bonds can be satisfied simultaneously. By using the normal modes of the inhomogeneous zigzag chain, we compute numerically the spin couplings, and show that for $\phi_{j_0,j_0+1}=0$, we obtain an antiferromagnetic  coupling (Fig.~\ref{spin_couplings}{\bf (c)}), whereas  an alternating sign arises for  $\phi_{j_0,j_0+1}=\pi/2$ (Fig.~\ref{spin_couplings}{\bf (d)}).

\section{ Experimental Considerations}
\label{experimental}

In  previous sections, we discussed the regimes of validity of the spin models by both analytic and numerical methods. Here, we analyze the capability of  ion-trap experiments to meet the required conditions, considering current  technology limitations and possible  sources of error in ion-trap experiments. This study  is supported by initial experiments.

\subsection{Trap design study}

A crucial condition for  the frustrated quantum spin mo\-dels is   $\omega_y\gg\omega_x\geq \omega_z$, which leads to the clustering of vibrational branches exploited in Sec.~\ref{section_model}. This condition implies that the  rotational symmetry of the trap potential in the  $xy$-plane, which is a common property of linear Paul traps, must be explicitly broken. In this section, we discuss two possible strategies to achieve this goal, and present supporting evidence based on numerical and experimental results.

\begin{figure}
\centering
\includegraphics[width=0.9\columnwidth]{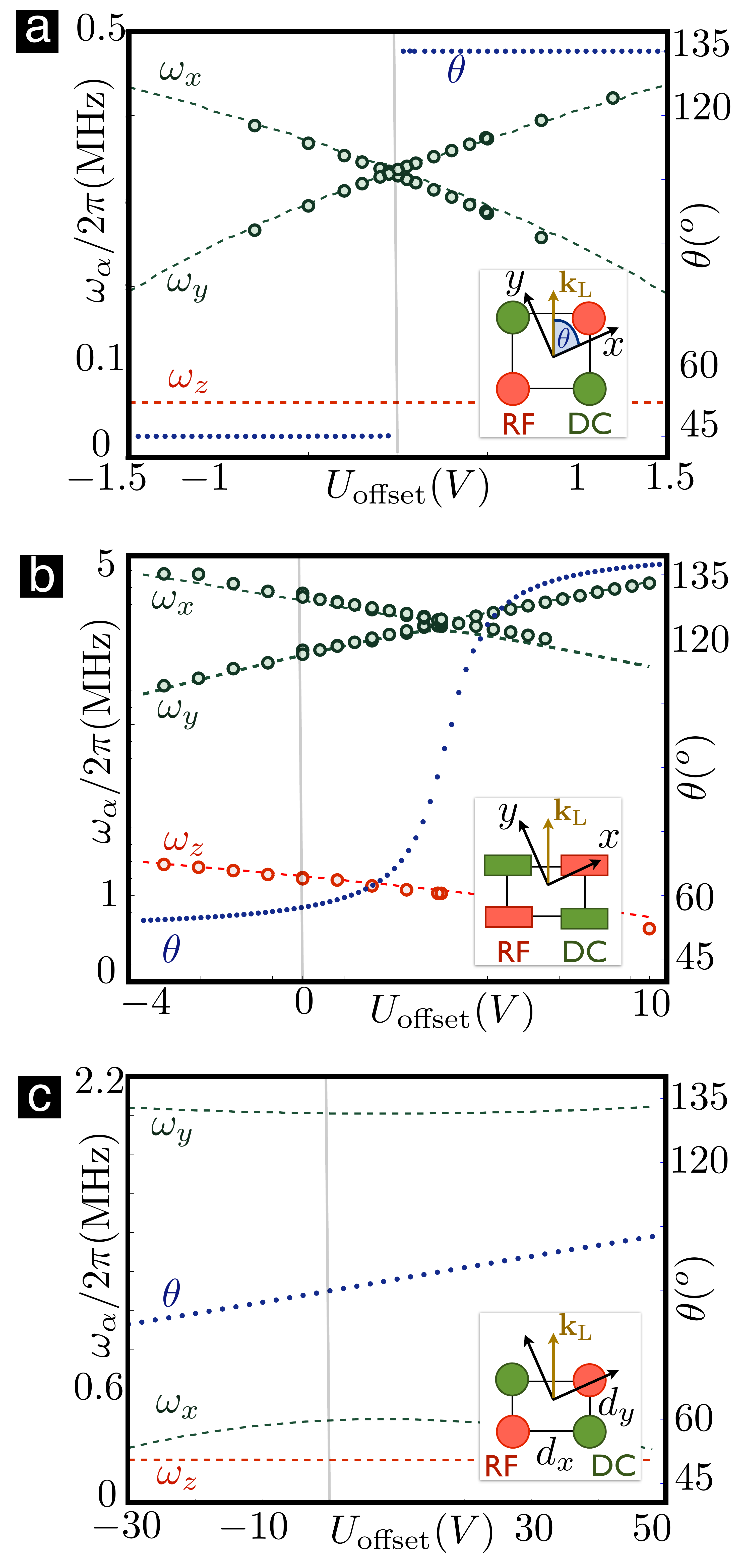}
\caption{{\bf Experimental determination and numerical simulations of the secular frequencies in different trap geometries:} Measured radial frequencies $\omega_x,\omega_y$ (green dots), calculations (green dashed lines), measured axial frequency $\omega_z$ (red dots) and calculated values (red dashed lines) for three different trap geometries. Furthermore the angle $\theta$ (blue dots) is shown. The insets in {\bf (a)-(c)} show the trap electrodes and axis seen from a radial cut through the trap. All experimental para\-meters match those of the simulation. Due to technical difficulties in determining the exact RF voltage we adapted its value by $\leq 10\%$. {\bf (a)} Quadrangular alignment ($d_x=d_y=0.82$ mm) of the four electrodes, which leads to degenerate radial frequencies for no offset-voltage and only two possible values for $\theta$. {\bf (b)} Microtrap-design as in~\cite{microtrap_details} with $d_x=125\mu$m, and $d_y=400\mu$m. {\bf (c)} Simulated values for the new trap study, as described in the text.}
\label{frequencyVSOffset}
\end{figure}

\begin{figure}
\centering
\includegraphics[width=1\columnwidth]{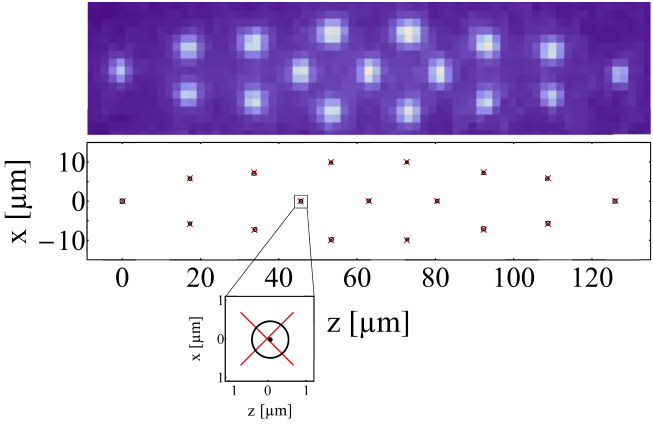}
\caption{{\bf Three-leg ion ladder:} {\bf (a)} Fluorescence of a 17 ion-crystal in the trap with $\omega_{x,y,z}/(2\pi)=\left\{260,390,111\right\} \text{ kHz}$ depicted in Fig.~\ref{frequencyVSOffset} {\bf(a)}, inset, imaged on a CCD camera. {\bf(b)} Comparison between calculated (red cross) and measured (black circle) positions show realtive errors well below one percent.}
\label{measuredCrystal}
\end{figure}

The first approach can be implemented in any linear Paul trap, such as the symmetric electrode arrangement  shown in the inset of Fig.~\ref{frequencyVSOffset}{\bf (a)}. By applying a positive offset voltage $U_{\rm offset}$ on the DC electrodes, the trapping potential becomes stronger along the dia\-gonal direction joining the DC electrodes, and the trapping frequencies fulfill $\omega_y>\omega_x$ (and vice versa for a negative offset voltage). For the experimental results shown in Fig.~\ref{frequencyVSOffset}{\bf (a)}, resonant radio-frequency radiation was used to excite the different modes, such that the induced ion motion was observed on a CCD camera. This method allows for the estimation of the trap frequencies, which show a clear agreement with numerical ion-trajectory simulations~\cite{numerical_tools}(main panel of Fig.~\ref{frequencyVSOffset}{\bf (a)}). Note that the anisotropy of the radial  frequencies $\omega_x/\omega_y$ can be tailored by the offset voltage, which also rotates the trap axes, and effectively changes  the angle $\theta$ of the laser wavevector ${\bf k}_{\rm L}$ with the $x$-axis (see Fig.~\ref{frequencyVSOffset}{\bf (a)}).  This pinpoints the possibility of shaping the spin frustration according to Eqs.~\eqref{ising_strengths}-\eqref{eq19} by modifying the electrode voltages,  avoiding  thus the more demanding modification of the laser-beam arrangement. In Fig.~\ref{measuredCrystal}, we show the measured positions for a $N=17$ ion crystal. We show how the trap-frequency anisotropy can be exploited to synthesize a particular 3-leg ladder whose equilibrium positions  match perfectly  the numerical predictions.

The second strategy is to exploit  a trap design with a non-quadrangular arrangement of the electrodes. This   breaks directly the rotational symmetry of the trapping potential (see the inset of Fig.~\ref{frequencyVSOffset}{\bf (b)}). Both the experimental data, as measured by laser spectroscopy, and the numerics  indicate a sizable splitting of the radial frequencies.  Note that the dependence of the angle $\theta$ on the offset voltage depends strongly on the trap geometry. Unfortunately, in the present case, the region with the largest tunability of  $\theta$  still coincides with the lowest radial anisotropies. 

In order to optimize  both effects, we have designed a new trap (Fig.\ref{frequencyVSOffset}{\bf (c)}) that meets with the  special  requirements of the QS. While the three-ion case-study may be realized with any state-of-the-art  experimental setup, the larger scale QS with planar spin systems with about 50 ions, as shown in Fig.~\ref{ladder_geometries}, will require a special trap design optimized according to the following conditions: (a) the trapping potential should be highly anisotropic  $\omega_y\gg\omega_x$,  (b) the trap should allow for the confinement of a large number of ions $N<$ 100,  (c) trap frequencies should be high such that an initialization of the ion crystal in a low thermal motional state is possible, (d) the trap geometry should reduce excess micromotion, and (e) optical access should allow for readout of the spin state. We now discuss the methods to achieve these requirements.

\begin{figure}
\centering
\includegraphics[width=0.9\columnwidth]{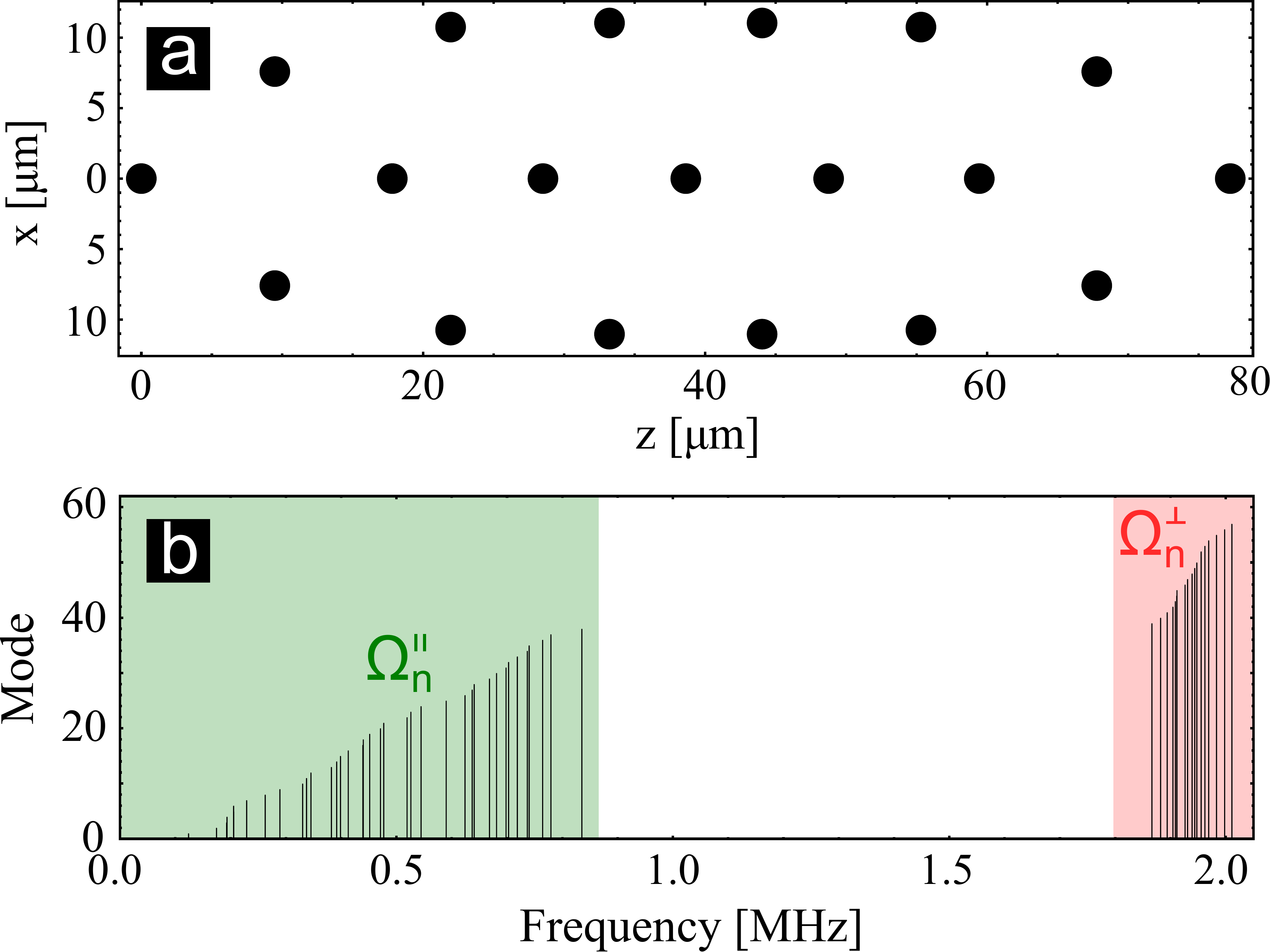}
\caption{{\bf Vibrational modes and equilibrium positions of a three-legged ladder:} {\bf (a)} Equilibrium positions for a Coulomb crystal of $N=19$ ions in a Paul trap with the axial frequency $\omega_z/2\pi=229\text{ kHz}$ and the radial frequencies $\omega_{x}/2\pi=434$~kHz,$\omega_y/2\pi= 2011\text{ kHz}$. The frequencies are generated by a special trap (see Fig.~\ref{frequencyVSOffset}{\bf (c)})design, presented in the text. {\bf (b)} Vibrational modes for the same $N=19$ ion-crystal. The transverse vibrational frequencies $\Omega_n^{\bot}$ are clearly seperated from the planar vibrational frequencies $\Omega_n^{\shortparallel}$. Note the high accord with the general description in Fig.~\ref{ladder_scheme_phonons}.}
\label{modesAndPositions19Ions}
\end{figure}

The geometry for such a trap device is sketched in the inset of Fig.~\ref{frequencyVSOffset}{\bf (c)}, where four round bars ($r=0.625 \text{mm}$) form the radial trapping potential. As we choose a non-quadrangular arrangement with distances $(d_x, d_y) = (0.42\text{ mm, } 2.90\text{ mm})$, the two simulated radial frequencies directly become  non-degenerate. With a trapping drive frequency  of $\Omega_{\rm rf}/2\pi=22$~MHz, and amplitude of 2.5~kV, we find a sufficiently high anisotropy $\omega_{x}/2\pi=0.43$~MHz$<\omega_{y}/2\pi=2.01$~MHz along the  requirement (a). The axial confinement is generated by two endcaps seperated by a distance of 25~mm, which lead to $\omega_z/2\pi=0.23$~MHz with 2~kV applied on the endcaps. In this potential, a $N=4$ ion crystal undergoes the first structural transition to the zigzag configuration, and for $N>9$ the second structural transition is obtained. We have calculated the equilibrium  positions and vibrational modes of a three-legged ladder of $N=19$ ions for these particular trap frequencies (Fig.~\ref{modesAndPositions19Ions}). We note that by lowering the axial DC voltage, we can increase  $N$ for  different crystal structures, fulfilling the requirement (b). Let us now address condition (c). By setting the trap frequency to $\omega_y/2\pi=$2.01~MHz,  the mean phonon numbers after Doppler cooling lie below $\bar{n}_y<5$. In order to have a thermal error below $1\%$ for these phonon numbers (see Eq.~\eqref{thermal_error_app}), the spin-phonon coupling should be smaller than $\Omega_{\rm L}=0.02|\delta_y|/\eta_y$. However, this reduces the spin-spin interactions and magnetic-field noise may affect the dynamics at the corresponding long timescales. Therefore, multi-mode EIT cooling \cite{EIT_cooling} to $\bar{n}_y=0.1$ shall allow us to keep the error rate to $1\%$, while maintaining spin-spin interactions in the kHz-regime (i.e. $\Omega_{\rm L}=0.15|\delta_y|/\eta_y$). Let us stress that the thermal error may be minimized by considering evolution times that are multiples of the detuning of the closest vibrational mode. In order to estimate the effects of the excess micromotion according to the point (d), we have estimated the $\xi_i$ values to be $\xi_{2i}=0$ and $|\xi_{1i}|<0.05 \pi$ with the same beam angles as in Eq.~(\ref{beamGeometry}). This should lead to relative errors on the order of  4-5\%.  Finally, the inter-ion distances are about 10~$\mu$m for the central ions, such that we can meet with requirement (e) via high numerical aperture optics allowing for single-site readout. 

With the above parameters, the relative orientation of the ion crystal and the laser wavevector ${\bf k}_{\rm L}$ yields an angle of $\theta=\pi/2$. As discussed above, we can vary $\theta$ and tune the spin frustration by applying an offset to the DC electrodes, which  rotates the crystal with respect to the fixed wavevector ${\bf k}_{\rm L}$. As shown in the simulations presented in Fig.~\ref{frequencyVSOffset}{\bf (c)}, where the new trap design allows for a smooth tunability of $\theta$, while preserving a strong anisotropy $\omega_x/\omega_y$. With $U_{\rm offset}=-4.9$~V, we find  $\omega_x/\omega_y=0.21$, while the angle becomes $\theta=0.49\pi$,  already leading to a change of sign of the spin-spin coupling strength $J^{\rm eff}_{ij}$ in Eq.~\eqref{eq19}, if we assume $\lambda_{\rm L}= 400$~nm.

\subsection{Tailoring the ladder geometry}

 In Figs.~\ref{ladder_geometries}{\bf (b)}, {\bf (c)} and {\bf (f)}, the ions self-organize naturally in a geometry of bond-sharing triangles. By controlling the number of legs in such triangular ladders, the QS is already capable of exploring a variety of interesting cooperative phenomena (see Secs.~\ref{section_scope} and~\ref{section_j1_j2}). However, it would be highly  desirable to have a   method to modify these geometries,  widening thus the applicability of our quantum simulator.

Since the  geometry of the  spin model~\eqref{quantum_ising} is determined by the couplings $J_{ij}^{\rm eff}$  between the spins $\sigma_i^z\leftrightarrow \sigma_j^z$, a possibility to tailor the geometry  is  to switch on/off some of these interactions. A possible route that allows for such a selective coupling is a type of { laser-beam hideout}. Thanks to the relatively large distances between the ions $l_{z}\approx $1-10 $\mu$m, it is possible to address them individually with laser light~\cite{individual_adressing} (note that the tools used so far only require global addressing).  The idea is that, after the global optical pumping to the state $\ket{\psi_{\rm s}}=\otimes_i\ket{\downarrow_i}$,   focused   laser radiation will selectively transfer the population to a different level $\ket{h_i}$  that is not coupled to the spin-dependent force. Therefore, the ion gets  effectively hidden. This may be achieved by polarization selection rules, or alternatively,  by the ac-Stark shift of highly-detuned laser beams. Using this idea, it is possible to  construct a simple ladder of corner-sharing triangles (see Fig.~\ref{ladder_geometries}{\bf (c)}), or a stripe of the two-dimensional Kagome lattice (see Fig.~\ref{ladder_geometries}{\bf (d)}). Additionally, it opens the possibility of introducing  defects in the lattice in order to study the role of disorder. As discussed in Sec.~\ref{section_scope}, this tool opens many possibilities for our QS.

\subsection{ Imperfections and noise in the quantum simulator}

 Let us now address the possible sources of error in the QS. We place a  special emphasis on  the ion micromotion, which has not been discussed in previous QSs~\cite{ising_ions,frustration_monroe,ising_monroe}, since it can be cancelled for linear  chains. We also discuss other  error sources shared with the linear-chain QS, such as thermal fluctuations and heating of the phonons, dephasing of the spins, and photon scattering. We  stress that, provided that the following constraints  to minimize the micromotion errors are fulfilled, the proposed  ladder QS should not present more limitations than its linear-chain counterparts~\cite{ising_ions,frustration_monroe,ising_monroe}.

{\it (i) Sources of dephasing:} Usually, the most important important source of error  in the experiments is the dephasing of the electronic states. These might be  caused by fluctuating Zeeman shifts on the magnetic-field sensitive states required to implement the spin-dependent dipole force~\eqref{dipole_force}; or  by fluctuations in the laser-beam intensities leading to uncompensated  ac-Stark shifts. A reasonable estimate of the typical coherence times  $T_2\approx 5$-10 ms~\cite{ion_trap_reviews} shows that these terms are important error sources in the timescales of interest $t_{\rm f}\propto1/J_{\rm eff}$, where $J_{\rm eff}/2\pi\approx 1$ kHz. As shown in Fig.~\ref{spin_dynamics}, the desired coherent dynamics described by the effective Ising model~\eqref{quantum_ising} dominates the behavior of the system for $t_{\rm f }\approx 1$-$2$ ms. This sets the timescale for the QS of frustrated magnetism to be in the millisecond range, after which the dephasing is so strong that the quantum simulator is no longer faithful.  

An interesting protocol to overcome both sources of dephasing simultaneously is based on the concept of continuous dynamical decoupling already applied to trapped ions~\cite{timoney}. In our case, it could be implemented   by a strong driving of the carrier transition~\cite{sideband_decoupling}. Hence, instead of relying on the spin-dependent force~\eqref{dipole_force}, it suffices to combine a red-sideband term with the strong driving of the carrier. In order to obtain a quantum Ising model, one should  introduce an additional microwave that provides a Zeeman shift oscillating at the Rabi frequency of the  strong carrier driving. The phase diagram can be explored adiabatically by an intermediate spin-echo pulse that refocuses the fast oscillations due to the strong driving.

{\it (ii) Role of the ion  micromotion:} The initial Hamiltonian in Eq.~\eqref{coulomb_ham}, upon which the derivation of the effective quantum spin ladders~\eqref{qi_ladder} has been built,  is based on the so-called  pseudo-potential approximation~\cite{ion_trap_reviews}. This approximation neglects the effects of the ion {micromotion}: a fast  motion of frequency  $\Omega_{\rm rf}$ that is synchronous to the radio-frequency (r.f.) field  of the Paul trap.  This allowed us to focus directly on the slow secular motion of the trapped ions, which is described  by an effective harmonic potential~\eqref{coulomb_ham} with frequencies $\omega_{\alpha}\ll \Omega_{\rm rf}$. The validity of this assumption is well justified for single trapped ions, where there are experimental methods to minimize the effects of the micromotion~\cite{ion_trap_reviews}. For linear ion chains, the micromotion is also minimized when the equilibrium positions lie along the trap axis.  Conversely, for the ion ladders considered in this work, the equilibrium positions necessarily lie off the trapping axis,  leading to an additional  micromotion that cannot be compensated.  

In Appendix~\ref{appendix_micromotion}, we present a detailed discussion of the conditions under which  this micromotion does not affect the  QS. Let us comment on the   possible sources of imperfections, and the  conditions to  minimize them. The r.f. heating of the transverse phonon modes responsible of the spin couplings is negligible when
\begin{equation}
\kappa_y^{1/2}|\mathcal{\tilde{V}}^{yy}_{ij}|\ll \frac{\Omega_{\rm rf}}{\omega_z},
\end{equation}
a condition easily verified for the parameters in this work. In addition, during the cooling stage of the transverse vibrational modes, the laser frequency must be carefully tuned to avoid heating caused by the micromotion-induced broadening of the transition, or to additional micromotion sidebands~\cite{micromotion_wineland}.

We have also considered how the micromotion may affect the spin-dependent dipole force~\eqref{dipole_force} and the spin-phonon coupling~\eqref{transverse_pushing}. To neglect the associated contributions, it is necessary to consider that the laser beams lie almost parallel to the $y$-axis, and are far off-resonant (see Fig.~\ref{level_scheme}), namely $\omega_{\rm L}/2\pi\approx\omega_y/2\pi\approx$10 MHz$\ll \Omega_{\rm rf}/2\pi\approx 0.1$ GHz$\ll \Delta/2\pi\approx10$ GHz. Besides, the Rabi frequencies of the laser beams should fulfill
\begin{equation}
|\Omega_{l,\uparrow}|,|\Omega_{l,\downarrow}|\ll\Delta,\hspace{3ex} \frac{|\Omega_{1,\uparrow}\Omega_{2,\downarrow}^*|}{2\Delta}\ll\omega_{\rm L}\ll\Omega_{\rm rf}.
\end{equation} 
Finally, some care must also be placed in order not to induce two-photon transitions to different states $\ket{a_i}$ of the atomic ground-state manifold. To avoid these processes, the relative Zeeman shifts of such transitions $\delta_{a,s}$, and the associated Rabi frequencies must be controlled such that 
\begin{equation}
|\Omega_{1,a}\Omega_{2,s}^*|\ll |\delta_{a,s}-\omega_{\rm L}\pm\Omega_{\rm rf}|.
\end{equation}
We thus conclude that, provided that the above restrictions on the parameters are fulfilled, the unavoidable excess micromotion of the ions in a ladder geometry should not affect our QS.

{\it (iii) Thermal motion and heating of the ions:} In Sec.~\ref{section_zz}, we have supported the validity of the QS based on numerical results for ground-state cooled ion crystals. However, ever since the early schemes for phonon-mediated  gates,  the thermal fluctuations of the ions have been  identified as a potential source of errors that must be carefully considered. 

In Appendix~\ref{heisenberg}, we estimate  the thermal error of our QS by both analytic and numerical methods. We solve exactly the Heisenberg equations of motion for the spin-dependent dipole force~\eqref{time_indep}, and derive a scaling law for the thermal contribution to the $T=0$  spin dynamics, which is then supported numerically. We find that the relative thermal error, defined as $\epsilon_T=|\langle\sigma_i^x\rangle_T-\langle\sigma_i^x\rangle_{T=0}|/|\langle\sigma_i^x\rangle_{T=0}|$, has an upper bound
\begin{equation}
\label{thermal_error}
\epsilon_{T}\leq\epsilon_{\rm th}=4\frac{|\Omega_{\rm L}|^2\eta_y^2}{\delta_y^2}\bar{n}_y,
\end{equation}
where  $\bar{n}_y$ represents the mean number of transverse phonons in the center-of-mass mode. By considering the parameters used in section~\ref{section_zz}, the thermal fluctuations have a small contribution (1\%) for $\bar{n}_y\approx0.1$  (see Fig.~\ref{thermal_error_fig}),  which shows that perfect ground-state cooling is not required to implement the  QS accurately.
 At this point, we should mention that the above arguments are valid for a vanishing transverse field $h=0$. We note, however, that the contribution to the thermal error due to $h>0$ in the regime of interest $h\approx J_{\rm eff}$ has been shown~\cite{ising_porras} to be negligible with respect to the estimate in Eq.~\eqref{thermal_error}. 

In Appendix~\ref{heisenberg}, we also present a phenomenological master equation that models possible heating mechanisms in the ion trap. We show that the relative heating error, defined as  $
\epsilon_{{\rm h}}=|\langle\sigma_i^x\rangle_{\Gamma_{\rm h}}-\langle\sigma_i^x\rangle_{\Gamma_{\rm h}=0}|/|\langle\sigma_i^x\rangle_{\Gamma_{\rm h}=0}|$, where $\Gamma_{\rm h}$ is the heating rate, only yields a small contribution (1\%) to the overall error of the QS for heating times above 5 ms/ phonon (see Fig.~\ref{heating_error_fig}).

{\it (iv) Spontaneous photon scattering:} The spontaneous emission  can severely limit the advantage of quantum-information protocols~\cite{decoherence_plenio,photon_scattering,photon_scattering_2}. In our case, the  use of two electronic ground-states as the effective spins (see Fig.~\ref{level_scheme}) makes the direct spontaneous emission  $\ket{\uparrow_i}\rightsquigarrow\ket{\downarrow_i}$ negligible. However, due to the Lambda-scheme responsible for the dipole force, there can be photon scattering events from the  intermediate excited state $\ket{r_i}$, which must be carefully   considered.

In Appendix~\ref{appendix_dipole}, we describe in detail  the derivation of an effective master equation ${\rm d}\rho/{\rm d}t=-\ii[H_{\rm d},\rho(t)]+\mathcal{D}_{\rm eff}(\rho)$, where $H_{\rm d}$ corresponds to the dipole force~\eqref{dipole_force}, and 
\begin{equation}
\mathcal{D}_{\rm eff}(\rho)=\sum_{n}\left(L^{\rm eff}_n\rho (L^{\rm eff}_n)^{\dagger}-\half (L^{\rm eff}_n)^{\dagger}L^{\rm eff}_n\rho-\half\rho (L^{\rm eff}_n)^{\dagger}L^{\rm eff}_n\right),
\end{equation}
accounts for the two possible decoherence channels. These are the so-called Raman and Rayleigh scattering of photons, and are contained in the following effective jump operators
\begin{equation}
\label{jumps}
\begin{split}
L_1^{\rm eff}=\frac{\sqrt{\Gamma}}{\Delta}&\left(\Omega_{1,\downarrow}\ee^{\ii({\bf k}_1\cdot{\bf r}-\omega_1t)}+\Omega_{2,\downarrow}\ee^{\ii({\bf k}_2\cdot{\bf r}-\omega_2t)}\right)\ket{{\downarrow}}\bra{{\downarrow}}+\\
+\frac{\sqrt{\Gamma}}{\Delta}&\left(\Omega_{1,\uparrow}\ee^{\ii({\bf k}_1\cdot{\bf r}-\omega_1t)}+\Omega_{2,\uparrow}\ee^{\ii({\bf k}_2\cdot{\bf r}-\omega_2t)}\right)\ket{{\downarrow}}\bra{{\uparrow}},\\
L_2^{\rm eff}=\frac{\sqrt{\Gamma}}{\Delta}&\left(\Omega_{1,\uparrow}\ee^{\ii({\bf k}_1\cdot{\bf r}-\omega_1t)}+\Omega_{2,\uparrow}\ee^{\ii({\bf k}_2\cdot{\bf r}-\omega_2t)}\right)\ket{{\uparrow}}\bra{{\uparrow}}+\\
+\frac{\sqrt{\Gamma}}{\Delta}&\left(\Omega_{1,\downarrow}\ee^{\ii({\bf k}_1\cdot{\bf r}-\omega_1t)}+\Omega_{2,\downarrow}\ee^{\ii({\bf k}_2\cdot{\bf r}-\omega_2t)}\right)\ket{{\uparrow}}\bra{{\downarrow}},\\
\end{split}
\end{equation}
where we have assumed that the scattering rate is much smaller than the laser detuning $\Gamma\ll\Delta$. Note that the first term of each jump operator does not  change the spin state (i.e. Rayleigh scattering), being responsible for pure dephasing. Conversely, the second term flips  the spin state (i.e. Raman scattering), leading thus to damping of the spin populations.

  A conservative estimate of the photon scattering is to consider the individual  effective rates~\eqref{jumps}, which scale as $\Gamma_{\rm eff}=\Gamma(|\Omega_{l,s}|/\Delta)^2\approx |\Omega_{\rm L}|(\Gamma/\Delta)$. By using the parameters introduced in Sec.~\ref{section_zz}, we find that $\Gamma_{\rm eff}/J_{\rm eff}\approx 10^3(\Gamma/\Delta)$. Hence, by considering a sufficiently large detuning, it is possible to keep the scattering rates below $\Gamma_{\rm eff}/J_{\rm eff}<10^{-1}$, so that they do not compromise the accuracy of the effective description~\eqref{quantum_ising}. For instance, for typical decay rates $\Gamma/2\pi\approx$1-10 MHz, one must consider detunings in the range $\Delta/2\pi\approx$10-100 GHz.
  
  {\it (v) Spatial dependence of the laser-beam profile:} The effective spin couplings in Eq.~\eqref{eq19} assume a Rabi frequency that is constant along the ion crystal. For large crystals, however, the characteristic gaussian  profile for the laser intensities may lead to weaker Rabi frequencies on the boundaries of the crystal. This effect will give rise to inhomogeneous spin-spin couplings that must be added to the inhomogeneities caused by the varying inter-ion distance in Coulomb crystals. Rather than considering these terms as an error, they can be seen as a gadget that makes the many-body problem even more interesting. In particular, they will lead to  inhomogeneous critical points which may be responsible of interesting effects~\cite{zigzag_inhomogeneous}.

\subsection{ Efficient detection methods}

 A crucial part of a QS is the ability to perform measurements  that yield information about the Hamiltonian under study. For the frustrated quantum Ising ladders~\eqref{qi_ladder}, a QS would  start by preparing the so-called paramagnetic state $\ket{\rm P}=\otimes_i\ket{{\rightarrow}}_i$ with $\ket{{\rightarrow}}_i=(\ket{{\uparrow}}_i+\ket{{\downarrow}_i})/\sqrt{2}$, which is the ground-state of the model in the absence of spin interactions. Such a separable state can be accurately prepared by means of optical pumping, followed by  $\frac{\pi}{2}$-pulses globally addressed to the whole ion crystal~\cite{ion_trap_reviews}. This step is followed by the adiabatic modification of the Hamiltonian parameters $J_{\rm eff},\tilde{J}_{\rm eff},h$, in order to connect the paramagnet to other phases of the quantum Ising ladder (see Secs.~\ref{section_scope} and~\ref{section_j1_j2} below). 

 A direct approach to the  measurement of these phases  is the so-called quantum state tomography, more precisely, the full determination of the   state of the system~\cite{state_tomography}. However, this approach becomes highly inefficient for many-body systems due to the exponential growth of the composite Hilbert space, and  alternative schemes must be studied. An interesting possibility for state estimation are the methods based on  matrix-product representations of the states~\cite{efficient_tomography}.  Another alternative that does  not require  full quantum state tomography is the measurement of  order parameters characterizing the phases.

One of the advantages of trapped-ion experiments with respect to other platforms is their ability to perform highly-accurate measurements  at the single-particle level~\cite{ion_trap_reviews}.   The technique of state-dependent fluorescence allows for the measurement of single and joint probability distributions of the electronic states $P_{i}^{\uparrow},P_{ij}^{\uparrow\uparrow}$. From the spatially resolved  fluorescence, it is possible to infer   local expectation values, such as  the magnetization $\langle\sigma_i^z\rangle=\frac{1}{2}(P_i^{\uparrow}-1)$, or  two-body  correlators $\langle\sigma_i^z\sigma_{j}^{z}\rangle=\frac{1}{4}[1-2(P_i^{\uparrow}+P_{j}^{\uparrow})+4P_{ij}^{\uparrow\uparrow}]$. As discussed  in Sec.~\ref{section_j1_j2}, these observables usually contain all the relevant information  about the different phases. In particular, one could study the dependence of the correlator with the distance~\eqref{corr_scaling}, or infer the magnetic structure factor $S_{zz}(q)=\sum_{ij}\langle\sigma_i^z\sigma_j^z\rangle\ee^{\ii q(i-j)}$.

Let us now comment on the possibility of recovering some of these magnitudes from  global properties of the fluorescence spectrum. By measuring the probability to find a fraction of $n$-ions in the excited state $P^{\uparrow}({n})$~\cite{ising_monroe}, one  obtains the total magnetization without single-site resolution $m_z=\frac{1}{N}\sum_i\langle\sigma_i^z\rangle$. To measure correlators, note that the fluorescence of an ensemble of emitters may carry information about their correlations~\cite{collective_fluorescence}. In our case,  the resonance fluorescence associated to the cycling transition  
 depends on the collective properties of the ion ensemble~\cite{spin_correlations}. In fact,  the power spectrum in a particular detection direction, $\hat{\bf r}$, is related to the structure factor 
 \begin{equation}
 \mathcal{S}_{\hat{\bf r}}(\omega)\propto\sum_{ij}\ee^{\ii \frac{2\pi}{\lambda}{\bf \hat{r}}\cdot({\bf r}^0_i-{\bf r}^0_j)}\langle(1+\sigma_i^z)(1+\sigma_j^z)\rangle\propto S_{zz}(q),
 \end{equation}
  where $\lambda$ is the wavelength of the emitted light.  Even if the ion spacing is  much larger than the optical wavelength $l_z\gg\lambda$, one may compensate it by setting the photodetector almost orthogonal to the plane defined by the ladder, and by using photodetectors with a very good angular resolution. We finally note that such magnetic structure factors yield a lower bound on the entanglement  without the need of state tomography~\cite{ent_bound}.

\section{ Scope of the Quantum Simulator}
\label{section_scope}

Once  the validity of the  spin-ladder Hamiltonian~\eqref{qi_ladder} has been addressed by both analytic and numerical methods  (Secs.~\ref{section_model} and~\ref{section_zz}), and its experimental viability discussed (Sec.~\ref{experimental}),  we can now focus on the many-body models to be explored with the QS. We place a special emphasis on the range of collective phenomena that are not fully understood, or have not been addressed so far to the best of our knowledge. These would directly benefit from the advent of such a QS.

\subsection{ $J_1$-$J_2$ quantum Ising model} 

The first many-body model that may be targeted with the proposed QS is the so-called axial next-to-nearest neighbor Ising  model~\cite{classical_annni}, supplemented by quantum fluctuations (see~\cite{itf_glass} and references therein). It has the  Hamiltonian
\begin{equation}
\label{qannni}
H_{\bigtriangleup\hspace{-1.1ex}\bigtriangledown\hspace{-1.1ex}\bigtriangleup}=J_1\sum_i \sigma_{i}^z\sigma_{i+1}^z+J_2\sum_{i}\sigma_i^z\sigma_{i+2}^z-h\sum_i\sigma_i^x,
\end{equation}
which consists of nearest neighbor ($J_1<0$) and next-to-nearest neighbor ($J_2>0$) couplings, and the transverse field $h$. In spite of the mild looking appearance of this Hamiltonian, it has  a rich phase diagram with some features that  are still controversial.   We refer to this model as the $J_1$-$J_2$ {quantum Ising model} ($J_1$-$J_2$QIM), a paradigm of frustrated  magnetism. 

The Hamiltonian of our QS~\eqref{qi_ladder} corresponds to the $J_1$-$J_2$QIM for the simplest  possible geometry, namely, the two-leg zigzag ladder~\cite{frustration_ulm}. The indices of this ladder, $\gamma=1,2$, $i_{\rm s}=1,\cdots,N/2$ (Fig.~\ref{qim_ladder}), are mapped onto a one-dimensional chain by the relation $i=2(i_{\rm s}-1)+\gamma$ (Fig.~\ref{spin_dynamics}{\bf (a)}). Then, the analogy with the $J_1$-$J_2$QIM follows directly provided that 
\begin{equation}
J_1=\tilde{J}^{12}_{i_{\rm s},i_{\rm s}}=\frac{J_{\rm eff}\cos\phi_{i,i+1}}{|\tilde{{\bf r}}_i^0-\tilde{{\bf r}}_{i+1}^0|^3},\hspace{1ex}J_2=J^{1}_{i_{\rm s},i_{\rm s}+1}=\frac{J_{\rm eff}}{|\tilde{{\bf r}}_i^0-\tilde{{\bf r}}_{i+2}^0|^3},
\end{equation}
Note that due to the particular laser-beam arrangement presented in Sec.~\ref{section_model}, it is possible to tailor the ratios $J_2/J_1$ and $h/J_1$ experimentally. We emphasize that, even if a solid-state material is found to be described by such a model,  a similar  microscopic control of the couplings seems rather difficult to achieve. Therefore, the trapped-ion platform is ideal to explore the different regions of the  phase diagram~\cite{note_j1j2}. Of particular relevance is the region around the frustration point $f_{\rm c}=J_2/|J_1|=1/2$, which is characterized by a macroscopically-degenerate ground-state. Therefore, even a small amount of quantum fluctuations due to the  transverse field may lift the classical degeneracy leading to a variety of magnetic phases.

 Such a rich phase diagram is studied numerically in Sec.~\ref{section_j1_j2}, where we identify some additional features caused by  the dipolar range of interactions present in  the trapped-ion QS. Let us briefly mention that these long-range interactions introduce incompatible sources of frustration capable of stabilizing a  new order that complements the  ferromagnetic, dimerized antiferromagnetic, and floating phases that are also present in the short-range model~\eqref{qannni}. Hence, our QS will be of the utmost interest to explore the interplay between frustration, quantum fluctuations, and long-range interactions. Besides, the proposed QS shall be able to address some open questions  about the phase diagram  that  are still a subject of  controversy~\cite{itf_glass}, such as the extent of the floating phase and the existence of a multi-critical Lifshitz point. 

\subsection{Dimensional crossover and quantum dimer models}

 An ambitious  enterprise is the understanding of the dimensional crossover from the two-leg quantum Ising ladder onto the two-dimensional (2D) triangular quantum Ising model
\begin{equation}
\label{2d}
\begin{split}
H_{\rm TQIM}=&\sum_{mn}J_2\sigma_{m,n}^z\sigma_{m+1,n}^z+J_1\sigma_{m,n}^z\sigma_{m,n+1}^z+J_1\sigma_{m+1,n}^z\sigma_{m,n+1}^z\\
-&\sum_{mn}h\sigma_{m,n}^x,
\end{split}
\end{equation}
where the spins are labelled according to the 2D Bravais lattice vectors ${\bf r}_{m,n}=m{\bf a}_1+n{\bf a}_2$ (see Fig.~\ref{qim_ladder}), and we consider anisotropic couplings $J_1<0$ and $J_2>0$. The correspondence with our trapped-ion QS~\eqref{qi_ladder} is straightforward
if one considers that $m$ labels the spins within each leg of the ladder coupled by $J_2\leftrightarrow J$, and $n$ the different legs coupled by $J_1\leftrightarrow \tilde{J}$. Note that, following recent experimental efforts~\cite{penning,ising_bollinger}, the triangular QIM may also be realized with ions in Penning traps~\cite{2d_porras}. However, the study of the ladders and the crossover phenomena seems to be better suited to ions in Paul traps.

The triangular classical  Ising model can be considered as the backbone of frustrated magnetism~\cite{ising_af_triangular}. Already in the absence of quantum fluctuations, there are  suggestive questions that deserve a careful consideration. For instance, the frustration point of the  two-leg zigzag ladder $f_{\rm c}=1/2$ flows to the isotropic point $f_{\rm c}=1$ in the 2D model. Moreover, the macroscopic ground-state degeneracies of these models yield different ground-state entropies. We believe that it would be fascinating to explore these topics with trapped ions, which allow for the  consecutive increase of the number  legs (Figs.~\ref{ladder_geometries}{\bf (c)},{\bf (f)}).

The dimensional crossover is even more exotic when quantum fluctuations are included. Let us note that the dimerized antiferromagnet of 
the two-leg zigzag ladder corresponds to two possible classical dimer coverings  of the ladder (see Sec.~\ref{section_j1_j2}), where each dimer corresponds to a nearest-neighbor bond that is not satisfied due to the frustration. Since the ground-state tries to minimize the number of dimers, each site of the dual lattice belongs to only one dimer, and the covering consists of the dimer arrangement along the rungs of the ladder. Note that each spin is connected to three satisfied bonds, and only one broken bond. The situation gets  more interesting 
for the 2D quantum Ising model, since there, a single spin may be connected to the same number of satisfied and broken bonds. In this case, the transverse field can flip the spin and produce a resonating effect for neighboring dimers~\cite{qim_frustration}, providing a beautiful connection to the so-called {\it quantum dimer models} (see~\cite{qdm} and references therein).

Quantum dimer models were introduced~\cite{qdm_rk}   in the context oh high-temperature superconductivity. Here, these models provided a neat playground where to study the quantum spin liquid phases of the undoped cuprates,  which were conjectured to play a key role in the onset of superconductivity  upon doping~\cite{rvb_anderson}. However, they have evolved into an independent subject displaying  exotic  effects, such as topologically ordered phases and fractional excitations.  Although originally introduced for Heisenberg antiferromagnets, where the dimers correspond to spin singlets, there is also a link to frustrated quantum Ising models by mapping the dimers to the broken magnetic bonds due to the frustration~\cite{qim_frustration}.   

Their connection to the isotropic triangular quantum Ising model~\cite{qim_frustration} has allowed to predict  a quantum version of the phenomenon of 'order by disorder'~\cite{od}, which yields an ordered phase out of the classically disordered frustration point $f_{\rm c}=1$ when quantum fluctuations are switched on. In contrast,  by switching the transverse field on  the  frustration point $f_{\rm c}=1/2$ of  the two-leg zigzag ladder, one only obtains   a disorder paramagnetic phase (see Sec.~\ref{section_j1_j2}). Therefore, the trapped-ion QS offers a unique playground to study how the phenomenon of  'order by disorder' sets in as the number of legs is increased, and how the anisotropy and the long range of the interactions  affects it.  

\subsection{Quantum spin liquid phases}

 In the  models considered so far, strong quantum fluctuations trigger a phase transition connecting an ordered phase to the uninteresting disordered paramagnet. However, this behavior  does not exhaust all possibilities. Quantum fluctuations may be responsible of  stabilizing more exotic  phases that do not break any symmetry of the Hamiltonian, the so-called quantum spin liquids~\cite{spin_liquids}.

We have discussed how the two-leg zigzag ladder consisting of bond-sharing triangles allows for the QS of the paradigm of  FQIM. However, this model only accounts for a disordered paramagnet. There exists another simple two-leg ladder which, although not so widely known,  has been argued to provide the simplest instance of a frustration-induced quantum spin liquid~\cite{qim_frustration_sawtooth} (see also~\cite{qim_frustration}). This is the so-called {\it sawtooth quantum Ising model}, which consists of corner-sharing triangles described by the following Hamiltonian 
\begin{equation}
\label{sawtooth}
H_{\bigtriangleup\hspace{-0.55ex}\bigtriangleup}=J_1\sum_i \sigma_{i}^z\sigma_{i+1}^z+J_2\sum_{i}\sigma_{2i-1}^z\sigma_{2i+1}^z-h\sum_i\sigma_i^x.
\end{equation}
Here, $J_1<0$ represents the coupling along the diagonal rungs, and $J_2>0$ stands for the interactions along the lower leg of the ladder. In the isotropic frustration point $f_{\rm c}=1$, the ground state is found by minimizing the number of broken bonds in each triangle independently, which leads to a macroscopic ground-state degeneracy. The numerical results in~\cite{qim_frustration_sawtooth}, which are based on the exact diagonalization of small ladders and perturbative expansions,  support  the absence of any symmetry breaking as the transverse field is increased. This effect has been coined as 'disorder by disorder', and would provide a testbed for a disordered quantum spin liquid state~\cite{qim_frustration}.

In order to perform a quantum simulation of this model, the geometry of a three-leg triangular ladder must be modified according to the method presented in Sec.~\ref{experimental} (see Fig.~\ref{ladder_geometries}{\bf (d)}). Then, the model Hamiltonian~\eqref{sawtooth} would follow directly from the QS Hamiltonian~\eqref{qi_ladder} with the usual identifications  $J_2\leftrightarrow J$, and  $J_1\leftrightarrow \tilde{J}$. In addition to the possibility of reaching larger spin ladders  to test the conclusions of~\cite{qim_frustration_sawtooth}, the trapped-ion QS may explore of the effects of  $f_2=J_2/|J_1|\neq f_{\rm c}$ and long-range interactions, hopefully leading to a rich phase diagram which, to the best of our knowledge, still remains  unexplored.

At this point, we should remark that quantum spin liquid phases can also be realized in  (quasi) one-dimensional  Heisenberg magnets~\cite{lechemitant_book,heisenberg_materials}. A much harder task is to find their higher-dimensional counterparts~\cite{spin_liquids}. By using the same technique to modify the geometry of the trapped-ion ladder, one may construct three-leg ladders such as the one displayed in Fig.~\ref{ladder_geometries}{\bf (e)}. Note that this amounts to a single stripe of the well-known two-dimensional (2D) Kagome lattice. Hence, our QS allows for the exploration of the dimensional crossover towards the {\it Kagome quantum Ising model}. In this case, the interplay between frustration and quantum fluctuations has been argued to give rise to a 2D quantum spin liquid phase~\cite{qim_frustration}, which could be targeted with the proposed QS. Besides, the study of the dimensional crossover and the long-range interactions is likely to  introduce a variety of interesting effects.

Before closing this section, let us remark that in addition to the aforementioned static phenomena, the trapped-ion QS is also capable of addressing dynamical many-body effects. In particular, the microscopic parameters of the above Hamiltonians can be tuned dynamically across the different quantum phase transitions. The breaking of the adiabatic approximation associated to a quantum critical point has been the subject of recent interest for different quantum systems (see e.g.~\cite{kz}).

\section{A Detailed Case: the $J_1$-$J_2$ Quantum Ising Model}
\label{section_j1_j2}
In this Section, we focus on the many-body physics of the $J_1$-$J_2$QIM. For the sake of completeness, we first review the properties of the short-range model, and then discuss the changes introduced by the dipolar range of the interactions, as realized in the trapped-ion quantum simulator.

\subsection{ Short-range $J_1$-$J_2$ quantum Ising model}

The essence of the QS is captured by the idealized next-to-nearest neighbor QIM~\eqref{qannni}, rewritten here for convenience
\begin{equation}
\label{nnn_j1j2}
H_{\bigtriangleup\hspace{-1.1ex}\bigtriangledown\hspace{-1.1ex}\bigtriangleup}=-|J_1|\left(\sum_i \sigma_{i}^z\sigma_{i+1}^z-f_2\sum_{i}\sigma_i^z\sigma_{i+2}^z+g\sum_i\sigma_i^x	\right),
\end{equation}
where the ratios $f_2=J_2/|J_1|$ and $g=h/|J_1|$ can be experimentally tailored. 
The original $J_1$-$J_2$ Ising model~\cite{classical_annni} considers competing ferromagnetic and antiferromagnetic interactions (see Fig.~\ref{spin_couplings}{\bf(b)}), namely ${\rm sign}(J_2)=-{\rm sign}(J_1)=+1$.  According to the Toulouse-Villain criterion~\eqref{frustration_criteria}, this leads to a frustrated quantum spin model $\mathcal{F}_{\triangle}=-1$, whereby  the 
ratio $f$ controls the frustration, and   $g$  the quantum fluctuations.  

The classical $J_1$-$J_2$ Ising chain, obtained by setting $g=0$ in the above Hamiltonian, can be solved exactly  (see~\cite{classical_annni_review} and references therein). Such a solution yields two possible phases. For $f_2< \half$, one lies in a ferromagnetic (F) phase whose ground-state manifold is two-fold degenerate
\begin{equation}
\ket{{\rm F}}\in{\rm span}\left\{\left |^{} _ {\uparrow} {}^{\uparrow} _{} {}^{} _ {\uparrow} \cdots ^{} _ {\uparrow} {}^{\uparrow} _{} {}^{} _ {\uparrow}\right \rangle, \left |^{} _ {\downarrow} {}^{\downarrow} _{} {}^{} _ {\downarrow} \cdots ^{} _ {\downarrow} {}^{\downarrow} _{} {}^{} _ {\downarrow}\right \rangle  \right\},
\end{equation}
where the spins have been arranged according to the trapped-ion zigzag layout. Conversely, for $f_2>\half$, the ground-state is a dimerized antiferromagnet (dAF) with four-fold degeneracy
\begin{equation}
\begin{split}
\ket{{\rm dAF}}\in{\rm span}&\left\{\left  |^{} _ {\uparrow} {}^{\uparrow} _{} {}^{} _ {\downarrow}{}^{\downarrow} _ {} \cdots ^{} _ {\uparrow} {}^{\uparrow} _{} {}^{} _ {\downarrow}{}^{} _ {}\right \rangle, \left |^{} _ {\downarrow} {}^{\downarrow} _{} {}^{} _ {\uparrow} {}^{\uparrow} _{} \cdots ^{} _ {\downarrow} {}^{\downarrow} _{} {}^{} _ {\uparrow}\right \rangle,  \right.\\
&\hspace{1.4ex}\left. \left |^{} _ {\downarrow} {}^{\uparrow} _{} {}^{} _ {\uparrow} {}^{\downarrow} _{} \cdots ^{} _ {\downarrow} {}^{\uparrow} _{} {}^{} _ {\uparrow}\right \rangle, \left |^{} _ {\uparrow} {}^{\downarrow} _{} {}^{} _ {\downarrow} {}^{\uparrow} _{} \cdots ^{} _ {\uparrow} {}^{\downarrow} _{} {}^{} _ {\downarrow}\right \rangle\right\}. 
 \end{split}
\end{equation}
Note that the degeneracy of the ferromagnetic manifold is related to the global spin-flipping  $\mathbb{Z}_2$ symmetry of the Hamiltonian  $U=\bigotimes_i\sigma_i^x$, such that $\mathbb{Z}_2=\{\mathbb{I},U\}$ is the smallest cyclic Abelian group. On the other hand,  the doubling of the dimerized-antiferromagnet degeneracy  is accidental (i.e. not related to symmetries). Such an accidental degeneracy becomes  more important at the  point $f_{\rm c}=\half$, where  the  frustration leads to a macroscopically  degenerate ground-state. In fact,  the degeneracy has been shown to scale exponentially with the number of spins $d_{\rm c}\propto\varphi^N$, where $\varphi=\half(1+\sqrt{5})$ is the golden ratio~\cite{classical_redner}. This yields the  hallmark of frustrated magnets, namely, a non-vanishing ground-state entropy.  

\begin{figure}
\centering
\includegraphics[width=0.9\columnwidth]{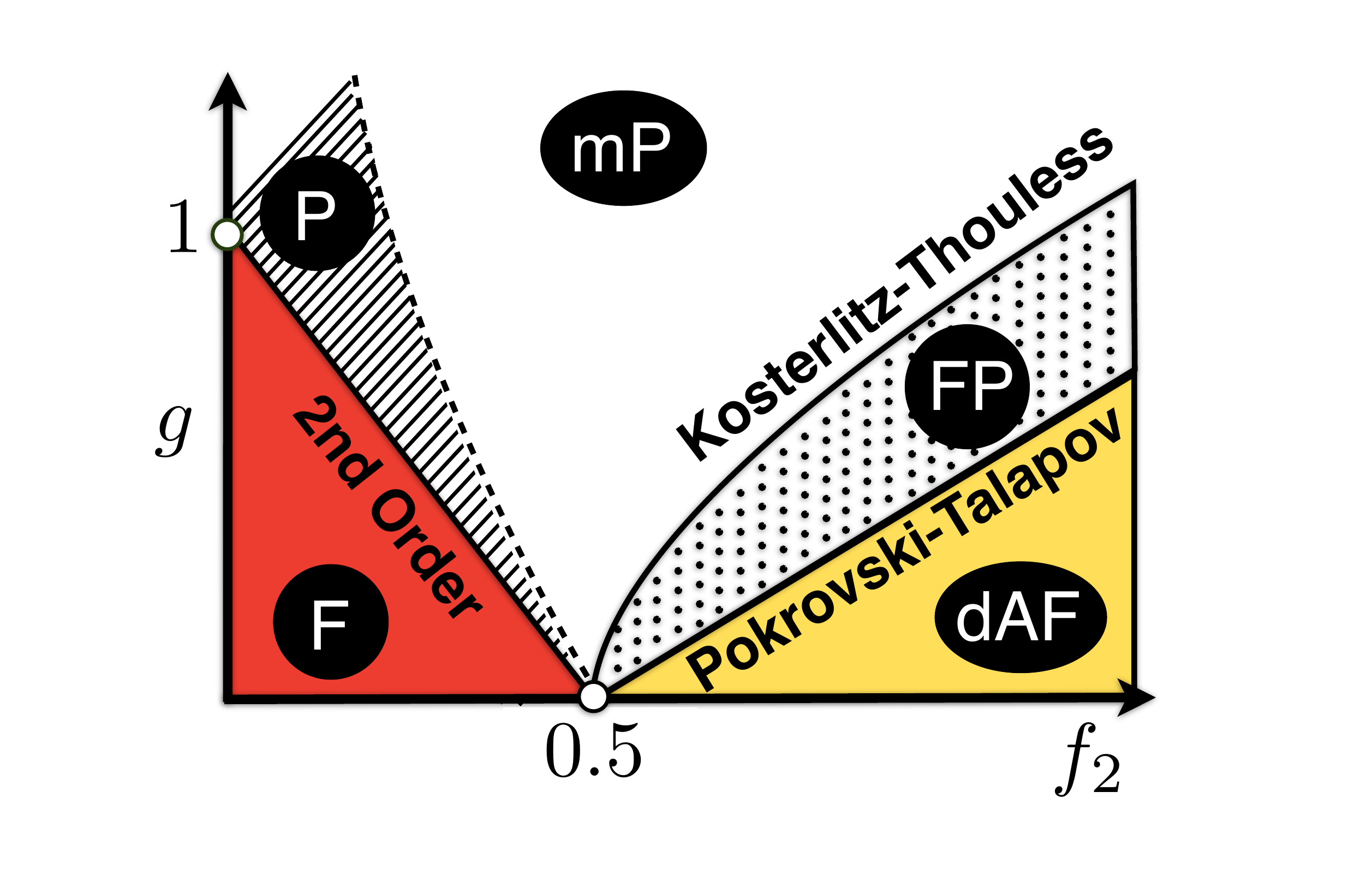}
\caption{ {\bf Schematic phase diagram of the $J_1$-$J_2$QIM:}    Ferromagnetic (F), dimerized anti-ferromagnetic (dAF), paramagnetic (P), modulated paramagnetic (mP), and floating (FP) phases. These states of matter are separated by different types of quantum phase transitions (Second Order, Kosterlitz-Thouless, and Pokrovsky-Talapov), and by the disorder line (dashed line). } 
\label{scheme_phases}
\end{figure}

  We are interested on the impact that quantum fluctuations may have on these degeneracies.  For  $g\gg 1\gg f_2$, the ground state corresponds to a single paramagnetic (P) state with all spins pointing towards the direction of the transverse field
\begin{equation}
\ket{{\rm P}}=\left |^{} _ {\rightarrow} {}^{\rightarrow} _{} {}^{} _ {\rightarrow}{}^{} _ {} \cdots ^{} _ {\rightarrow} {}^{\rightarrow} _{} {}^{} _ {\rightarrow}{}^{} _ {}\right \rangle,  
\end{equation}
where  $\ket{{\rightarrow}}=(\ket{{\uparrow}}+\ket{{\downarrow}})/\sqrt{2}$. The situation gets more interesting for intermediate fields, whereby  additional exotic phases and a variety of quantum phase transitions occur. Since the model  is no longer integrable, the analysis of the full phase diagram has been a big challenge,  requiring the combination of a variety of techniques. For instance, the mapping to a
classical 2D model~\cite{annni_hamiltonian} yields a direct link to  commensurate-incommensurate thermal phase transitions~\cite{classical_annni_review,annni_comensurability}. Hence, it is possible to use the methods developed in this area in order to understand the magnetic phases of the quantum model, being numerical Monte Carlo~\cite{annni_montecarlo} and free-fermion approximations~\cite{annni_fermionization} two representative examples. Together with the more recent application of bosonization techniques~\cite{allen_bosonization} and   numerical renormalization group  methods~\cite{anni_dmrg}, these studies yield the rich phase diagram  represented in Fig.~\ref{scheme_phases}. Note that in addition to the aforementioned phases, there is a  modulated paramagnetic phase (mP), and a highly-debated incommensurate floating phase (FP). Since all these different phases meet at  $f_{\rm c}=\half$, this  macroscopically-degenerate  point is also known as the multi-phase point. There are several critical lines emerging from the multi-phase point, which give rise to quantum phase transitions of second order, Kosterlitz-Thouless~\cite{kt_transition}, or Pokrovski-Talapov~\cite{pt_transition} type.

Notwithstanding these big efforts, we emphasize that there still exists  some controversy about the floating phase. In particular, the extent of the floating phase  is still a question of debate. Whereas some results point towards a FP  that prolongs towards $f_2\gg1$, other treatments predict a finite region that terminates in a multi-critical Lifshitz point (see~\cite{itf_glass} and references therein).
This makes a QS of the utmost interest to settle down these discrepancies. Moreover, as we  discuss in detail below, the introduction of long-range interactions   leads to additional open questions that have not been previously addressed to the best of our knowledge. From our numerical survey, we conjecture that the dipolar-range of the couplings leads to the splitting of the multi-phase point and the appearance of an intermediate  phase (see Fig.~\ref{scheme_dipolar}).

\subsection{Dipolar-range $J_1$-$J_2$ quantum Ising model}

 Trapped ions are an ideal platform to test the effects of long-range Ising interactions. For non-frustrated  systems, even if the  model belongs to the same universality class as the nearest-neighbor case, these long-range interactions may shift the critical point, and  favor long-distance quantum correlations~\cite{ising_porras}. For the frustrated systems under study, the effect of  long-range interactions is expected to be more significant. 

\begin{figure}
\centering
\includegraphics[width=0.8\columnwidth]{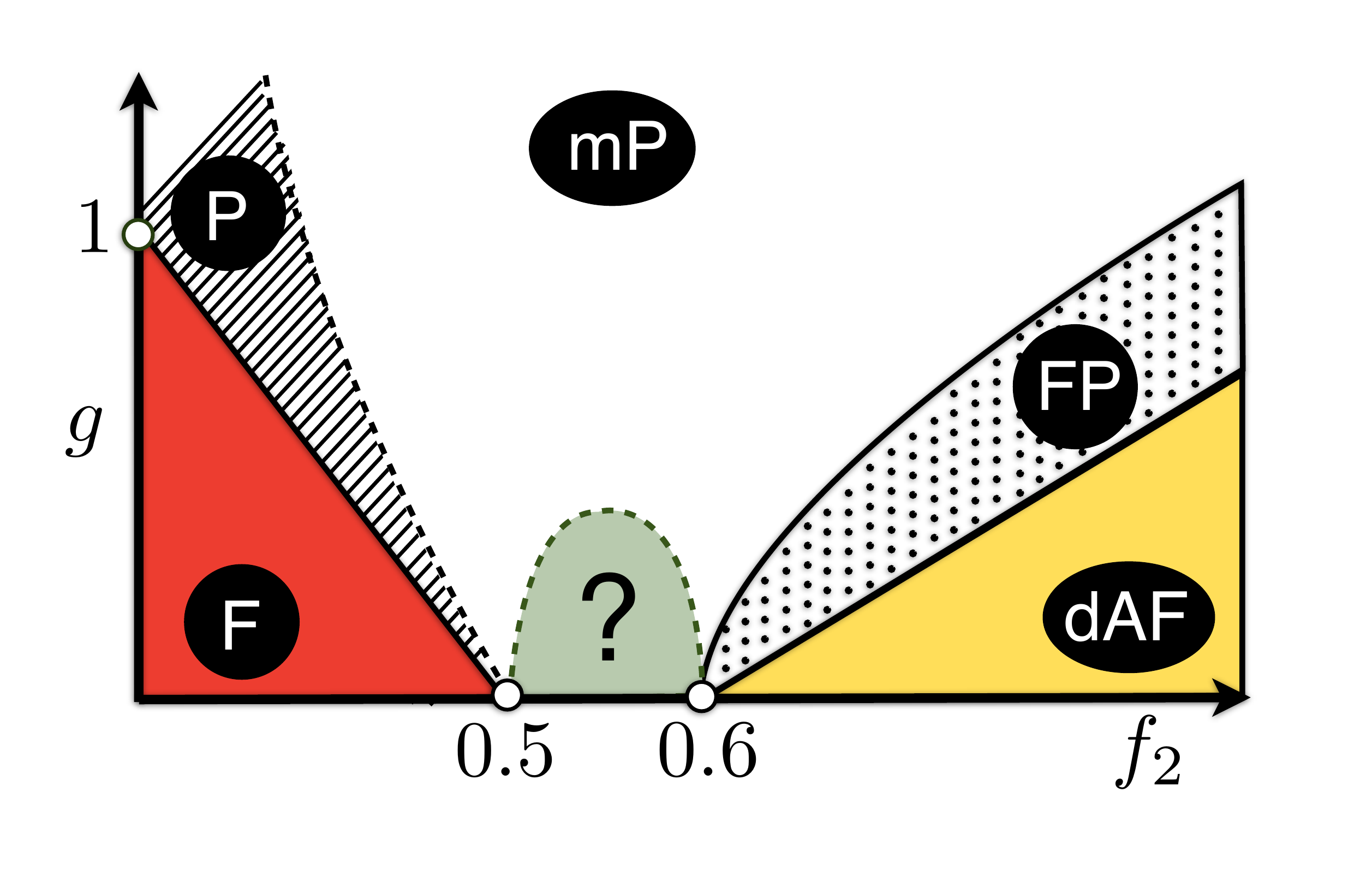}
\caption{ {\bf Conjectured phase diagram of the dipolar $J_1$-$J_2$QIM:} The longer-range dipolar Ising couplings, typical of Coulomb-mediated interactions, lead to the splitting of the multiphase point, and the appearance of a different ordered phase in between.  } 
\label{scheme_dipolar}
\end{figure}

By considering the dipolar range of the trapped-ion ladder Hamiltonian~\eqref{qi_ladder}, the $J_1$-$J_2$QIM~\eqref{nnn_j1j2}  must be modified  to
\begin{equation}
\label{dipolar_j1j2}
\begin{split}
H_{\bigtriangleup\hspace{-1.1ex}\bigtriangledown\hspace{-1.1ex}\bigtriangleup}=-|J_1|\bigg(&\sum_i\sum_{\delta\in {\rm odd}} f_{\delta}\sigma_{i}^z\sigma_{i+\delta}^z-\\
&-\sum_i\sum_{\delta\in {\rm even}} f_{\delta}\sigma_{i}^z\sigma_{i+\delta}^z+\sum_ig\sigma_i^x	\bigg),
\end{split}
\end{equation}
where we have introduced the ratios $f_{\delta}=J_{\delta}/|J_1|$, such that $J_{\delta}$ account for the dipolar decay of the interactions. In this notation, the
frustration ratio of the nearest-neighbor chain is  $f_2=J_2/|J_1|$, and we can study the additional frustration coming from the dipolar range $f_{3}, f_4...$. We have analyzed
numerically the phase diagram of such a long-range frustrated spin
model by means of an optimized Lanczos algorithm that allows us to
reach efficiently ladders with $L=24$ spins. All the results showed
below have been performed including the $f_3$ and $f_4$ terms of the
dipolar tail, in addition to the competing nearest-neighbor ($J_1$) and
next-to-nearest-neighbor ($J_2$) couplings. We have checked that the effect of
including longer ranged interactions (i.e., those corresponding to
$f_5$ and $f_6$ couplings) does not affect qualitatively the results.

In order to distinguish the phases, we focus on the
two-body correlators, which show the following scaling when
$|i-j|\gg1$
\begin{equation}
\label{corr_scaling}
\begin{split}
&\langle\sigma_i^z\sigma_j^z\rangle_{\rm F{\phantom{AF}}}\sim \hspace{1ex} m_0^2\cos(q_{\rm F}(i-j)),\\
&\langle\sigma_i^z\sigma_j^z\rangle_{\rm dAF}\sim \hspace{1ex} m_0^2\cos(q_{\rm dAF}(i-j)),\\
&\langle\sigma_i^z\sigma_j^z\rangle_{\rm P{\phantom{AF}}}\sim \hspace{1ex} m_0^2\cos(q_{\rm P}(i-j))\ee^{-|i-j|/\xi_{\rm P}},\\
&\langle\sigma_i^z\sigma_j^z\rangle_{\rm mP{\phantom{ii}}}\sim \hspace{1ex} m_0^2\cos(q_{\rm mP}(i-j))\ee^{-|i-j|/\xi_{\rm mP}},\\
&\langle\sigma_i^z\sigma_j^z\rangle_{\rm FP{\phantom{F}}}\sim \hspace{1ex} m_0^2\cos(q_{\rm FP}(i-j)){|i-j|^{-\eta}},\\
\end{split}
\end{equation}
where $0<m_0<1$, $\xi_{\rm P},\xi_{\rm mP}$ stand for the correlation lengths of the paramagnets, $\eta>0$ characterizes the algebraic decay of the correlations in the gapless floating phase, and the different modulation parameters $q_{[\dots]}$ have been listed in Table~\ref{modulation}. 

\begin{figure}
\centering
\includegraphics[width=0.75\columnwidth]{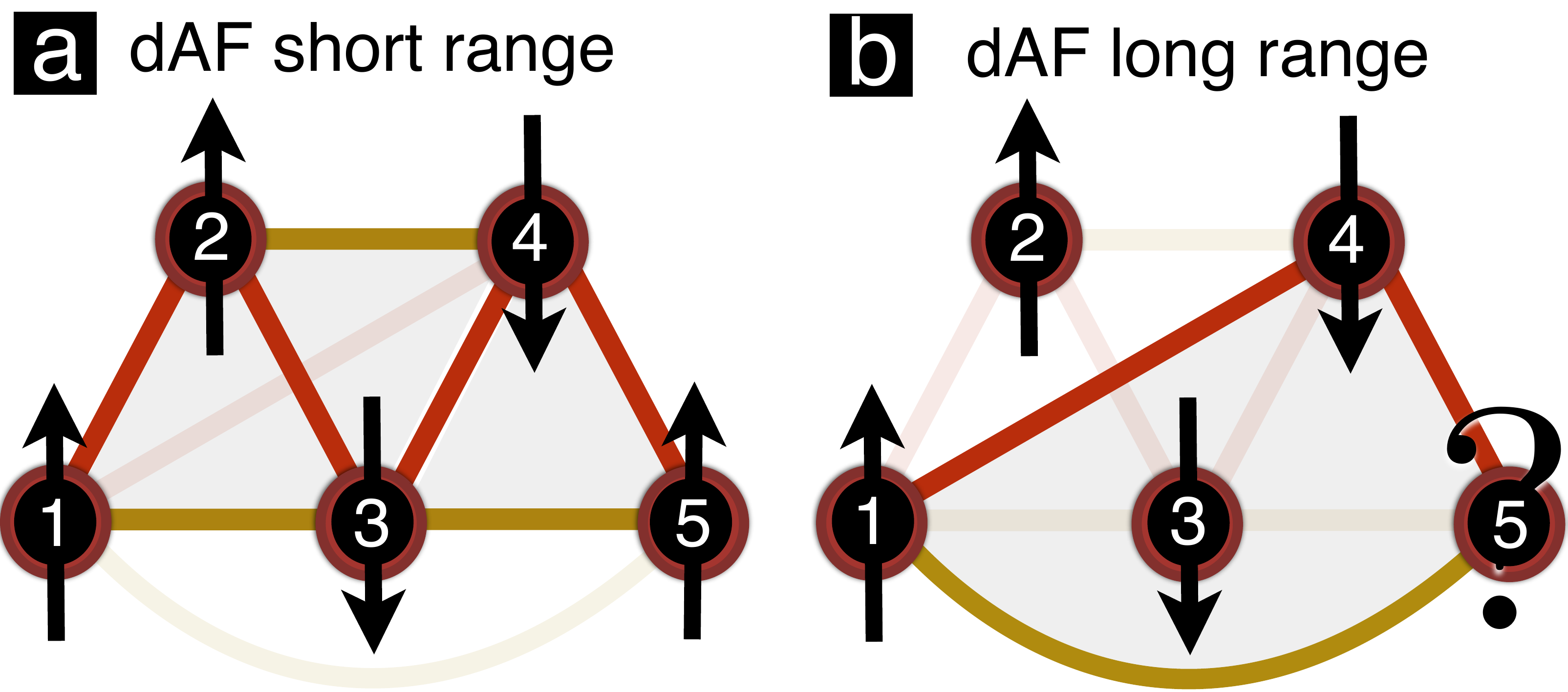}
\caption{ {\bf Incompatible sources of partial frustration:} {\bf (a)} The partial frustration for $f_2<\half$ in the short-range model can be minimized by one of the four possible dAF ground-states. {\bf (b)} The introduction of longer-range couplings yields additional sources of frustration, which are incompatible with the dAF ground-state, and thus responsible of the splitting of the multi-phase point.}
\label{frustration_dipolar}
\end{figure}

\begin{table}[!hbp]
  \centering 
   \caption{{\bf Magnetic modulation parameters}  }
  \begin{tabular}{ c  c c c c c c }
\hline
\hline
$\hspace{2ex}q_{\rm F}\hspace{2ex}$ & $\hspace{2ex}q_{\rm dAF}\hspace{2ex}$ & $\hspace{2ex}q_{\rm P}\hspace{2ex}$ & $\hspace{2ex}q_{\rm mP}\hspace{2ex}$& $\hspace{2ex}q_{\rm FP}\hspace{2ex}$ & $\hspace{2ex}q_{?}\hspace{2ex}$ \\
\hline
$0$ & $\frac{\pi}{2}$ & $0$ & $\hspace{2ex}q_{\rm mP}(g,f_2)\hspace{2ex}$ & $\hspace{2ex}q_{\rm FP}(g,f)\hspace{2ex}$& ? \\
\hline
\hline
\end{tabular}
  \label{modulation}
\end{table}

According to these expressions, the F and d-AF display  long-range magnetic order with different periodicities, whereas the P and mP phases are disordered. Finally, the FP phase has quasi-long range  order with a modulation parameter that flows with the ratios $f_2,g$ (hence the adjective floating),  and is generally incommensurate with the underlying lattice.

An observable capable of capturing the periodic modulations of the long-range ordered phases, and thus the features of the phase diagram, is the so-called magnetic structure factor. It is defined as the Fourier transform of the spin correlations
\begin{equation}
\label{structure_factor}
S_{zz}(q)=\sum_{ij}\langle\sigma_i^z\sigma_j^z\rangle\ee^{\ii q(i-j)},
\end{equation}
with $q\in[0,2\pi)$, and should attain a maximum at the different values shown in Table~\ref{modulation} for each of the phases. In order to evaluate the magnetic structure factor numerically, we have considered periodic boundary conditions, which shall  capture the bulk properties in the center of the trapped-ion ladders.

According to the scheme depicted in 
fig.~\ref{scheme_dipolar}, the main features of phase diagram of the
dipolar $J_1$-$J_2$QIM can be divided into three different regions: the
region where the ferromagnetic order prevails ($f_2<0.5$), the region with the dimerized-AF order ($f_2>0.6$), and the new intermediate region (roughly, $f_2 \in (0.5,0.6)$) that appears due to the competition of different long-ranged frustration mechanisms. In Figs.~\ref{frustration_dipolar}{\bf (a)-(b)}, we show how the dAF phase is destabilized by these competing mechanisms, which accounts schematically for the splitting of the multi-phase point. We will show below  how the numerical evaluation of the structure factor  supports  this division. We note that the structure factor has been normalized to unity.

\begin{figure}
\includegraphics[width=1\columnwidth]{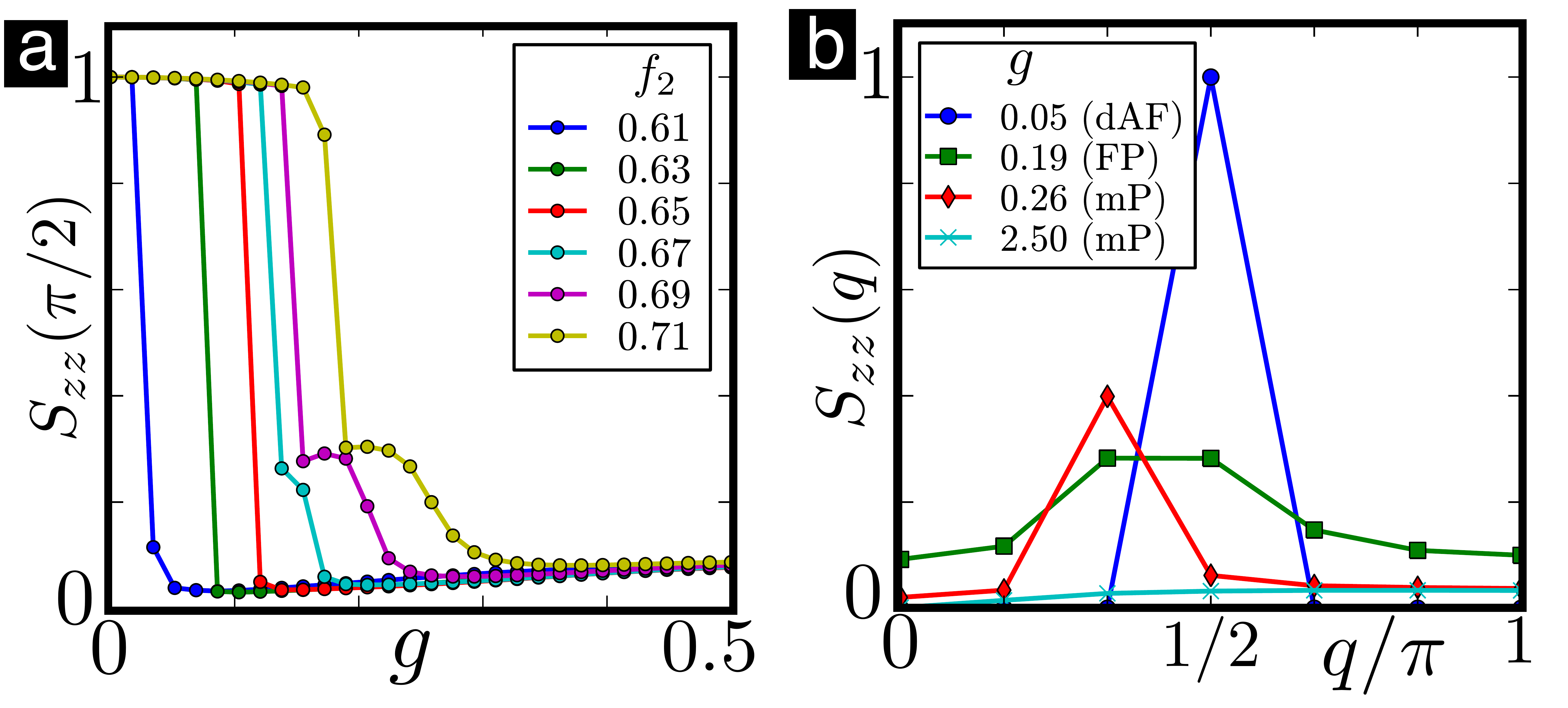}
\caption{ \textbf{Dipolar $J_1$-$J_2$QIM (dAF region, L=24)}: \textbf{(a)} Antiferromagnetic order parameter $S_{zz}(\pi/2)$. \textbf{(b)} Structure factor $S_{zz}(q)$ for $f_2=0.69$.}
\label{alm:aferro_transitions}
\end{figure}

In Fig.~\ref{alm:aferro_transitions}{\bf (a)}, we plot the order parameter
$S_{zz}(\frac{\pi}{2})$ corresponding to the dAF phase. It can be observed
that both phase transitions, namely the discontinuous  transition (Pokrovski-Talapov type)
separating the dAF phase from the FP, and  the continuous
transition (Kosterlitz-Thouless type) between the FP and the mP
phases, reflect themselves clearly in the structure factor as two
consecutive  jumps  with increasing transverse field. The fact that
only one transition is observed for lower values of $f_2$ hint at the
possibility that the multicritical point  has been shifted from $g=0$ to some finite value. Whether this shift  survives for  larger
lattices, or corresponds to a finite-size effect,   is an open question that cannot be addressed due to
the limitations of the diagonalization routine. 

The dependence of the
structure factor with the lattice momentum $q$ can be seen in
Fig.~\ref{alm:aferro_transitions}{\bf (b)}. Here, we represent the
evolution of this magnitude as we increase $g$ for a
fixed value of $f_2$. We have chosen some representatives for
each phase: for low fields $g$, we are well into the dAF phase, whose
periodicity is given by a single $q=\frac{\pi}{2}$ component. Increasing the
transverse field $g$, the state undergoes a transition to the FP, showing a
($f_2$-, $g$-dependent) incommensurate modulation. Upon further
increase of the field, we reach the mP phase. Note that for moderate $g$, the modulation of this phase is apparent in the peaked form
of its structure factor. Increasing further the magnetic field
polarizes the spins in the transversal direction yielding a flat
vanishing structure factor.

\begin{figure}
\includegraphics[width=1\columnwidth]{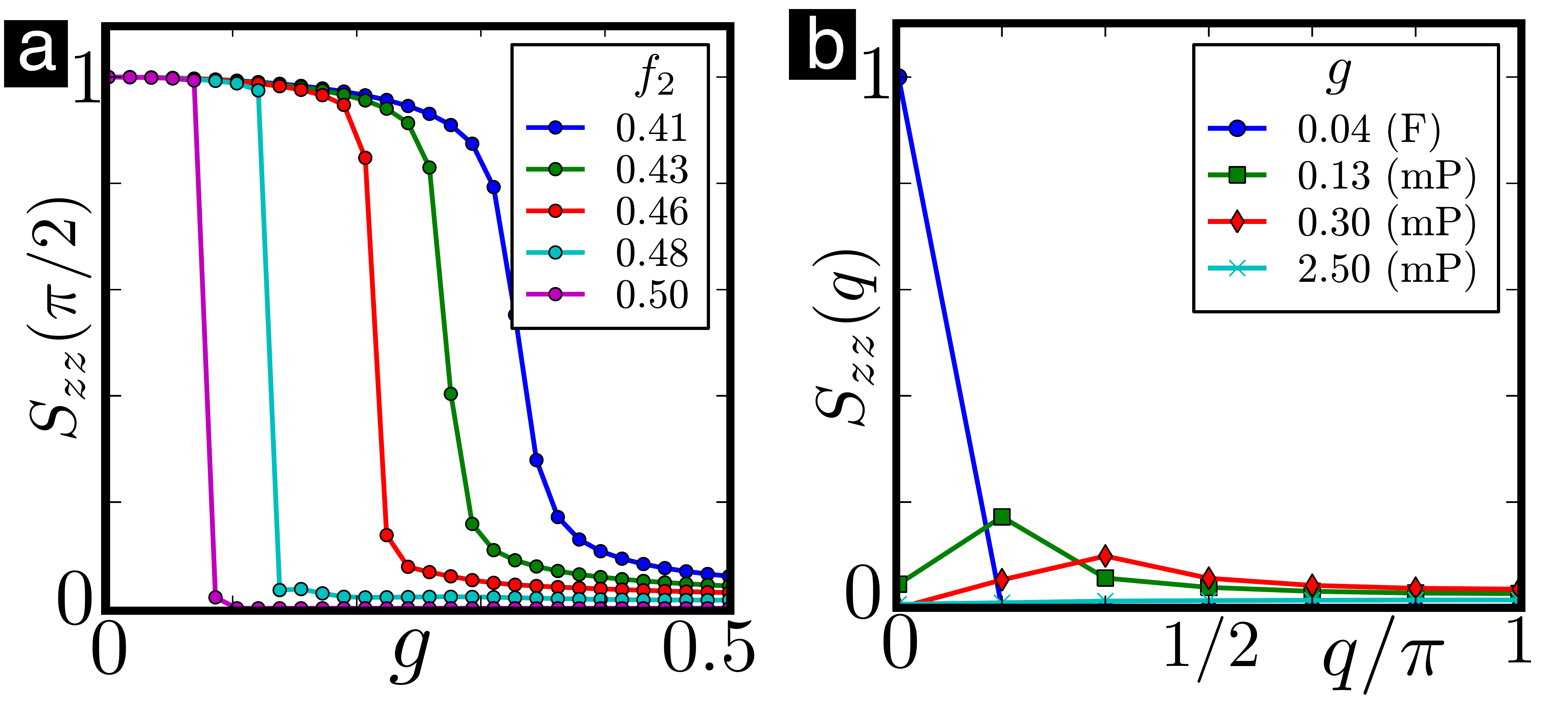}
\caption{ \textbf{Dipolar $J_1$-$J_2$QIM (F region, L=24)}: \textbf{(a)} Ferromagnetic order parameter $S_{zz}(0)$. \textbf{(b)} Structure factor $S_{zz}(q)$ for $f_2=0.45$}
\label{alm:ferro_transitions}
\end{figure}

The same analysis has been carried out in the ferromagnetic region. In
Fig.~\ref{alm:ferro_transitions}{\bf (a)}, we have plotted the $S_{zz}(0)$
component characteristic of a ferromagnet. In this case, however, only
one clear phase transition arises  as we
increase the transverse field $g$ for a fixed value of $f_2$. Also in
accordance with this observation, the modulation of the ground-state in Fig.~\ref{alm:ferro_transitions}{\bf (b)}, shows a shift between a pure ferromagnetic order for low values of $g$, to a modulated paramagnetic one with increasing
field, and finally to the completely polarized paramagnet for $g\gg 1$. This suggests the possibility that the unmodulated
paramagnetic phase observed in the $J_1$-$J_2$QIM is not stable upon the
effect of further dipolar couplings, or that its extent in the phase
diagram is small enough to prevent to be accurately captured with
finite-sized lattices.

\begin{figure}
\includegraphics[width=1\columnwidth]{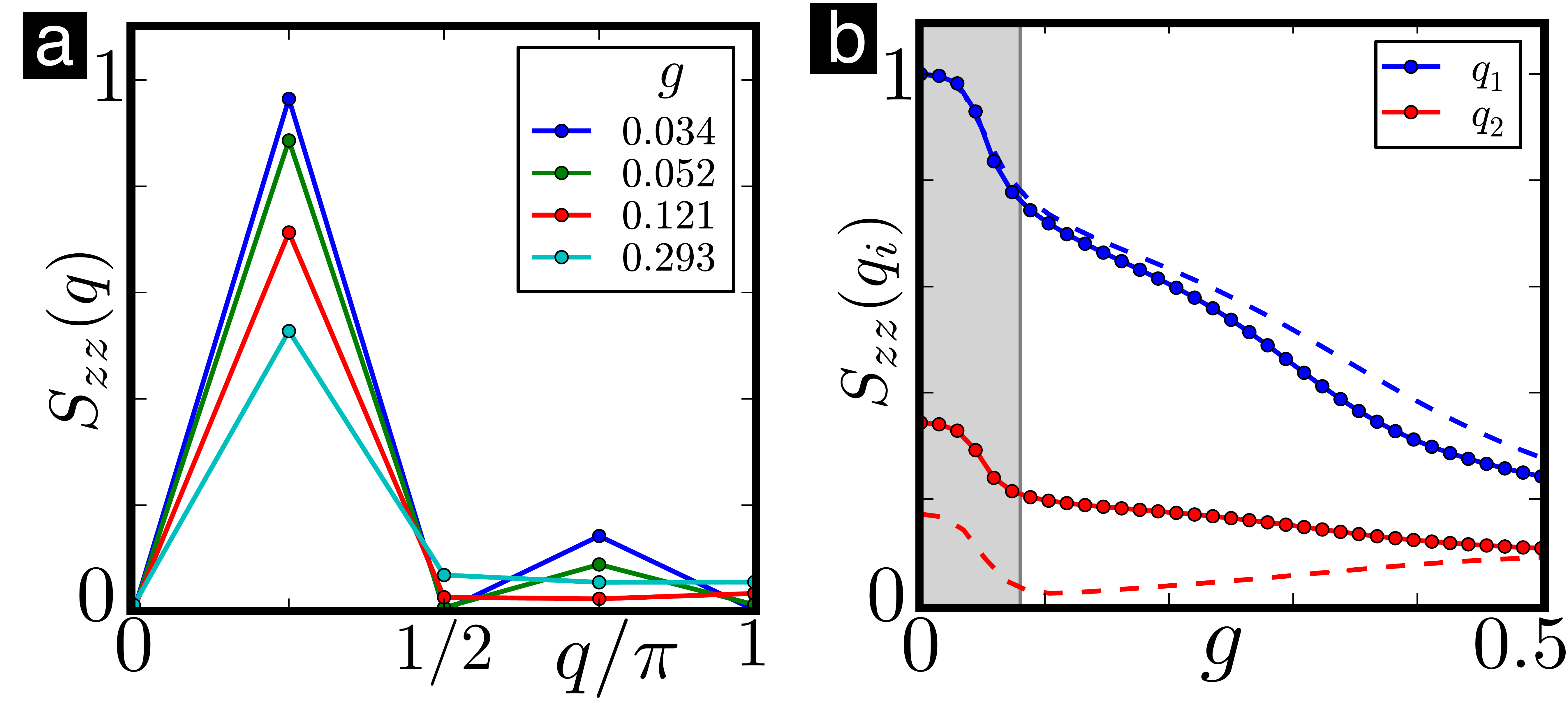}
\caption{ \textbf{Dipolar $J_1$-$J_2$QIM (intermediate region)}:
\textbf{(a)} Structure factor computed in a
  lattice of size $L=16$. \textbf{(b)} Amplitude of the two
  differentiated peaks in the intermediate region. The accuracy of its
  location is conditioned by the size of the lattice: (solid lines) $L=24$,  $q_1=\frac{\pi}{3}$ and $q_2=\frac{5\pi}{3}$
  (dashed lines) $L=16$, $q_1=\frac{\pi}{4}$, $q_2=\frac{3\pi}{4}$. }
\label{alm:conjectured_phase_peaks}
\end{figure}

In the $J_1$-$J_2$QIM, there exists a one-point boundary between the
ferromagnetic and dimerized-antiferromagnetic phases precisely located
at the muticritical point $(g=0, f_2
=0.5)$. Interestingly enough, in the dipolar $J_1$-$J_2$QIM this is no
longer true and both phases are separated by a finite region. In
Fig.~\ref{alm:conjectured_phase_peaks}{\bf (a)}, we have plotted the structure
factor $S_{zz}(q)$ for  the lattice size $L=16$
 along a fixed value of  $f_2$  in between the F-
and dAF-phases. In this graph, a new type of modulation shows up in the form of two
diferentiated peaks with momenta $q_1=\frac{\pi}{4}$ and $q_2=\frac{3\pi}{4}$. However, the precision in the
determination of such peaks is limited by the size of our
lattice. We have carried out additional numerics for $L=24$, which allow us to bound these peaks
between  $q_1 \in [\frac{\pi}{6}, \frac{\pi}{3}]$ and $q_2 \in [\frac{2\pi}{3}, \pi]$.  We have checked
that the relative amplitude of these peaks does not depend on the ratio between the
intra- and inter-leg couplings. In Fig.~\ref{alm:conjectured_phase_peaks}{\bf (b)}, we 
plot the amplitude of these two characteristic peaks with
increasing $g$. The shaded area represents the range where both peaks coexist, i.e, the extent of the new ordered state. It would be very interesting to explore the
precise origin of these modulations via the trapped-ion QS.

We remark that 
the modulations $q_1$ and $q_2$ coexist for low transverse fields. As we increase $g$, only the modulation with lower lattice momentum survives
as a differentiated peak. Indeed, the momentum of this surviving
modulation is compatible with that found  in the mP phase for intermediate
fields (compare the light-blue curve in
Fig.~\ref{alm:conjectured_phase_peaks}{\bf (a)} with the  red ones in
Figs.~\ref{alm:ferro_transitions}{\bf (b)} and~\ref{alm:aferro_transitions}{\bf (b)}). Whether this  is a
crossover or a quantum phase transition between the new conjectured phase
and the mP, goes beyond the limitations of our
numerical tools. 

\begin{figure}
\includegraphics[width=0.8\columnwidth]{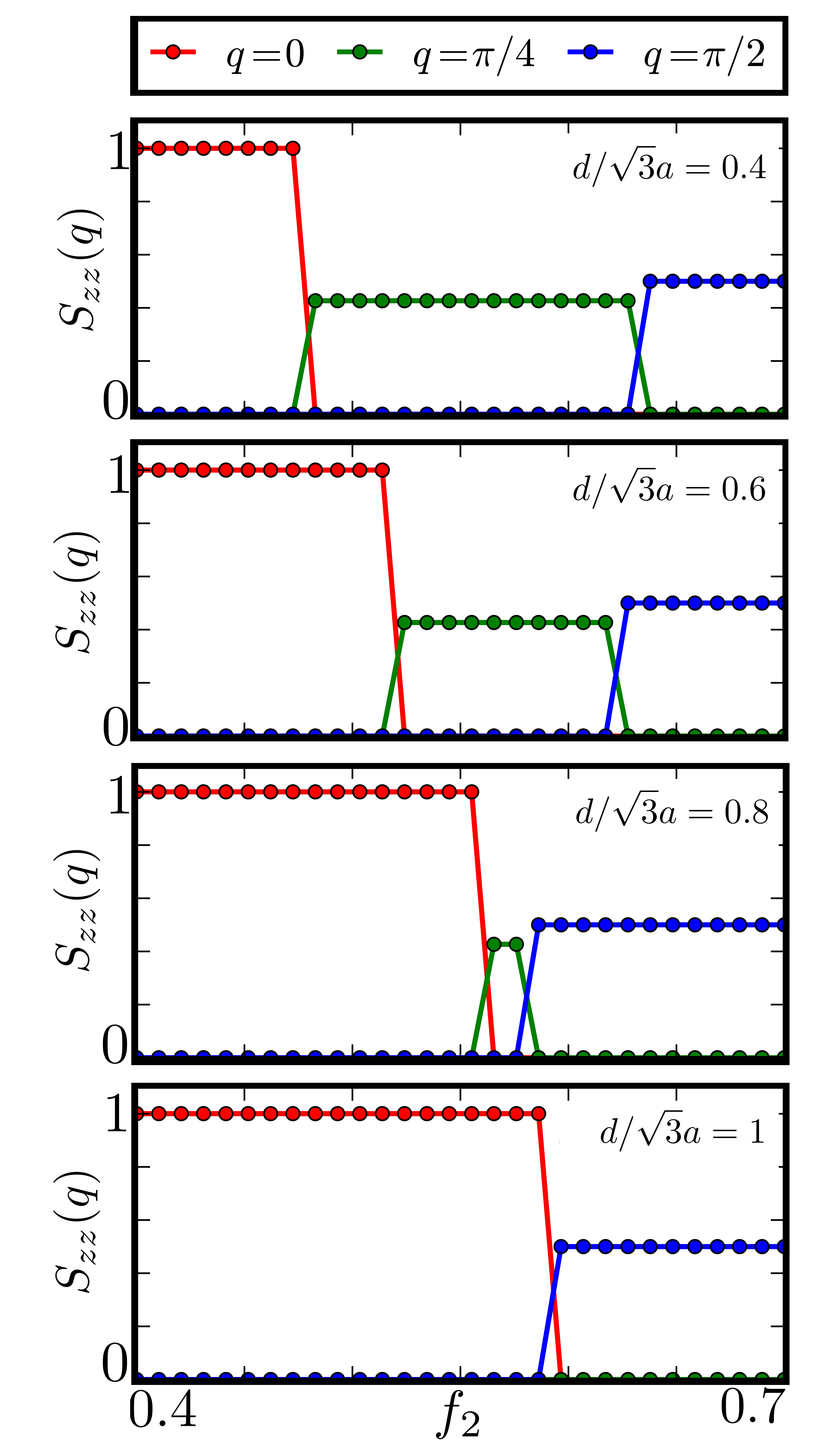}
\caption{ \textbf{Splitting of the multiphase critical point}: order
  parameters corresponding to the F ($S_{zz}(0)$) intermediate
  ($S_{zz}(\pi/4)$)and dAF ($S_{zz}(\pi/2)$) phases computed in a
  $L=16$ lattice. Each figure corresponds to a fixed anisotropy ratio
  $d/a$. The homogeneous equilateral zigzag lattice corresponds
  to $d/a=\sqrt{3}$ (figure at the bottom).}
\label{alm:conjectured_phase_anisotropy}
\end{figure}

All the above results have been computed in ladders
\begin{equation}
\label{equilib}
\tilde{x}_i^0=\frac{d}{2}(-1)^i, \hspace{1ex}\tilde{y}_i^0=0,\hspace{1ex} \tilde{z}_i^0=(i-\half N)a,
\end{equation}  
 with lattice
parameters $d=a$ . Note that a ladder
composed of equilateral triangles would correspond to
$d=\sqrt{3}a$. The extent of the new intermediate phase in the dipolar
$J_1$-$J_2$QIM is indeed strongly dependent on the geometry of the
triangular plaquettes. A straightforward way of measuring the extent of this phase
is computing the distance between the F and dAF phases along the line
$g=0$. In Fig.~\ref{alm:conjectured_phase_anisotropy}, we show the
order parameters $S_{zz}(0)$, $S_{zz}(\pi/4)$  and $S_{zz}(\pi/2)$ for different values of the anisotropy ratio $d/a$, which
correspond to the F phase, the new intermediate phase, and the dAF
phase, respectively. From these graphs
(qualitatively similar results are obtained with $L=24$), it is apparent
that the splitting of the multiphase  point is strongly
enhanced in anisotropic lattices.

\section{Conclusions and Outlook}

\label{section_conclusions}

Low-dimensional quantum Ising magnets are often representative models for a variety of emergent cooperative phenomena. Unfortunately, in contrast to their Heisenberg counterparts, the identification of materials accurately described by these models has turned out to be a much more difficult task. Moreover, whenever the models include anisotropic interactions and tunable ladder structures, the prospects of realizing them diminish  even further. A promising alternative to overcome such difficulties are the so-called quantum simulators.

In this work, we have shown  that cold ion crystals are promising candidates for the quantum simulation of a variety of quantum spin-ladder compounds. This avenue of research will allow for the  study of collective phenomena due to the combination of frustration, quantum fluctuations, long-range interactions, and dimensional crossover. In particular, these ion crystals can be used  to explore paradigmatic, yet controversial, models of  quantum Ising ladders  whereby analytical and numerical techniques seem to disagree. Moreover, they may also allow for the study of previously unexplored features of the models, such as the effects of  long-range  interactions. 

 First, we have shown how to control the geometry of different self-assembled trapped-ion ladders. Based on this possibility, we have presented a protocol to tailor  the anisotropy of the  magnetic interaction  mediated by the transverse phonons, which couple pairs of distant ions indirectly.  When these ions belong to different legs of the ladder, it is possible to tune both the sign and the magnitude of the spin-spin couplings by manipulating the laser-beam arrangement. The validity of this technique is supported by a detailed discussion of the possible sources of error in current ion-trap experiments, and by  numerics  showing an excellent agreement with our predictions. 
 
 This tool opens a vast amount of possibilities for trapped-ion-based QS of cooperative magnetic phenomena. For instance, we have presented a thorough description of the QS for the cornerstone of frustrated quantum Ising magnets, the  $J_1$-$J_2$ quantum Ising model. Moreover, we have also discussed how the QS has the potential of realizing quantum Ising ladders with connections to the exotic quantum dimer models introduced in the context of high-temperature superconductivity, and address the intriguing dimensional crossover phenomena. Finally, we have also pointed out how this QS may yield a route  towards the long-sought  quantum spin liquid phases. 

\acknowledgements

This work was supported by the EU STREP projects HIP, PICC, AQUTE, QESSENCE, and by the Alexander von Humboldt Foundation.
A.B. thanks FIS2009-10061, and QUITEMAD S2009-ESP-159.
\appendix

\section{Spin-dependent dipole forces and dissipation}
\label{appendix_dipole}

In this Appendix, we present a detailed derivation of the effective laser-ion interaction~\cite{sd_force,sd_force_review} in Eq.~\eqref{dipole_force}. By taking into account the spontaneous decay from the excited level (see Fig.~\ref{level_scheme}), we can discuss the regime where the spin-dependent dipole forces arise, and also analyze the sources of error due to photon scattering in the experiments. We consider the master equation for the Lambda scheme in  Fig.~\ref{level_scheme}, namely 
\begin{equation}
\label{master_equation}
\frac{{\rm d}\rho}{{\rm d}t}=-\ii[H_0+V,\rho(t)]+\mathcal{D}(\rho(t)).
\end{equation}
 The coherent part of the evolution is given by 
\begin{equation}
\label{lambda}
H_0=\!\!\!\!\sum_{m=r,\uparrow,\downarrow}\!\!\!\!\!\epsilon_{m}\ket{m}\bra{m},\hspace{1ex}V=\!\!\!\sum_{l=1,2}\sum_{s=\uparrow,\downarrow}\!\half \Omega_{l,s}\ket{r}\bra{s}\ee^{-\ii\omega_lt}+\text{H.c.},
\end{equation}
where we have introduced the energies of the internal states $\epsilon_{r},\epsilon_{\uparrow},\epsilon_{\downarrow}$,   the  Rabi frequencies of the transitions $\Omega_{l,s}$, and the laser frequencies $\omega_l$.  The dissipator  describing the spontaneous decay from the excited state is of the Lindblad form
\begin{equation}
\mathcal{D}(\rho)=\sum_{n=1,2}\left(L_n\rho L_n^{\dagger}-\half L_n^{\dagger}L_n\rho-\half\rho L_n^{\dagger}L_n\right),
\end{equation}
with the following jump operators  $L_1=\sqrt{\Gamma}\ket{{\downarrow}}\bra{r}$, and $L_2=\sqrt{\Gamma}\ket{{\uparrow}}\bra{r}$, where $\Gamma$ is the spontaneous decay rate from the excited state back to the spin manifold (see Fig.~\ref{level_scheme}). Let us now define the  detunings for all possible transitions
\begin{equation}
\delta_{l,s}=\epsilon_r-\epsilon_{s}-\omega_l.
\end{equation}
As announced in the main text, when these detunings are much larger than the Rabi frequencies and the decay rate, namely $\delta_{l,s}\gg\Omega_{l,s},\Gamma$, it is possible to adiabatically eliminate the excited state from the dynamics, and obtain an effective master equation within the spin manifold. We use the formalism introduced in~\cite{eff_jumps}, and obtain the  master equation
\begin{equation}
\label{effective_master_equation}
\begin{split}
\frac{{\rm d}\rho}{{\rm d}t}&=-\ii[H_{\rm eff},\rho(t)]+\\
&+\sum_{n}\left(L^{\rm eff}_n\rho (L^{\rm eff}_n)^{\dagger}-\half (L^{\rm eff}_n)^{\dagger}L^{\rm eff}_n\rho-\half\rho (L^{\rm eff}_n)^{\dagger}L^{\rm eff}_n\right),
\end{split}
\end{equation}
where the  Hamiltonian is $H_{\rm eff}=H_{\rm s}+H_{\rm r}+H_{\rm d}$, such that
\begin{equation}
H_{\rm s}=(\epsilon_{\downarrow}+\Delta\epsilon_{\downarrow})\ket{{\downarrow}}\bra{{\downarrow}}+(\epsilon_{\uparrow}+\Delta\epsilon_{\uparrow})\ket{{\uparrow}}\bra{{\uparrow}}
\end{equation}
includes the following ac-Stark shifts
\begin{equation}
\label{stark_shift}
\Delta\epsilon_{s}=-\sum_{l=1,2}\frac{|\Omega_{l,s}|^2\delta_{l,s}}{4\delta_{l,s}^2+\Gamma_{\rm t}^2},
\end{equation}
and we have introduced the sum of the two decay rates $\Gamma_{\rm t}=2\Gamma$. Additionally, we get the following two-photon stimulated Raman transitions
\begin{equation}
H_{\rm r}=\sum_{l,l'}\half\Omega^{\rm r}_{l,l'}\sigma^-\ee^{\ii(\omega_l-\omega_{l'})t}+{\rm H.c.},
\end{equation}
where we have introduced $\sigma^-=\ket{{\downarrow}}\bra{{\uparrow}}=(\sigma^+)^{\dagger}$, and the effective Rabi frequencies for the different Raman transitions
\begin{equation}
\Omega^{\rm r}_{l,l'}=-\frac{\Omega_{l,\downarrow}^*\Omega_{l',\uparrow}(\delta_{l,\downarrow}+\delta_{l',\uparrow})}{(2\delta_{l,\downarrow}+\ii\Gamma_{\rm t}) (2\delta_{l',\uparrow}-\ii\Gamma_{\rm t})},
\end{equation}
which involve the absorption of a photon from the beam $l'=1,2$ and posterior photon emission into the beam $l=1,2$. The last term of the effective Hamiltonian  is the one responsible for the spin-dependent dipole forces
\begin{equation}
\label{Hd}
H_{\rm d}=\sum_{s=\uparrow,\downarrow}\half\Omega^{\rm d}_{s}\ket{s}\bra{s}\ee^{\ii(\omega_1-\omega_2)t}+\text{H.c.},
\end{equation}
where we have introduced the following Rabi frequencies
\begin{equation}
\Omega^{\rm d}_{s}=-\frac{\Omega_{1,s}^*\Omega_{2,s}(\delta_{1,s}+\delta_{2,s})}{(2\delta_{1,s}+\ii\Gamma_{\rm t})(2\delta_{2,s}-\ii\Gamma_{\rm t})}.
\end{equation}
Now, depending on the particular laser frequencies $\omega_{l}$, it is possible to select whether the laser-ion interaction leads to a stimulated Raman transition (i.e. $\omega_{\rm L}:=\omega_1-\omega_2\approx\epsilon_{\uparrow}-\epsilon_{\downarrow}=:\omega_0$), or to a spin-dependent dipole force (i.e. $\omega_{\rm L}\ll \omega_0$). The effects of the spin-dependent dipole force are more transparent by rewriting Eq.~\eqref{Hd} as follows
\begin{equation}
\label{app_dipole_force}
H_{\rm d}=\half\tilde{\Omega}_{\rm L}\mathbb{I}\ee^{-\ii\omega_{\rm L}t}+\half{\Omega}_{\rm L}\sigma^z\ee^{-\ii\omega_{\rm L}t}+\text{H.c.},
\end{equation}
where we have introduced $\sigma^z=\ket{{\uparrow}}\bra{{\uparrow}}-\ket{{\downarrow}}\bra{{\downarrow}}$, and 
\begin{equation}
\label{differential_rabi}
\tilde{\Omega}_{\rm L}=\half \big(\Omega^{\rm d}_{\uparrow}+\Omega^{\rm d}_{\downarrow}\big)^*,\hspace{1ex}
{\Omega}_{\rm L}=\half \big(\Omega^{\rm d}_{\uparrow}-\Omega^{\rm d}_{\downarrow}\big)^*.
\end{equation}
Once the coherent part of the effective master equation has been derived, one must obtain the effective jump operators. In our setup, they can be expressed as follows
\begin{equation}
\label{effective_jump_operators}
\begin{split}
L_1^{\rm eff}&=\sqrt{\Gamma}\left(\frac{\Omega_{1,\downarrow}\ee^{-\ii\omega_1t}}{\delta_{1,\downarrow}-\ii\Gamma_{\rm t}}+\frac{\Omega_{2,\downarrow}\ee^{-\ii\omega_2t}}{\delta_{2,\downarrow}-\ii\Gamma_{\rm t}}\right)\ket{{\downarrow}}\bra{{\downarrow}}+\\
&+\sqrt{\Gamma}\left(\frac{\Omega_{1,\uparrow}\ee^{-\ii\omega_1t}}{\delta_{1,\uparrow}-\ii\Gamma_{\rm t}}+\frac{\Omega_{2,\uparrow}\ee^{-\ii\omega_2t}}{\delta_{2,\uparrow}-\ii\Gamma_{\rm t}}\right)\ket{{\downarrow}}\bra{{\uparrow}},\\
L_2^{\rm eff}&=\sqrt{\Gamma}\left(\frac{\Omega_{1,\uparrow}\ee^{-\ii\omega_1t}}{\delta_{1,\uparrow}-\ii\Gamma_{\rm t}}+\frac{\Omega_{2,\uparrow}\ee^{-\ii\omega_2t}}{\delta_{2,\uparrow}-\ii\Gamma_{\rm t}}\right)\ket{{\uparrow}}\bra{{\uparrow}}+\\
&+\sqrt{\Gamma}\left(\frac{\Omega_{1,\downarrow}\ee^{-\ii\omega_1t}}{\delta_{1,\downarrow}-\ii\Gamma_{\rm t}}+\frac{\Omega_{2,\downarrow}\ee^{-\ii\omega_2t}}{\delta_{2,\downarrow}-\ii\Gamma_{\rm t}}\right)\ket{{\uparrow}}\bra{{\downarrow}}.\\
\end{split}
\end{equation}
Note that the first term of each of the effective jump operators corresponds to the so-called Rayleigh photon scattering, which takes place without modifying the internal spin state. In this formulation, it becomes clear why the Rayleigh scattering will only introduce dephasing when the amplitudes (i.e. terms between brackets) of each jump operator are different, as observed in recent experiments~\cite{photon_scattering_2}. The second term of each jump operator corresponds to the Raman scattering, whereby the spin state is changed after the  emission of the photon.

\begin{figure}
\label{diff_rabi}
\includegraphics[width=0.8\columnwidth]{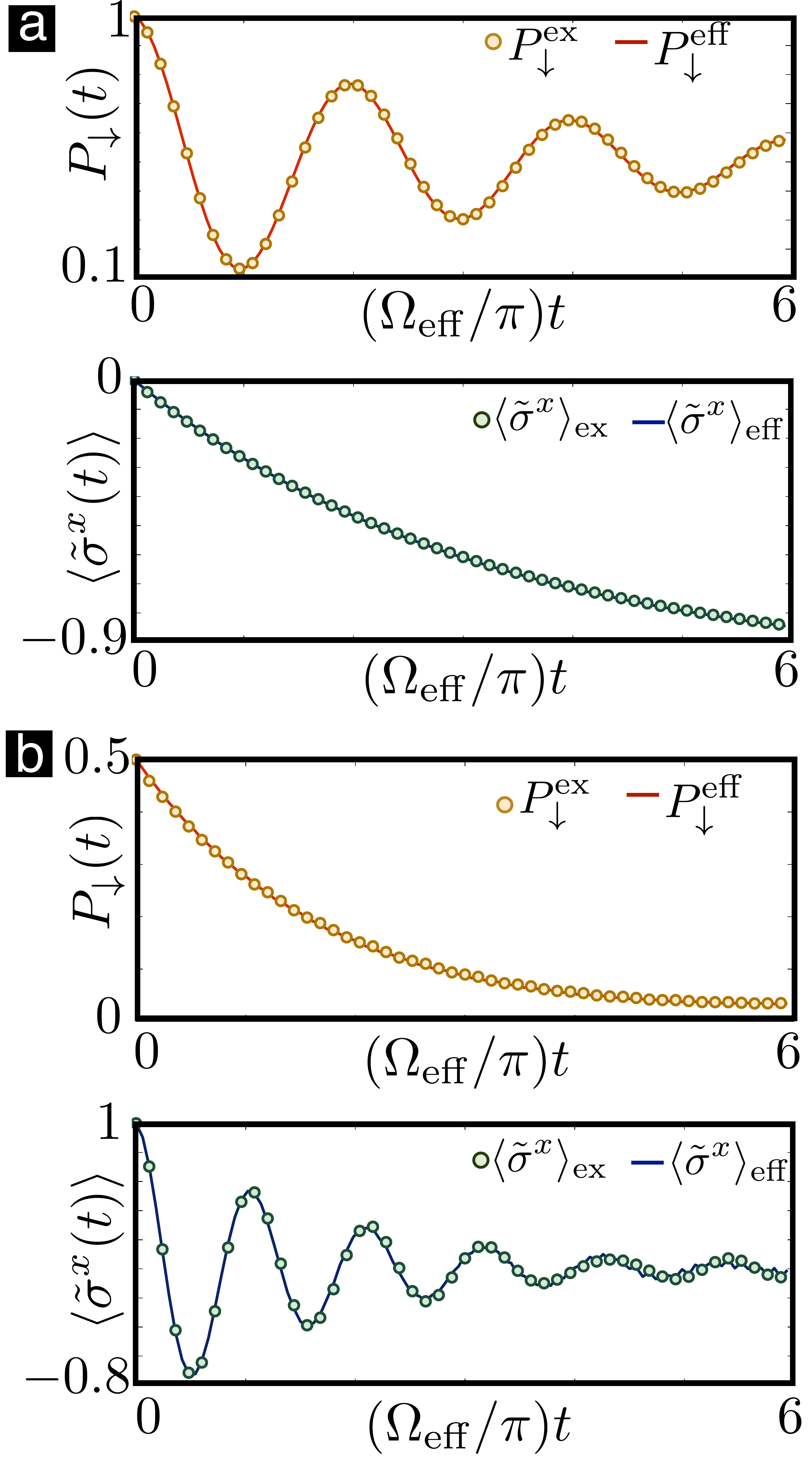}
\caption{ \textbf{Effective dissipative dynamics}: {\bf (a)} Population $P_{\downarrow}(t)={\rm Tr}\big\{\ket{{\downarrow}}\bra{{\downarrow}}\rho(t)\big\}$, and coherences $\langle\tilde{\sigma}^x\rangle={\rm Tr}\big\{\ket{{\downarrow}}\bra{{\uparrow}}\ee^{-\ii\omega_0t}\rho(t)\big\}+\text{c.c.}$ in the regime of two-photon stimulated Raman transitions. {\bf (b)} Same observables in the regime of spin-dependent dipole forces.}
\label{ad_elimination}
\end{figure}

In order to study the accuracy of this effective description~\eqref{effective_master_equation}, we confront it with the exact numerical integration of the original master equation~\eqref{master_equation}.   In Fig.~\ref{ad_elimination}{\bf (a)}, we compare both predictions for $ \epsilon_{\uparrow}/\epsilon_{r}=0.1, \epsilon_{\downarrow}/\epsilon_{r}=0.05$, and setting the laser parameters such that $\delta_{1,\uparrow}/\epsilon_{r}=0.5$, $\omega_{\rm L}=\omega_0$,  $\Omega_{2,\uparrow}/\epsilon_r=\Omega_{1,\downarrow}/\epsilon_r=0.05$, and $\Omega_{1,\uparrow}=\Omega_{2,\downarrow}=0$. This set of parameters leads to the regime of stimulated two-photon transitions, so that we expect to find periodic Rabi oscillations in the populations with a frequency $\Omega_{\rm eff}/\epsilon_r=2.5\cdot 10^{-3}$  when the initial state is  $\ket{\psi_0}=\ket{{\downarrow}}$. As shown in the upper panel of Fig.~\ref{ad_elimination}{\bf (a)}, these oscillations get damped due to the spontaneous decay  $\Gamma/\epsilon_r=0.05$, where a clear agreement of the exact and effective dynamics can be observed. In the lower panel, we represent the time-evolution of the coherences, which are damped due to the photon scattering.

More interesting to our purposes are the results displayed in Fig.~\ref{ad_elimination}{\bf (b)}, where we have kept the same parameters as above, but set $\omega_{\rm L}=10^{-3}\omega_0$. This guarantees the absence of two-photon Raman transitions. By switching on the additional laser beams, such that  $\Omega_{2\downarrow}/\epsilon_r=-\Omega_{1\uparrow}/\epsilon_r=0.05$, the non-vanishing differential Rabi frequency in Eq.~\eqref{differential_rabi} should lead to the $\sigma^z$ dipole force~\eqref{app_dipole_force}. Accordingly, we expect to find damped  Rabi oscillations in the coherences for $\ket{\psi_0}=(\ket{\uparrow}+\ket{\downarrow})/\sqrt{2}$, whereas the populations should only show an exponential damping. This agrees with the results displayed in Fig.~\ref{ad_elimination}{\bf (b)}, which also support the accuracy of the effective master equation in this regime.

To recover the action of the lasers on the vibrational degrees of freedom, one should substitute  $\Omega_{l,s}\to\Omega_{l,s}\ee^{\ii{\bf k}_l\cdot {\bf r}}$ in all the expressions above. Besides, in the regime of interest
\begin{equation}
\label{detuning}
|\Omega^{\rm r}_{l,l'}|\ll\omega_0\ll\epsilon_r-\epsilon_{\uparrow},\hspace{1ex}{\rm and}\hspace{1ex} \Gamma_{\rm t}\ll\delta_{l,s}\approx\delta_{1,\uparrow}=:\Delta,
\end{equation}
and summing over all the trapped ions, one recovers the dipole force term in Eq.~\eqref{dipole_force} with the effective Rabi frequency
$\Omega_{\rm L}=\frac{1}{2\Delta}\big(\Omega_{1,\downarrow}\Omega_{2,\downarrow}^*-\Omega_{1,\uparrow}\Omega_{2,\uparrow}^*\big)$. Let us note that the running wave of strength $\tilde{\Omega}_{\rm L}$  that only couples to the vibrational excitations, together with the ac-Stark shifts contained in~\eqref{stark_shift}, must be  compensated experimentally. Finally, the expression for the effective jump operators is the following
\begin{equation}
\label{effective_jump_operators}
\begin{split}
L_1^{\rm eff}&=\frac{\sqrt{\Gamma}}{\Delta}\left(\Omega_{1,\downarrow}\ee^{\ii({\bf k}_1\cdot{\bf r}-\omega_1t)}+\Omega_{2,\downarrow}\ee^{\ii({\bf k}_2\cdot{\bf r}-\omega_2t)}\right)\ket{{\downarrow}}\bra{{\downarrow}}+\\
&+\frac{\sqrt{\Gamma}}{\Delta}\left(\Omega_{1,\uparrow}\ee^{\ii({\bf k}_1\cdot{\bf r}-\omega_1t)}+\Omega_{2,\uparrow}\ee^{\ii({\bf k}_2\cdot{\bf r}-\omega_2t)}\right)\ket{{\downarrow}}\bra{{\uparrow}},\\
L_2^{\rm eff}&=\frac{\sqrt{\Gamma}}{\Delta}\left(\Omega_{1,\uparrow}\ee^{\ii({\bf k}_1\cdot{\bf r}-\omega_1t)}+\Omega_{2,\uparrow}\ee^{\ii({\bf k}_2\cdot{\bf r}-\omega_2t)}\right)\ket{{\uparrow}}\bra{{\uparrow}}+\\
&+\frac{\sqrt{\Gamma}}{\Delta}\left(\Omega_{1,\downarrow}\ee^{\ii({\bf k}_1\cdot{\bf r}-\omega_1t)}+\Omega_{2,\downarrow}\ee^{\ii({\bf k}_2\cdot{\bf r}-\omega_2t)}\right)\ket{{\uparrow}}\bra{{\downarrow}}.\\
\end{split}
\end{equation}
Since we are considering that $|\Omega_{l,s}|\ll\Delta$, it becomes clear that the effective scattering rates scale as $\Gamma_{\rm eff}=\Gamma(|\Omega_{l,s}|/\Delta)^2$.

\section{Analysis of the ion micromotion}
\label{appendix_micromotion}

In this Appendix, we take into account  the micromotion by considering the radio-frequency (r.f.) trapping fields rather than the effective pseudo-potential  in Eq.~\eqref{coulomb_ham}~\cite{ion_trap_book}. More precisely, the trapping potential is
\begin{equation}
\begin{split}
V_{\rm t}(\{ {\bf r}_j\})=&-\frac{e}{2}V_0\cos(\Omega_{\rm rf}t)\sum_i\left(1+\frac{1}{r_0^2}\left(x_i^2-y_i^2\right)\right)\\
&-e\kappa_{\rm g} U_0\sum_i\frac{1}{z_0^2}\left(z_i^2+\frac{1}{2}\left(x_i^2+y_i^2\right)\right),
\end{split}
\end{equation}
where $V_0,U_0$ are the a.c. and d.c. potentials of the trap, $r_0,z_0$ are the distances from the trap center to the electrodes and end-caps respectively, and $\kappa_{\rm g}<1$ is a geometric factor. Together with the Coulomb interaction, this trapping potential leads to a set of coupled Mathieu equations
\begin{equation}
\label{mathieu}
\frac{{\rm d}^2r_{i\alpha}}{{\rm d}t^2}+\frac{\Omega_{\rm rf}^2}{4}[a_{\alpha}+2q_{\alpha}\cos(\Omega_{\rm rf}t)]r_{i\alpha}-\frac{e^2}{m}\sum_{j\neq i}\frac{r_{i\alpha}-r_{j\alpha}^{\phantom{3/3}}}{|{\bf r}_{i}-{\bf r}_{j}|^{3}}=0,
\end{equation}
where we have defined  $a_{x}=4e\kappa_{\rm g} U_0/mz_0^2\Omega_{\rm rf}^2=a_y=-a_z/2$, and $q_x=-2eV_0/mr_0^2\Omega_{\rm rf}^2=-q_y,q_z=0$. In the majority of the experimental setups, these parameters fulfill $a_{\alpha},q_{\alpha}^2\ll 1$, so that one finds the following solution 
\begin{equation}
\label{micromotion}
r_{i\alpha}=r_{i\alpha}^0\left(1+\half q_{\alpha}\cos(\Omega_{\rm rf}t)\right)+\Delta r_{i\alpha}(t)
\end{equation}
where $\Delta {\bf r}_{i}(t)$ stand for the secular vibrations of the ions and frequency shifts due to micromotion~\cite{Ferdiand_exp,Landa_micromotion} that shall lead to the phonon Hamiltonian~\eqref{phonon_ham}, and ${\bf r}_{i}^0$ are the equilibrium positions of the ion crystal. To lowest order in $a_{\alpha},q_{\alpha}^2$, these  are found from the static part of Eq.~\eqref{mathieu}, after solving
\begin{equation}
\frac{m}{4}\Omega_{\rm rf}^2\left(a_{\alpha}+\frac{1}{2}q_{\alpha}^2\right) r_{i\alpha}^0-e^2\sum_{i\neq j}\frac{r^0_{i\alpha}-r^0_{j\alpha}}{|{{\bf r}^0_{i}-{\bf r}^0_{j}|^3}}=0,
\end{equation}
which is equivalent to Eq.~\eqref{system} after identifying the effective trapping frequencies $\omega_{\alpha}=\half\Omega_{\rm rf}(a_{\alpha}+\half q_{\alpha}^2)^{1/2}$. From Eq.~\eqref{micromotion}, one identifies two sources of radial micromotion. The second term corresponds to the micromotion associated to the small-amplitude secular vibrations of the ions, whereas the first term stands for an additional micromotion connected to the equilibrium positions lying off the trap axis. The former can be minimized by laser cooling, whereas the latter is a driven motion that cannot be cooled, and is  intrinsically linked to the planar structure of the ladder geometries (i.e. for linear ion chains, this micromotion can be compensated by aligning the ions along the trap axis). Let us remark that for the regimes of interest, $q_{x}\approx 0.1$-$0.2$, $l_{ z}\approx 1$-$10$ $\mu$m,  this micromotion can largely exceed that created by the  secular oscillations, and must be thus considered as a potential source of error in our QS. The discussion below focuses on this type of micromotion $r_{i\alpha}^0(t)=r_{i\alpha}^0\left(1+\half q_{\alpha}\cos(\Omega_{\rm rf}t)\right)$, which is usually referred to as the excess micromotion~\cite{micromotion_wineland}. 

\vspace{1ex}
{\it Micromotion heating:} One possible consequence of the micromotion is undesired heating, either due to the r.f. field of the ion trap, or to the additional lasers used for cooling. In particular, we focus on the transverse phonons since they are responsible for the spin-spin interaction.

One term in the effect of micromotion can be understood from the interplay between the excess micromotion and those terms in the vibrational Hamiltonian~\eqref{vib}  that do not conserve the number of vibrational excitations.  By working in the local vibrational basis used in Sec.~\ref{section_zz}, these terms amount to
\begin{equation}
\label{rf_heating}
\Delta H(t)=\frac{\omega_z}{2}\sum_{i,j}\kappa_{y}^{1/2}\mathcal{\tilde{V}}_{ij}^{yy}(t)a_{i,y}^{\dagger}a_{j,y}^{\dagger}\ee^{2\ii\omega_y t}+{\rm H.c.},
\end{equation}
where the couplings $\mathcal{\tilde{V}}_{ij}^{yy}(t)=\mathcal{{V}}_{ij}^{yy}(t)/(e^2/l_z^3)$ correspond to those of Eq.~\eqref{vs} after taking into account the micromotion
\begin{equation}
\label{micromotion_subs}
r_{i\alpha}^0\to r_{i\alpha}^0(t)=r_{i\alpha}^0\left(1+\half q_{\alpha}\cos(\Omega_{\rm rf}t)\right).
\end{equation}
In order to neglect  $\Delta H(t)$, which is responsible of the  r.f. heating, we expand Eq.~\eqref{rf_heating} to leading order in $q_{\alpha}\ll1$, and find that all the relevant terms can be neglected under a RWA if the following condition is fulfilled
\begin{equation}
\kappa_y^{1/2}|\mathcal{\tilde{V}}^{yy}_{ij}|\ll \frac{|2\omega_{y}\pm\Omega_{\rm rf}|}{\omega_z}\approx\frac{\Omega_{\rm rf}}{\omega_z}.
\end{equation}
 Since the ladder compounds present $\kappa_y\ll1$,  and $\Omega_{\rm rf}\gg\omega_z$, the validity of the RWA is easily fulfilled. 
 Thus the leading r.f. heating mechanism would be due to non linearities. This mechanism was analysed numerically in \cite{Schuessler_Heating,Drewsen_Heating}, where scaling laws for the r.f. heating rates were predicted. In these studies, the heating rates originate from non linearities and thus scale with the initial temperature of the crystal. For the temperatures that apply for our scheme, the r.f. heating would be much smaller than the anomalous heating and thus could be neglected in our analysis.

A different possibility is that of  laser heating. Note that the QS  requires  laser cooling of the transverse phonon modes, although not necessarily to the ground-state. As emphasized in~\cite{laser_micromotion_heating, micromotion_wineland},  depending on the ratio of the r.f. frequency to the decay rate of the cooling transition, either a broadening of the transition or the appearance of multiple micromotion sidebands may occur, which can lead to undesired heating even when the laser frequency is  tuned below the atomic resonance. To overcome this effect, one must carefully tune the laser frequency according to the regimes described in~\cite{ micromotion_wineland}.

\vspace{1ex}
{\it Micromotion contribution to the spin-dependent forces:}
 An important question to address is whether  the micromotion modifies   the spin-dependent dipole force, as derived in Appendix~\ref{appendix_dipole}. In the interaction picture, the Hamiltonian describing the laser coupling~\eqref{lambda} for the whole ion crystal becomes  
\begin{equation}
\label{raman}
V(t)=\sum_{l,s}\sum_i\frac{1}{2}\Omega_{l,s}\ket{r_i}\bra{s_i}\ee^{\ii{\bf k}_l\cdot {\bf r}_i}\ee^{\ii\xi_{li}\cos(\Omega_{\rm rf}t)}\ee^{\ii\delta_{l,s}t}+\text{H.c.},
\end{equation}
after including the micromotion. Here, we have introduced $\xi_{li}=q_x {\bf k}_{l}\cdot{\bf r}_i^0/2$, which represents the ratio of the radial excess micromotion~\eqref{micromotion} to the wavelength of the laser radiation. 

To carry on with the analysis, we need to specify a particular laser-beam arrangement (see the inset of Fig.~\ref{ladder_scheme_phonons}{\bf (b)}). We parametrize the laser wavevectors  as follows
\begin{equation}
{\bf k}_{l}=\frac{2\pi}{\lambda_{\rm sp}}(\cos\alpha_l{\bf e}_x+\sin\alpha_l{\bf e}_y),
\end{equation}
 where $\alpha_l$ determines their angle with respect to the $x$-axis, and $\lambda_{\rm sp}$ is the wavelength of the $n^{2}S_{1/2}$-$n^{2}P_{3/2}$ transition (see Fig.~\ref{level_scheme}). Let us recall that in order to  control the anisotropy of the  spin interactions~\eqref{eq19}, $\tilde{J}^{\rm eff}_{ij}\propto \cos(\phi_{ij})$, the corresponding angles must span the range $\phi_{ij}\in[0,2\pi]$.  This implies that the laser-beam arrangement must fulfill $\alpha_1=-\alpha_2-\Delta\alpha$, where $\Delta\alpha\ll|\alpha_l|$. One particular choice is the following 
 \begin{equation}
 \alpha_2=\frac{\pi}{2},\hspace{1ex}\alpha_1=-\frac{\pi}{2}-\Delta\alpha,\hspace{3ex}\Delta\alpha\approx\frac{\lambda_{\rm sp}}{2d}\ll1,
 \label{beamGeometry}
 \end{equation}
where $d$ is the inter-leg distance (see Fig.~\ref{qim_ladder}). With this choice, we find $\xi_{2i}=0$, and $\xi_{1i}=\pm\frac{\pi}{4}q_x$, both fulfilling $|\xi_{li}|\ll1$. This property will allow us to truncate  the following series
\begin{equation}
\label{bessel}
\ee^{\ii\xi_{li}\cos(\Omega_{\rm rf}t)}=\sum_{m\in\mathbb{Z}}\ii^mJ_m(\xi_{li})\ee^{\ii m \Omega_{\rm rf}t},
\end{equation}
where $J_m(x)$ are the Bessel functions of the first kind. By substituting the series~\eqref{bessel} in the laser-ion coupling~\eqref{raman}, we obtain a sum over all possible micromotion sidebands 
\begin{equation}
V(t)\propto\sum_{l,s,i}\sum_m\frac{1}{2}\Omega_{l,s}J_{m}(\xi_{li})\ket{r_i}\bra{s_i}\ee^{\ii{\bf k}_l\cdot {\bf r}_i}\ee^{\ii(\delta_{l,s}-m\Omega_{\rm rf})t}+\text{H.c.}.
\end{equation}
According to Eq.~\eqref{detuning}, we can set $\Delta\approx\delta_{l,s}$, such that for a sufficiently large detuning $\Omega_{\rm rf}/2\pi\approx 0.1$ GHz $\ll \Delta/2\pi\approx10$ GHz, we find that the micromotion sidebands only introduce a resonance for  $m\approx 10$. However, the contribution of such terms  is negligible since  $\xi_{li}\ll 1$, and $J_m(\xi_{li})\propto(\xi_{li})^m$.  As shown in \cite{Wineland_micromotion}, the correction to the leading term should be taken into account by adjusting the Rabi frequency.
Let us also note that the remaining non-resonant sidebands can also be neglected in a RWA, since $|\Omega_{l,s}|\ll\Delta$.
 Hence, we conclude that  the validity of the spin-dependent dipole force derived  in Appendix~\ref{appendix_dipole} is not compromised by the micromotion.
 
\begin{figure}
\includegraphics[width=0.8\columnwidth]{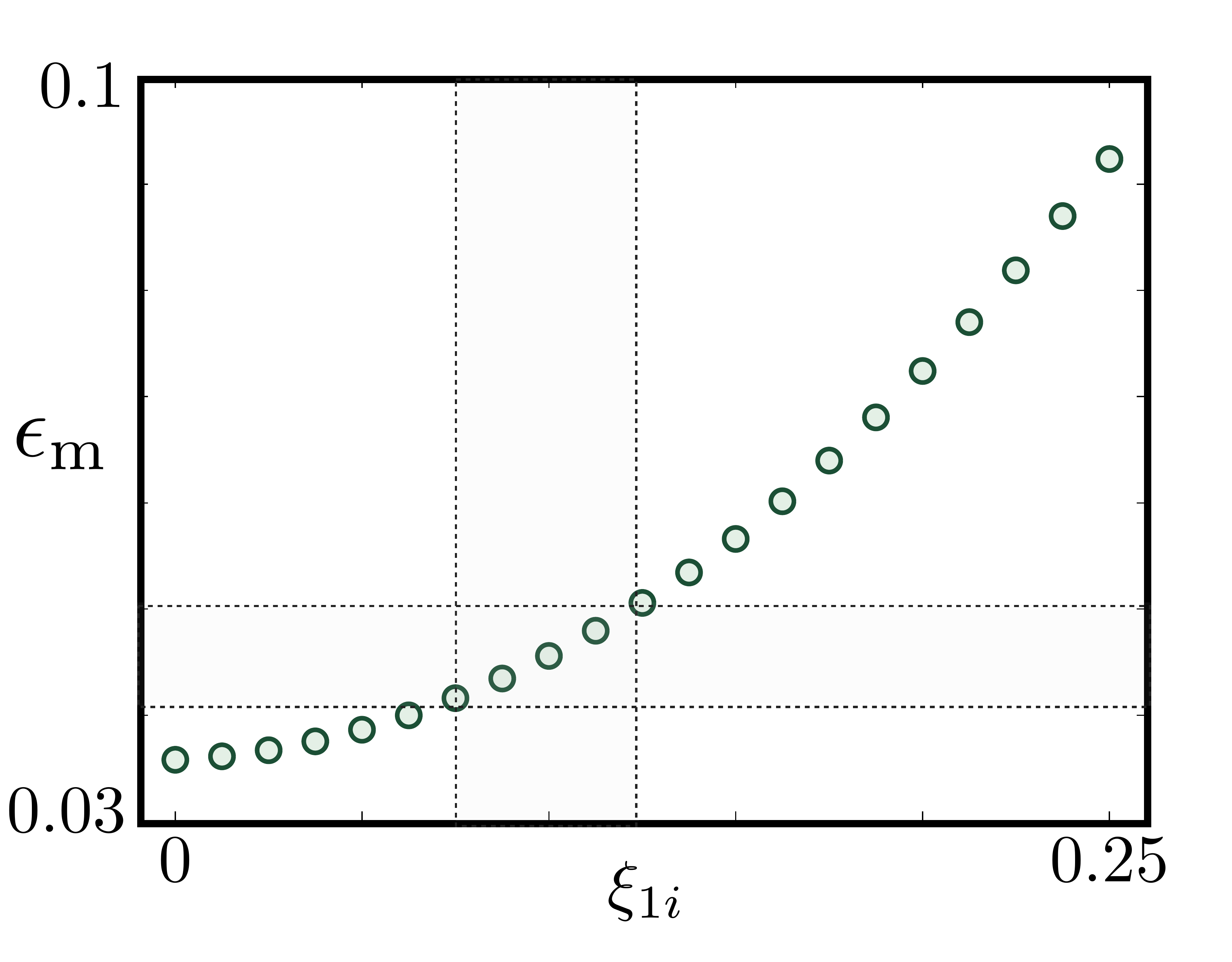}
\caption{ \textbf{Micromotion error}: Scaling of the micromotion error bound $\epsilon_{\rm m}$ with the ratio of the excess micromotion to the laser wavelength $\xi_{1i}$. The shaded area represents the parameters of interest.}
\label{micromotion_error_fig}
\end{figure}

  To quantify  the accuracy of this argument, we introduce
 \begin{equation}
 \epsilon_{{\rm m}}={\rm max}_t\left\{\big|\langle\tilde{\sigma}^x(t)\rangle_{\rm mic}-\langle\tilde{\sigma}^x(t)\rangle_{\rm eff}\big|,\hspace{0.5ex}t\in\left[0,\textstyle{\frac{6\pi}{\Omega_{\rm L}}}\right]\right\},
 \end{equation}
where $\langle\tilde{\sigma}^x(t)\rangle={\rm Tr}\big\{\ket{{\downarrow}}\bra{{\uparrow}}\ee^{-\ii\omega_0t}\rho(t)\big\}+\text{c.c.}$ represents the coherences. In this expression, the effective evolution under the dipole force $\langle\tilde{\sigma}^x(t)\rangle_{\rm eff}$, as represented in Fig.~\ref{ad_elimination}{\bf (b)}, is compared to the time evolution including the micromotion sidebands $\langle\tilde{\sigma}^x(t)\rangle_{\rm mic}$. Accordingly,  $\epsilon_{{\rm m}}$ sets an upper bound to the error of the dipole force~\eqref{dipole_force}. In Fig.~\ref{micromotion_error_fig}, we represent this error bound as a function of the relative micromotion amplitude $\xi_{1i}$. We use the same parameters as in Appendix~\ref{appendix_dipole}, and set the r.f. frequency to $\Omega_{\rm rf}/\epsilon_{r}=5\cdot 10^{-3}$, which is consistent with the constraint $\Omega_{\rm rf}\ll\Delta$.
We observe that the micromotion sidebands only contribute with a small error (4-5\%) for the regimes of interest $\xi_{1i}=\frac{\pi}{4}q_x,\xi_{2i}=0$ (shaded region).

Once this has been shown, we must consider the effects of the micromotion on the derivation of the spin-phonon term~\eqref{transverse_pushing}. Following a similar procedure, we obtain the micromotion contributions to the Lamb-Dicke expansion connecting the dipole force~\eqref{dipole_force} to the spin-phonon coupling~\eqref{transverse_pushing}
\begin{equation}
 \label{transverse_pushing_micromotion}
 H_{\rm d}=\frac{\Omega_{\rm L}}{2}\sum_{i,m,p}J_m(\xi_{{\rm L}i})\ee^{\ii{\bf k}_{\rm L}\cdot {\bf r}_i^0}\sigma_i^z f_{p}(\{a_n\})\ee^{\ii(m\Omega_{\rm rf}-\omega_{\rm L})t}+\text{H.c.},
 \end{equation}
 where we have introduced $\xi_{{\rm L}i}=q_x{\bf k}_{\rm L}\cdot{\bf r}_i^0/2$, and
 \begin{equation}
 f_{p}(\{a_n\})=\frac{1}{p!}\left(\ii\sum_n\eta_{n\bot}\mathcal{M}_{in}^{\bot}\left(a_n\ee^{-\ii\Omega_n^{\bot}t}+a_n^{\dagger}\ee^{\ii\Omega_n^{\bot}t}\right)\right)^{p}.
 \end{equation}
Note that the above expression~\eqref{transverse_pushing_micromotion} includes all the possible resonances between the secular and micromotion sidebands with the laser beatnote $\omega_{\rm L}=\omega_1-\omega_2$. The leading-order term for small Lamb-Dicke parameter $\eta_{n\bot}\ll1$ occurs for the secular resonance $m=0$, and $p=1$. This term leads directly to the desired spin-phonon coupling~\eqref{transverse_pushing} provided that 
\begin{equation}
\omega_{\rm L}\approx \Omega_{n}^{\bot}, \hspace{1ex}|\Omega_{\rm L}|\ll\omega_{\rm L}.
\end{equation}
Considering the different orders of magnitude in the problem
$\omega_{\rm L}/2\pi\approx\omega_y/2\pi\approx$10 MHz $\ll \Omega_{\rm rf}/2\pi\approx 0.1$ GHz,  the leading micromotion resonance would  occur for the term   $\Omega_{\rm L}\eta_{n\bot}^pJ_{m}(\xi_{{\rm L}i})\ee^{\ii(-p\Omega_{n}^{\bot}+m\Omega_{\rm rf}-\omega_{\rm L})t}$ with $m=1$ and $p\approx9$. Since  these terms scale as $(\eta_{n\bot})^{p}$  with  $\eta_{n\bot}\ll1$, they get exponentially suppressed and can be thus neglected. Let us note that the remaining off-resonant sidebands can also be neglected via a RWA in the regime of interest $\Omega_{\rm L}\ll\omega_{\rm L}\ll\Omega_{\rm rf}$. Therefore, we conclude that the micromotion does not modify the spin-phonon coupling~\eqref{transverse_pushing}.

\vspace{1ex}

{\it Micromotion contribution to unwanted transitions:}  Equation~\eqref{raman} implicitly assumes that the dynamics due to the Raman-beam configuration can be accurately described by only three levels $\ket{{\uparrow_i}},\ket{{\downarrow_i}},\ket{r_i}$. While this is always true for Zeeman ions, some special care must be taken for hyperfine ones, where the laser beams may excite some of the remaining states of the ground-state manifold due to the extra resonances introduced by the micromotion. Hence, the laser-ion  Hamiltonian in Eq.~\eqref{raman} must be supplemented $V+\Delta V$ by 
\begin{equation}
\label{raman_add}
\Delta V(t)=\sum_{l,i,a}\frac{\Omega_{l,a}}{2}\ket{r_i}\bra{a_i}\ee^{\ii{\bf k}_l\cdot {\bf r}_i}\ee^{\ii\xi_{li}\cos\Omega_{\rm rf}t}\ee^{\ii\delta_{l,a}t}+\text{H.c.},
\end{equation}
which includes all the additional states $\ket{a_i}$ of the ground-state manifold with energies $\epsilon_a$, and we have introduced the detunings $\delta_{l,a}=\epsilon_r-\epsilon_{a}-\omega_l$. These detunings  can be controlled by the Zeeman shifts of the ground-state manifold $\{\ket{a_i}\}$ caused by an external magnetic field. Note that the aforementioned unwanted transitions that take the state out of  the spin subspace $s\in\{\uparrow,\downarrow\}$ follow from two-photon processes evolving like $\Omega_{1,a}\Omega_{2,\uparrow}^*J_{m}(\xi_{1i})J_{m'}(\xi_{2i})^*\ee^{-\ii(\delta_{a,\uparrow}-\omega_{\rm L}- (m-m')\Omega_{\rm rf})t}$. Therefore, the Zeeman  splittings $\delta_{a,s}=\epsilon_a-\epsilon_s$ must be tuned  to
\begin{equation}
|\Omega_{1,a}\Omega_{2,s}^*|\ll |\delta_{a,s}-\omega_{\rm L}\pm\Omega_{\rm rf}|,
\end{equation}
such that these unwanted transitions become highly off-resonant and can be neglected in a RWA.

\section{Thermal fluctuations and phonon heating}
\label{heisenberg}

In this appendix, we discuss an alternative derivation of the effective spin models based on the Heisenberg equation of motion, which shall allow us to predict the effects of finite temperatures for the results presented in Sec.~\ref{section_zz}. The starting point is the time-independent  spin-phonon Hamiltonian in Eq.~\eqref{time_indep}, rewritten here for convenience 
 \begin{equation}
 \label{spin_phonon}
\tilde{H}_{\rm p}+\tilde{H}_{\rm d}=\sum_n\delta_n^{\bot}a_n^{\dagger}a_n^{\phantom \dagger}+\sum_{i,n}(\mathcal{F}_{in}\sigma_i^za_{n}^{\dagger}+\text{H.c.}),
 \end{equation}
where we have introduced $\mathcal{F}_{in}=\ii\frac{\Omega_{\rm L}}{2}\ee^{\ii{\bf k}_{\rm L}\cdot {\bf r}_i^0}\eta_{n\bot}\mathcal{M}_{in}^{\bot}$. In this picture, the evolution of the operators is given by the following system of coupled differential equations
 \begin{equation}
 \begin{split}
 \frac{{\rm d} \sigma_i^+(t)}{{\rm d}t}&=\sigma_i^+(t)\sum_n2\ii(\mathcal{F}_{in}a_n^{ \dagger}(t)+\mathcal{F}^*_{in}a_n^{\phantom \dagger}(t)),\\
   \frac{{\rm d}\hspace{0.1ex} a_n^{\phantom{+}}(t)}{{\rm d}t}&=-\ii\delta_n^{\bot}a_n^{\phantom \dagger}(t)-\ii\sum_i\mathcal{F}_{in}\sigma_i^z(t),\\
  \frac{{\rm d} \sigma_i^{z\phantom{i}}(t)}{{\rm d}t}&=0.
 \end{split}
 \end{equation}
 Thanks to the last  conserved quantity $\sigma_i^z(t)=\sigma_i^z(0)$, we can integrate this system of equations exactly. The evolution of the phonon operators corresponds to that of a forced quantum harmonic oscillator, namely
 \begin{equation}
 a_n^{\phantom \dagger}(t)= a_n^{\phantom \dagger}(0)\ee^{-\ii\delta_n^{\bot}t}+\sum_i\frac{\mathcal{F}_{in}}{\delta_n^{\bot}}\sigma_i^z(0)\left(\ee^{-\ii\delta_n^{\bot}t}-1\right).
 \end{equation}
 By substituting on the remaining equation, we find a homogeneous linear differential equation for  $\sigma_i^+(t)$, which can be integrated exactly. We are interested in deriving an estimate for the scaling of the error  at finite temperatures
  \begin{equation}
 \label{relative_error}
\epsilon_T=\frac{|\langle\sigma_i^x\rangle_T-\langle\sigma_i^x\rangle_{T=0}|}{|\langle\sigma_i^x\rangle_{T=0}|}, 
 \end{equation} 
 where $T$ is the temperature determining the phonon Gibbs state $\rho_{\rm th}=Z^{-1}\ee^{-\beta\sum_n\Omega_{n}^{\bot}a_n^{\dagger}a_n},$ such that $Z={\rm Tr}\{\ee^{-\beta\sum_n\Omega_{n}^{\bot}a_n^{\dagger}a_n}\}$ is the partition function, and $\beta=(k_{\rm B}T)^{-1}$ is expressed in terms of the Boltzman constant $k_{\rm B}$. By considering a separable initial state $\rho(0)=\ket{\psi_{\rm s}}\bra{\psi_{\rm s}}\otimes\rho_{\rm th}$, where $\ket{\psi_{\rm s}}$ is a pure spin state, we find the following expression for the relative thermal error
 \begin{equation}
 \epsilon_{T}=\left(1-{\rm Tr}\left\{\rho_{\rm th}\ee^{\sum_n\frac{2\mathcal{F}_{in}}{\delta_n^{\bot}}(\ee^{\ii\delta_n^{\bot}t}-1)a_n^{\dagger}-\rm {H.c.}}\right\}\right).
 \end{equation}
 This expectation value can be evaluated exactly to yield the following result
 \begin{equation}
 \label{relative_error}
 \epsilon_{T}=\left(1-\ee^{-\sum_m\frac{8|\mathcal{F}_{im}|^2}{(\delta_m^{\bot})^2}\big(1-\cos\delta_m^{\bot}t\big)\bar{n}_m^{\bot}}\right),
 \end{equation}
 where the effects of the zero-point motion have  been included in $\langle \sigma_i^x\rangle_{T=0}$, and we have introduced the mean phonon numbers for each of the vibrational modes $\bar{n}_m^{\bot}=\langle a_m^{\dagger}a_m\rangle$.  It is interesting to note that, as argued for the so-called quantum phase gates~\cite{sd_force_review}, the error can be minimized by considering evolution times that are multiples of the detuning of the closest vibrational mode $\delta^{\bot}_{m^*}$, namely $t_{\rm f}=2\pi n/\delta_{m^*}^{\bot}$, where $n\in\mathbb{Z}$. 
 
 \begin{figure}
\includegraphics[width=0.8\columnwidth]{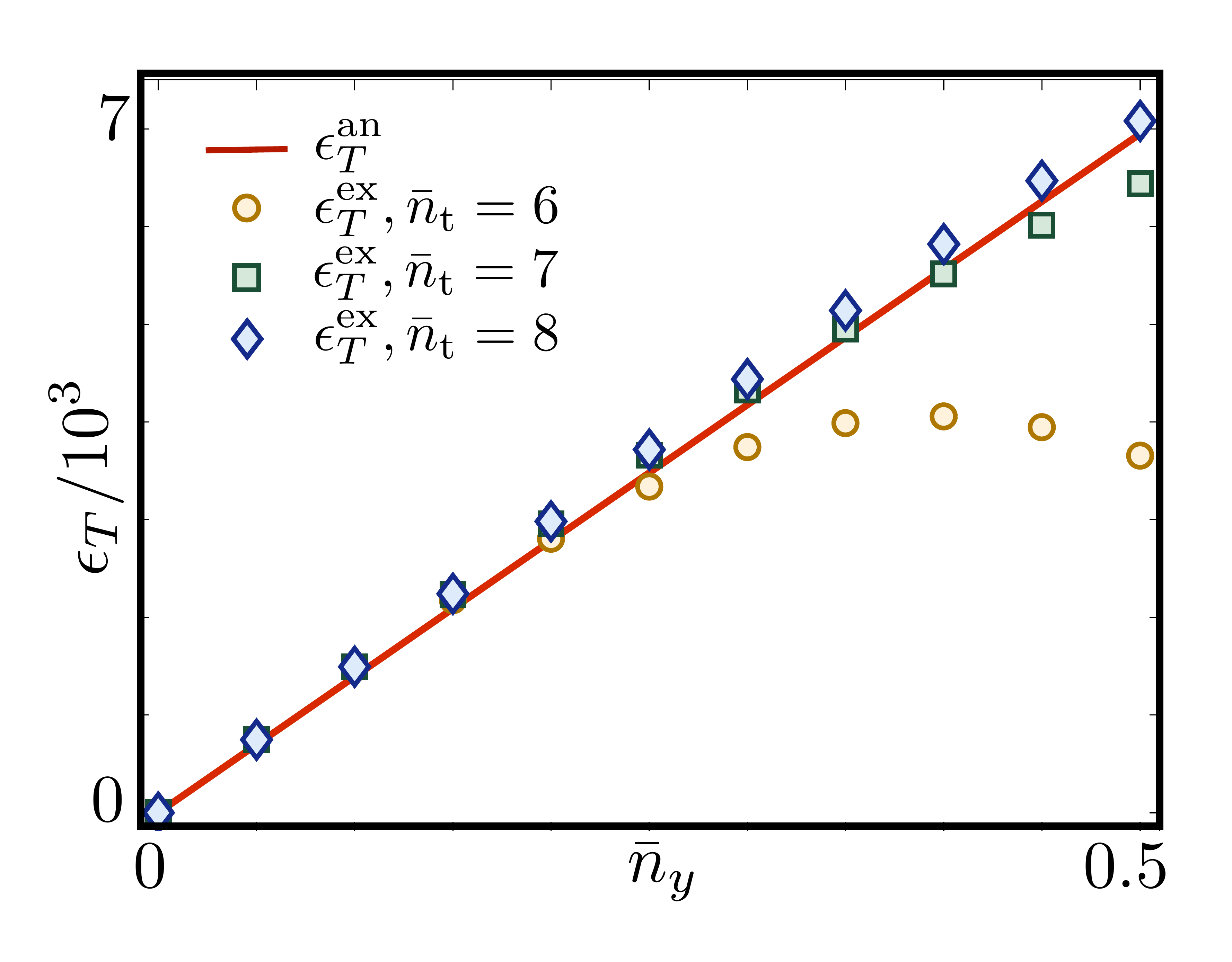}
\caption{ \textbf{Relative thermal error}: Scaling of the relative thermal error $\epsilon_T$ with the mean number of phonons in the center-of-mass mode $\bar{n}_y$.The analytical estimate $\epsilon_T^{\rm an}$ in Eq.~\eqref{relative_error} (red solid line) is compared to the exact numerical results  $\epsilon_T^{\rm ex}$ (doted lines) for different truncations of the phonon Hilbert space $\bar{n}_{\rm t}=6,7,8$. }
\label{thermal_error_fig}
\end{figure}
 
 In order to check the validity of our derivation, we have confronted the prediction~\eqref{relative_error} to the exact time evolution of the spin-phonon model in Eq.~\eqref{spin_phonon}. We use the parameters of Sec.~\ref{section_zz}, namely, the trap frequency  $\omega_y/2\pi=20$ MHz, and the laser parameters $\eta_y=0.1$,  $\omega_{\rm L}=1.1\omega_y$, $\Omega_{\rm L}=0.15|\delta_y|/\eta_y$, and ${\bf e}_x\cdot{\bf k}_{\rm L}=0$. In Fig.~\ref{thermal_error_fig}, we represent the exact results for the thermal error $\epsilon_{\rm T}^{\rm ex}$ (dotted lines) obtained by the numerical integration of the Liouville equation  ${\rm d}\rho/{\rm d}t=-\ii[\tilde{H}_{\rm p}+\tilde{H}_{\rm d},\rho]$ up to $t_{\rm f}=\pi/8J_{\rm eff}$ for the  Hamiltonian~\eqref{spin_phonon}. Note that we truncate the Hilbert space of each vibrational mode to $\bar{n}_{\rm t}=6$ (yellow circles), $\bar{n}_{\rm t}=7$ (green squares), and $\bar{n}_{\rm t}=8$ (blue diamonds). These results are compared to the analytical estimate $\epsilon_{\rm T}^{\rm an}$ (red solid line) in Eq.~\eqref{relative_error},  showing a remarkable agreement for a sufficiently large truncation of the phonon Hilbert space.

We now derive a useful expression for the scaling of the error in terms of  experimental parameters, in the limit of $|\mathcal{F}_{in}|\ll\delta_{n}^{\bot}$. We perform a Taylor expansion 
of Eq.~\eqref{relative_error} for $\kappa_y\ll1$, and take into account the orthonormality properties of the normal-mode displacements $\mathcal{M}_{in}^{\bot}$. For  low- and high-temperatures, we have $\bar{n}_n^{\bot}\approx\beta\Omega_n^{\bot}$ and $\bar{n}_n^{\bot}\approx(\beta\Omega_n^{\bot})^{-1}$  respectively, where $\beta=1/k_{\rm B}T$. In both regimes, we find the following  upper bound for the error
  \begin{equation}
  \label{thermal_error_app}
\epsilon_{T}\leq\epsilon_{\rm th}={\rm max}_t\{\epsilon_{T}\}=\left(1-\ee^{-\frac{4|\Omega_{\rm L}|^2\eta_y^2}{\delta_y^2}\bar{n}_y}\right),
  \end{equation}
  where we  have introduced  $\bar{n}_y=(\beta\omega_y)$, which corresponds to the mean phonon number for the center-of-mass mode. Finally, in the low-temperature regime, one finds
  \begin{equation}
\epsilon_{T}\leq\epsilon_{\rm th}=\frac{4|\Omega_{\rm L}|^2\eta_y^2}{\delta_y^2}\bar{n}_y  \end{equation}
   By controlling the experimental parameters, this error term should be minimized for the QS. Note that this linear scaling of the relative error with the mean number of phonons coincides with the results shown in Fig.~\ref{thermal_error_fig}.
  
 \begin{figure}
\includegraphics[width=0.8\columnwidth]{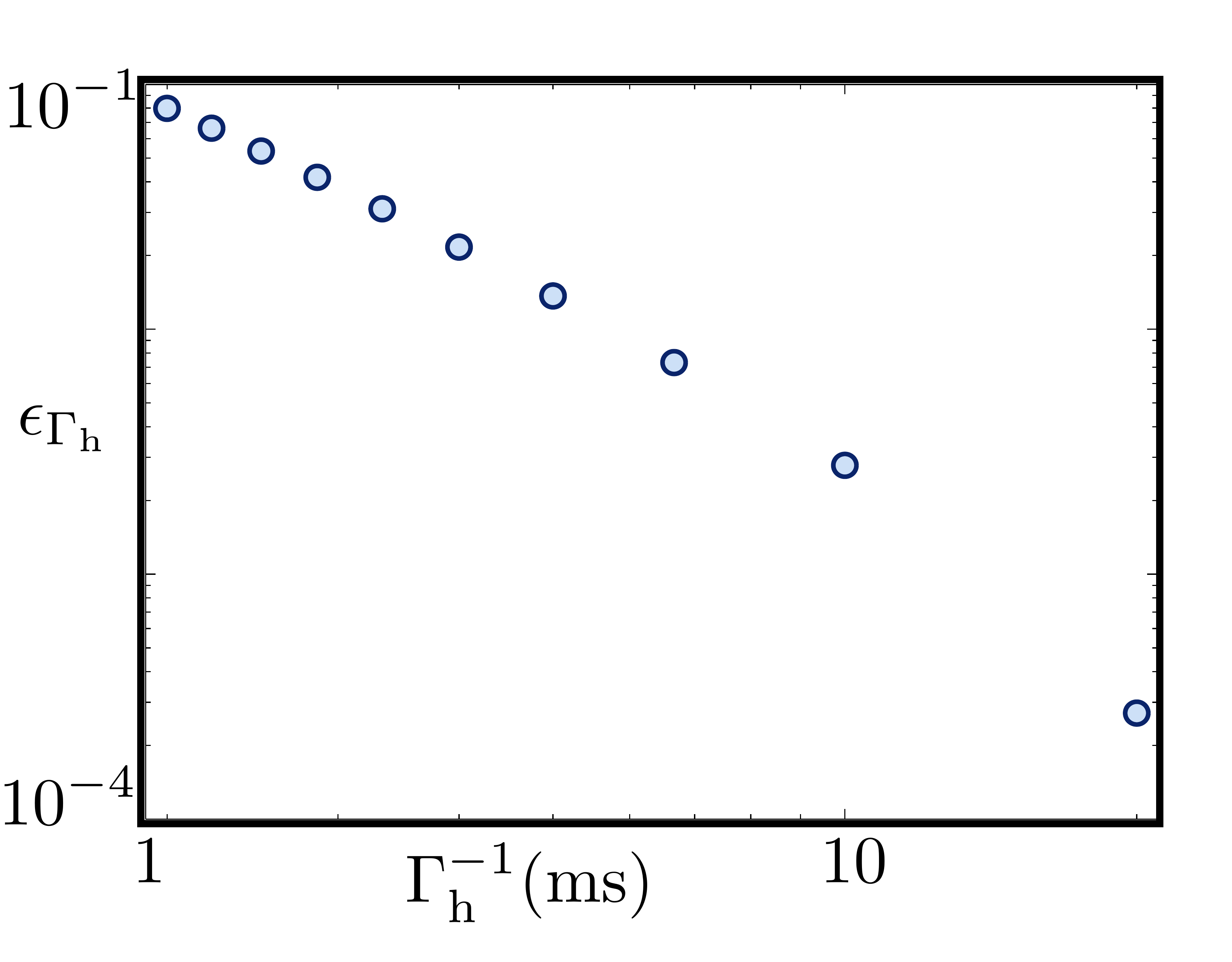}
\caption{ \textbf{Relative heating error}: Scaling of the relative heating error $\epsilon_{\Gamma_{\rm h}}$ with the inverse of the heating rate $\Gamma_{\rm h}^{-1}$ (logarithmic scale). }
\label{heating_error_fig}
\end{figure}

Another important question to address is the error caused by heating mechanisms. The heating of a particular vibrational mode may be induced by a variety of factors, such as the combination of stray electric fields and fluctuating trap parameters, elastic collisions with a background gas,  fluctuating patch fields in the trap electrodes, or non-linear static electric fields (see~\cite{ion_trap_reviews} for details). Therefore, we shall not focus on a particular microscopic model, but use instead a  phenomenological master equation 
\begin{equation}
\frac{{\rm d}\rho}{{\rm d}t}=-\ii[\tilde{H}_{\rm p}+\tilde{H}_{\rm d},\rho(t)]+\mathcal{D}_{\rm h}(\rho(t)),
\end{equation}
where the coherent dynamics is given by Eq.~\eqref{spin_phonon}, and the heating dissipator corresponds to
\begin{equation}
\label{heating_dissipator}
\mathcal{D}_{\rm h}(\rho)=\sum_{n}\Gamma_{\rm h}\left(a_n^{\dagger}\rho a_n-\half a_na_n^{\dagger}\rho-\half \rho a_na_n^{\dagger}\right).
\end{equation}
Here, we have considered that the heating rate $\Gamma_{\rm h}$ is equal for all the vibrational modes. In addition to Eq.~\eqref{relative_error}, the heating mechanism provides another source of error. In order to single out such a contribution,   we consider an initially ground-state-cooled ion crystal, and integrate numerically the above master equation. To estimate the relative error caused by the heating mechanism, we evaluate the following figure of merit
  \begin{equation}
 \label{relative_error_heating}
\epsilon_{{\rm h}}=\frac{|\langle\sigma_i^x\rangle_{\Gamma_{\rm h}}-\langle\sigma_i^x\rangle_{\Gamma_{\rm h}=0}|}{|\langle\sigma_i^x\rangle_{\Gamma_{\rm h}=0}|}.
 \end{equation} 
  By considering a small timescale $t_{\rm f}\ll(\Gamma_{\rm h})^{-1}$, the evolution of the mean number of phonons due to the dissipator~\eqref{heating_dissipator} yields a simple linear heating $\bar{n}^{\bot}_n(t)=\langle a_n^{\dagger}a_n\rangle\approx \Gamma_{\rm h}t$. We set $t_{\rm f}=1\pi/(8J_{\rm eff})$, in such a way that $\bar{n}^{\bot}_n(t)\ll\bar{n}_{\rm t}=2$. The remaining parameters are the same as above. In Fig.~\ref{heating_error_fig}, we represent the relative error~\eqref{relative_error_heating} as a function of the inverse heating rate $\Gamma_{\rm h}^{-1}$, which sets the timescale for the creation of one vibrational excitation. As can be seen in this figure, heating times above 5 ms/ phonon, only have a small contribution (1\%) to the overall error of the QS.



\begin{references}
\bibitem{note_optical}
For optical ions, such as $^{40}{\rm Ca}^+$, it is possible to use a single-photon scheme and implement the spin couplings via a pair of sidebands according to the M\o lmer-S\o rensen scheme. However, once the direction of the beams is fixed, the ratio of their wavelength to the mutual ion distance, which determines the anisotropy of the spin couplings, cannot be modified further.
\bibitem{comment}
 The canonical transformation in Eq.~\eqref{lang_firsov} resembles the so-called Lang-Firsov transformation customary of polaron physics (see~\cite{polaron_book} and references therein). This spin-dependent
displacement was first used in the field of paramagnetic salts with impurities~\cite{kanamori} (see also~\cite{elliot}). In the context of trapped ions, it has been exploited for quantum-logic~\cite{spin_displacement_wunderlich}, and quantum simulations purposes~\cite{ising_porras}.
\bibitem{note_j1j2}
  The full phase diagram of the $J_1$-$J_2$QIM is spanned by the ratios $J_2/J_1\in(-\infty,\infty)$, and $h/J_1\in(-\infty,\infty)$. However, in the absence of a longitudinal field, it suffices to restrict to $(J_2/J_1,h/J_1)\in(-\infty,0]\times(-\infty,0]$, such that the model consists of competing nearest-neighbor ferromagnetic interactions $J_1<0$, and next-to-nearest neighbor antiferromagnetic ones $J_2>0$. The magnetic phases with $J_1>0$ can be obtained by a $\mathbb{Z}_2$ inversion that only involves next-to-nearest neighboring spins.
\end{references}
\end{document}